\documentclass[UTF8]{article}
\usepackage{amsmath}
\usepackage{graphicx}
\usepackage{float}

\usepackage[left=2cm,right=2cm,top=3cm,bottom=3cm]{geometry}

\usepackage{a4wide}
\usepackage{amsmath}
\usepackage[
      colorlinks=true,
      linkcolor=blue,
      urlcolor=blue,
      filecolor=black,
      citecolor=blue,
            ]{hyperref}
\usepackage{color}
\usepackage{amsfonts}
\usepackage{amssymb}
\usepackage{epstopdf}

\begin{document}

\title{Magnetotransport  and Complexity of Holographic Metal-Insulator Transitions}
\author{\large
Yu-Sen An$^{a,b}$\footnote{anyusen@itp.ac.cn}~,
~~Teng Ji$^{a,b}$\footnote{jiteng@itp.ac.cn}~,
~~Li Li$^{a,b,c}$\footnote{liliphy@itp.ac.cn (corresponding author)}~,
\\
\\
\small $^a$CAS Key Laboratory of Theoretical Physics, Institute of Theoretical Physics, \\
\small Chinese Academy of Sciences, Beijing 100190, China\\
\small $^b$School of Physical Sciences, University of Chinese Academy of Sciences, \\
\small No.19A Yuquan Road, Beijing 100049, China\\
\small $^c$School of Fundamental Physics and Mathematical Sciences, Hangzhou Institute for Advanced Study, \\\
\small UCAS, Hangzhou 310024, China
}

\maketitle
\begin{abstract}
We study the magnetotransport in a minimal holographic setup of a metal-insulator transition in two spatial dimensions. Some generic features are obtained without referring to the non-linear details of the holographic theory. The temperature dependence of resistivity is found to be well scaled with a single parameter $T_0$, which approaches zero at some critical charge density $\rho_c$, and increases as a power law $T_0\sim|\rho-\rho_c|^{1/2}$ both in metallic $(\rho>\rho_c)$ and insulating $(\rho<\rho_c)$ regions in the vicinity of the transition. Similar features also happen by changing the disorder strength as well as magnetic field. By requiring a positive definite longitudinal conductivity in the presence of an applied magnetic field restricts the allowed parameter space of theory parameters. 
We explicitly check the consistency of parameter range for two representative models, and compute the optical conductivities for both metallic and insulating phases, from which a disorder-induced transfer of spectral weight from low to high energies is manifest. We construct the phase diagram in terms of temperature and disorder strength. The complexity during the transition is studied and is found to be not a good probe to the metal-insulator transition.
\end{abstract}

\newpage
\tableofcontents

\section{Introduction}

The mechanism of metal-insulator transition is one of the oldest, yet one of the fundamentally least understood problems in condensed matter physics. For a period of time in the past it was prevailingly believed that there is even no metal-insulator phase transition in two spatial dimensional systems in zero magnetic field, since all charge carriers are thought to be localized in an infinity large two-dimensional system~\cite{Abrahams:2000}. While it is an important problem, to understand the metal-insulator transition is difficult both from the practical and the conceptual points of view. It is obvious that a good metal and a good insulator are very different physical systems, and can be characterized by quite different elementary excitations. In particular, in the intermediate regime of the metal-insulator transition, there coexist different types of excitations and simple theoretical tools prove of little help. The research of metal-insulator transition came to the strong correlation era since the discovery of high temperature superconductivity. It is believed that strong correlation physics dominates, and physical pictures based on weak-coupling approaches prove insufficient or even misleading. Many theories have been proposed to understand the metal-insulator transition, such as phenomenological scaling hypothesis formulated for quantum criticality, scaling theories of disorder-driven transitions and order-parameter approaches to interaction-localization. Nevertheless, mechanisms toward the metal-insulator transition remain controversial and somewhat incomplete, see~\cite{Imada:1998,MIT:2011} for reviews for a general introduction to metal-insulator transitions.

On the other hand, holography has been providing towards the study of fascinating phenomena in strongly correlated systems~\cite{Jan:book,Ammon:book}. It has seen an increasing interest in applying the techniques of holography to probe the rich structure of strongly coupled quantum phases of matter, see \emph{e.g.}~\cite{Cai:2015cya,Hartnoll:2016apf,Landsteiner:2019kxb,Baggioli:2019rrs} for reviews in the context of condensed matter applications. Of particular interest are the transport properties as a function of temperature and other parameters, such as charge density and magnetic field. Holography provides a framework to deal with states of quantum matter without quasiparticle excitations, for which the transport properties deviate strongly from conventional approach described by Fermi liquid theory.\,\footnote{Holographic realization for the anomalous scalings of strange metals can be found \emph{e.g.} in~\cite{Kim:2010zq,Amoretti:2015gna,Kiritsis:2016cpm,Ge:2016sel,Blauvelt:2017koq,Cremonini:2018kla}.} In order to remove the unphysical divergence of transport due to translational invariance at finite charge density,  a crucial ingredient is to introduce the momentum dissipation. While there are many ways of introducing momentum relaxation in holography, the simplest and most convenient one is the so-called linear axion models~\cite{Andrade:2013gsa}, where translations are broken by a linear source that preserves a combination of translations and shift symmetry. Some results obtained from this mean field approach have been found to be qualitatively similar to those following from more generic mechanisms of momentum relaxation.\footnote{Despite a lot of work on linear axion models and generalizations, see \emph{e.g.}~\cite{Seo:2016vks,Amoretti:2018tzw,Baggioli:2014roa,Cremonini:2017usb,Donos:2019tmo,Baggioli:2019abx,Amoretti:2019kuf,Baggioli:2020edn}, the physical nature of the dual field theories is not well understood yet~\cite{Ammon:2019apj,Baggioli:2020nay}.}

In the spirit of effective field theory, a minimal holographic model of a disorder-driven metal-insulator transition was proposed in~\cite{Baggioli:2016oqk}. Such effective holographic theory describes the low energy physics of dual field theory that involves only two sectors: a charge current and a translation symmetry breaking sector that mimics the effects of disorder. After including a direct coupling between the charge sector and the translation breaking sector, the transport shows some interesting features. In particular, the DC electrical conductivity does not obey any lower bound and there is a disorder-driven metal-insulator transition at zero magnetic field. The authors of~\cite{Baggioli:2016oqk} considered a special model and studied its electric response in absence of magnetic field. Given the rich phenomenological features of this holographic setup, it is worth understanding the theory further and uncovering new transport behaviors. An important outcome of~\cite{Baggioli:2016oqk} is that the new cross-coupling must have a negative slop in order to obtain a metal-insulator transition. While it was checked explicitly by considering a simple case, a deeper understanding of this issue is necessary. To have a well-defined effective theory, one should concern the conditions under which a model is consistent and free from pathologies. The analysis of consistency was done in the decoupling limit with the metric kept frozen and in zero magnetic field. A further check on the constraints on the couplings is necessary. What's more, as a hallmark of a phase transition, the scaling of an appropriate physical observable in this metal-insulator transition has not been disclosed.

In this work we will address some issues mentioned above and will extend the study of the holographic metal-insulator transition by including a non-trivial background magnetic field. Firstly, magnetic field is a natural and experimentally relevant knob. It remains a challenge to understand how the continuum of quantum critical excitations responds to it. It is clear that a magnetic field will introduce qualitatively new features into transport. In particular, Hall conductivities are now allowed due to the breaking of time reversal invariance by the magnetic field. Secondly, unitarity as well as the second law of thermodynamics enforces the matrix of conductivities to be positive definite, which in turn requires the longitudinal conductivity to be non-negative. Considering the magnetotransport can give a non-trivial test on the constraints on the couplings proposed in~\cite{Baggioli:2016oqk}, or otherwise might impose further constraints. As we will show in the present study, it is indeed a more stringent restriction on the theory parameters for the model studied in~\cite{Baggioli:2016oqk}.\footnote{In the absence of magnetic field, the constraints proposed in~\cite{Baggioli:2016oqk} gaurantee that the electric DC conductivity is positive definite.} Last but not least, it is definitely interesting to see more the phenomenology of this transition and its generic features. We would like to uncover some universal behaviors during the metal-insulator transition and try to examine possible probes that can be used to characterize this transition. Indeed, we will understand why the new coupling plays the key role in the metal-insulator transition and will find universal scaling features near the phase transition. While it was called disorder-driven metal-insulator transition in~\cite{Baggioli:2016oqk}, we will show that the transition can also be driven by dialing the charge density and magnetic field.

The plan of this paper is as follows. In Section~\ref{Sec:Model}, we introduce the holographic theory and find the general dyonic black brane solutions analytically. We will discuss the constraints on the coupling functions of the theory. In Section~\ref{Sec:Transport} we present the main phenomenological features of transport behaviors. We will obtain the DC magnetotransport in terms of horizon data and then will discuss some generic features. The DC transport for two representative models will be studied in more details. In Section~\ref{Sec:optical} we study the optical conductivity for various phases and construct the temperature-disorder phase diagrams. We will check the behaviors of specific heat and charge susceptibility during the metal-insulator transition. Section~\ref{Sec:Complexity} is devoted to examining the behavior of complexity in the metal-insulator transition. We conclude with further discussions in Section~\ref{Sec:Conclusion} and with some technical details in Appendix~\ref{Sec:App}.

\section{Holographic Setup}
\label{Sec:Model}
We introduce the minimal holographic theory of metal-insulator transition. The action takes the form~\cite{Baggioli:2016oqk}
\begin{equation}\label{action}
\mathbf{\mathcal{S}}=\int d^4 x \sqrt{-g}\left[\frac{1}{2\kappa_N^2}(R-2\Lambda)-\frac{1}{4e^2}Y(X) F_{\mu\nu}F^{\mu\nu}-m^2 V(X)\right]\,,
\end{equation}
where $\kappa_N$ is the gravitational constant connected with $G_N$ by the relation $2\kappa_N^2=16\pi G_N$, 
$e$ stands for the U(1) charge that represents the unit of charge of the charge carriers, and $\Lambda$ describes the cosmological constant.  The charge degrees of freedom are encoded in the U(1) gauge field $A_\mu$ with its strength $F_{\mu\nu}=\partial_\mu A_\nu-\partial_\nu A_\mu$. $Y$ and $V$ are functions of the translation breaking sector $X=\sum_{I=1}^2\frac{1}{2}g^{\mu\nu}\partial_\mu\phi^I\partial_\nu\phi^I$. 

The disorder deformation is introduced  by the St\"{u}ckelberg field profiles $\phi^I\propto x^I$ with $x^I$ the spatial coordinate of the system, which captures an averaged description of disorder coming from a homogeneously distributed set of impurities. From an effective field theory point of view, such setup provides a holographic metal-insulator transition as minimal as possible, because there are no more dynamical ingredients in the dual field theory other than the translation breaking and the charge sector. While $V(X)$ does not couple directly to the charge carriers, $Y(X)$ captures the effects from charged impurities and is important to realize a metal-insulator transition. Along similar lines, more general models were introduced in~\cite{Gouteraux:2016wxj,Baggioli:2016pia}, where higher derivative corrections to an effective holographic action of homogeneous disorder have been discussed in absence of magnetic field.

The equations of motion for the theory~\eqref{action} are given by:
\begin{align}
\nabla_\mu \left[\left(\frac{Y'(X)}{4e^2}F_{\mu\nu}F^{\mu\nu} +m^2 V'(X)\right)\nabla^\mu\phi^I\right]&=0\,,\\
\label{eommax}\nabla_\mu [Y(X) F^{\mu\nu}]&=0\,,\\
\label{eomgravity}R_{\mu\nu}-\frac{1}{2}(R-2\Lambda)g_{\mu\nu}&=2\kappa_N^2\, T_{\mu\nu}\,,
\end{align}
with $T_{\mu\nu}$ the stress tensor
\begin{align}
T_{\mu\nu}=&-\frac{1}{2}g_{\mu\nu}\left[\frac{Y(X)}{4e^2} F_{\mu\nu}F^{\mu\nu}+m^2V(X)\right]\notag\\
&+\frac{1}{2}\left[\frac{Y'(X)}{4e^2} F_{\mu\nu}F^{\mu\nu}+m^2V'(X)\right]\sum_{I=1}^2(\partial_\mu\phi^I\partial_\nu\phi^I)+\frac{Y(X)}{2e^2} F_{\mu\rho}{F_\nu}^\rho\,.
\end{align}
Those equations of motion admit asymptotically AdS dyonic black brane solutions that are given by
\begin{align}\label{solution}
&d s^{2}=\frac{1}{u^{2}}\left[-f(u) d t^{2}+\frac{1}{f(u)} d u^{2}+d x^{2}+d y^{2}\right]\,,\notag\\
&A_\mu dx^\mu=A_{t}(u)\,dt+\frac{1}{2}B(x\,dy-y\,dx)\,,\notag\\
&\phi^I=\alpha \,x^I \ \ \ \ (x^1=x, x^2=y)\,,\\
&f(u)=u^3 \int_{u_h}^u \frac{1}{2} \left(\frac{B^2 \kappa_N^2 Y\left(\alpha ^2 \xi ^2\right)}{e^2}+\frac{\kappa_N^2 \rho ^2e^2}{
   Y\left(\alpha ^2 \xi ^2\right)}-\frac{6}{L^2\xi ^4}+\frac{2 \kappa_N^2 m^2 V\left(\alpha ^2 \xi ^2\right)}{\xi ^4}\right) \,d\xi\,,\notag\\
&A_t(u)=e^2\rho \int_{u}^{u_{h}} \frac{1}{Y\left(\xi^{2} \alpha^{2}\right)} d \xi\,, \notag
\end{align}
where we have taken the cosmological constant to be $\Lambda=-3/L^2$ with $L$ the AdS radius and $u_h$ stands for the location of the event horizon. $\rho$ and $B$ are the charge density and the magnetic field, respectively. The temperature of the background geometry is given by
\begin{align}\label{tem}
T=-\frac{f'(u_h)}{4\pi}=\frac{3}{4 \pi  L^2 u_h}-\frac{B^2 \kappa_N^2 u_h^3 Y\left(\alpha ^2 u_h^2\right)}{8 \pi  e^2}-\frac{\kappa_N^2 \rho ^2 e^2 u_h^3}{8 \pi Y \left(\alpha ^2 u_h^2\right)}-\frac{\kappa_N^2 m^2 V\left(\alpha ^2 u_h^2\right)}{4 \pi  u_h}\,.
\end{align}

The consistency of a theory imposes some constraints on the couplings that appear in the Lagrangian. For the present theory~\eqref{action}, it has been argued that $V(X)$ and $Y(X)$ should satisfy the following constraints~\cite{Baggioli:2016oqk}:
\begin{align}\label{oldconstraint}
V^{\prime}(X)>0, \quad Y(X)>0, \quad Y^{\prime}(X)<0\,.
\end{align}
It has been shown that $Y'(X)<0$ plays a key role in triggering a metal-insulator transition in the absence of a magnetic field.\,\footnote{See~\cite{Donos:2012js,Mefford:2014gia,Rangamani:2015hka,Baggioli:2016oju,Ling:2014saa,Cremonini:2017qwq,Andrade:2017cnc} for other holographic realizations of metal-insulator transitions driven via other mechanisms.} 
In the following discussion, we will fix the gravitational constant, the charge unit and the AdS radius to one, \emph{i.e.} $\kappa_N=e=L=1$.

\section{Magnetotransport}\label{Sec:Transport}
Having setup the holographic system, we study the transport properties in the presence of a magnetic field in this section. We obtain the analytic result for the electric DC conductivity and resistivity in terms of horizon data and discuss some generic features without being concerned with details of the holographic theory.  We will show that there is a non-trivial restriction on $Y(X)$ in order to avoid a negative diagonal conductivity in the magnetic field. We can also uncover some universal scaling behaviors near the phase transition.

\subsection{DC transport and constraint}\label{sub:const}

We now focus on the magnetotransport in the theory~\eqref{action}, for which an explicit formula in terms of horizon data can be obtained by following the method developed in~\cite{Donos:2014uba,Blake:2014yla}. In the presence of a magnetic field, the DC conductivity is described by a two dimensional matrix $\sigma_{ij}$ with the longitudinal and Hall components given by~\footnote{The derivation of this result is presented in Appendix~\ref{Sec:App}.}
\begin{align}
\label{sigxx}\sigma_{x x}=&\sigma_{y y}=\frac{\Omega Y[\Omega+ Y(B^2 Y^2+\rho^2)u_h^2]}{(\Omega+B^2Y^3 u_h^2)^2+B^2\rho^2Y^4 u_h^4}\,,\\
\label{sigxy}\sigma_{x y}=&-\sigma_{y x}=\frac{B\rho Y^3 u_h^2[2\Omega+ Y(B^2 Y^2+\rho^2)u_h^2]}{(\Omega+B^2Y^3 u_h^2)^2+B^2\rho^2Y^4 u_h^4}\,,\\
\Omega=&\alpha^2[m^2V' Y^2+\frac{u_h^4}{2}(B^2Y^2-\rho^2)Y']\,.\nonumber
\end{align}
The resistivity matrix $R_{ij}$ is obtained by inverting the conductivity matrix $\sigma_{ij}$:
\begin{align}
\label{omexx}R_{x x}=&R_{yy}=\frac{\sigma_{xx}}{\sigma_{xx}^2+\sigma_{yy}^2}=\frac{\Omega [\Omega+ Y(B^2 Y^2+\rho^2)u_h^2]}{Y
  [\left(\Omega+\rho^2Y u_h^2\right)^2+B^2 \rho ^2Y^4 u_h^4]}\,,\\
\label{omexy}R_{x y}=&-R_{y x}=-\frac{\sigma_{xy}}{\sigma_{xx}^2+\sigma_{yy}^2}=-\frac{B \rho Y u_h^2 [2\Omega+ Y(B^2 Y^2+\rho^2)u_h^2]}{\left(\Omega+\rho^2Y u_h^2
   \right)^2+B^2 \rho ^2Y^4 u_h^4}\,,
\end{align}
From now on all functions will be understood to be evaluated at the horizon $u=u_h$. One finds that $\sigma_{xx}$ and $\sigma_{xy}$ are controlled by two couplings $V$ and $Y$. Note that $u_h$ is in general a function of temperature $T$, as seen from~\eqref{tem}. So the magnetotransport above is implicitly temperature-dependent, while the dependence on the remaining scales in the system--the magnetic field $B$, the strength of disorder $\alpha$ as well as the charge density $\rho$--is explicitly visible. Notice that due to a scaling symmetry, only three of these four scales $(T, B, \alpha,\rho)$ are actually physical.

As a consistency check, we first turn off the magnetic field and then the DC conductivity reduces to a simple expression
\begin{equation}\label{noBdc}
\sigma_{DC}\equiv\sigma_{xx}(B=0)=Y(\alpha^2 u_h^2)+\frac{\rho^2 u_h^2}{\alpha^2\left(m^2 V'(\alpha^2 u_h^2)-\frac{\rho^2 Y'(\alpha^2 u_h^2) u_h^4}{2Y(\alpha^2 u_h^2)^2}\right)}\,,
\end{equation}
which precisely recovers the vanishing magnetic field result in~\cite{Baggioli:2016oqk}. Note that in absence of a magnetic field, $\sigma_{xy}=R_{xy}=0$ as a consequence of parity symmetry, and $R_{xx}=1/\sigma_{DC}$.

Another interesting case is to consider the clean limit by taking $\alpha\rightarrow 0$, which corresponds to no disorder or momentum dissipation at all.  For this limit in which the strength of disorder $\alpha$ is much smaller than any other scales $(T, \sqrt{B},\sqrt{\rho})$, the dimensionless quantity $X=u_h^2\alpha^2$ is a small quantity. Without loss of generality, we parametrize the couplings $Y$ and $V$ in the following expansion as $X\rightarrow 0$: 
\begin{equation}\label{YVexpansion}
Y(X)=1-k\,X+\mathcal{O}(X^2),\quad V(X)=\frac{1}{2 m^2}X+\mathcal{O}(X^2)\,,
\end{equation}
where $k\geqslant 0$ from the requirement~\eqref{oldconstraint}.\footnote{We point out that $Y'(X)$ can be zero. The most simple case is that $Y$ is a constant, for which there is no direct coupling between the charge sector and the translation breaking sector.} One immediately obtains from~\eqref{sigxx} and~\eqref{sigxy} that
\begin{equation}\label{clean}
\sigma_{xx}=0,\quad \sigma_{xy}=\frac{\rho}{B}\,,
\end{equation}
which are independent of the temperature as well as the details of the theory we are considering. This universal feature can be understood as a generic consequence of Lorentz invariance when $\alpha\rightarrow 0$, and can also be obtained from relativistic hydrodynamics. Note also that the conductivities do not divergent in the clean limit, which is due to a Lorentz force on the system that violates momentum conservation.

Including the leading correction coming from momentum dissipation, we find that the diagonal component of conductivity is given by
\begin{equation}\label{small alpha}
\sigma_{xx}=\frac{u_h^2}{2}[-k+B^{-2}(u_h^{-4}+\rho^2 k)]\alpha^2+\mathcal{O}(\alpha^4)\,.
\end{equation}
Since $\sigma_{xx}$ should not be negative, we must have 
\begin{equation}
-k+B^{-2}(u_h^{-4}+\rho^2 k)\geqslant 0\Rightarrow u_h^{-4}\geqslant(B^2-\rho^2)k\,,
\end{equation}
no matter the value of $B$ and $\rho$ are. Meanwhile, one finds from~\eqref{tem} that
\begin{equation}
T=\frac{3}{4\pi u_h}-\frac{(B^2+\rho^2)u_h^3}{8\pi}+\mathcal{O}(\alpha^2)\,,
\end{equation}
and the requirement of $T\geqslant 0$ demands
\begin{equation}
\frac{3}{4\pi u_h}-\frac{(B^2+\rho^2)u_h^3}{8\pi}\geqslant 0 \Rightarrow u_h^{-4}\geqslant \frac{B^2+\rho^2}{6}\,.
\end{equation}
Therefore, there is no regime for a negative longitudinal conductivity provided the following inequality is satisfied for a general choice of $B$ and $\rho$.
\begin{equation}\label{constraint}
\frac{B^2+\rho^2}{6}\geqslant(B^2-\rho^2)k\,,
\end{equation}
which is guaranteed only for~\footnote{When talking about pseudo-spontaneous breaking of translational invariance and holographic pinning mechanism (see \emph{e.g.}~\cite{Baggioli:2019abx,Baggioli:2020edn}), it is convenient to keep $m^2$ explicitly in the action~\eqref{action} and a typical choice of graviton mass term $V(X)$ is given by
\begin{equation}
V(X)=\frac{1}{2}X+\frac{\beta}{m^2}X^n,\quad n>5/2 \nonumber\,.
\end{equation}
Using a simple scaling argument, it is easy to show that for the above case the allowed parameter space becomes $0\leqslant k\leqslant m^2/6$ ($-m^2/6\leqslant Y'(0)\leqslant 0$) instead of~\eqref{constk}. So it should be careful if one wants to consider pseudo-spontaneous regime where $m^2\ll 1$ and $\beta\gg m^2$ for a non-trivial $Y(X)$.}
\begin{equation}\label{constk}
0\leqslant k\leqslant1/6\quad\Rightarrow -1/6\leqslant Y'(0)\leqslant 0\,.
\end{equation}
%

The key message is that by taking into account magnetic field, we are able to give a generic constraint on $k$, without referring to the non-linear details of the coupling functions $Y$ and $V$.\,\footnote{In contrast, for the case without magnetic field, the inequality~\eqref{constraint} is satisfied automatically since $k\geqslant 0$. Thus one is not able to obtain any constraint from~\eqref{constraint} when $B=0$.}  To obtain the constraint~\eqref{constk}, we have only considered a very generic expansion~\eqref{YVexpansion} and demanded $\sigma_{xx}\geqslant0$ in the weak disorder limit.  As one can see from~\eqref{sigxx} and~\eqref{sigxy} that the behaviors of the magnetotransport depend on the details of couplings in a theory. Although we have obtained a strong constraint on $k$~\eqref{constk} from the weak disorder analysis, it does not guarantee that $\sigma_{xx}\geqslant 0$ away from the weak disorder regime. We still need to check explicitly whether the longitudinal conductivity is positive definite or not for a given model. 

\subsection{Metal-insulator transition and scaling behavior}\label{sub:scaling}

Before proceeding to specific examples, let's understand why a metal-insulator transition or crossover can be triggered in the present setup. We adopt the following working and phenomenological definition of a metal versus an insulator behavior:
\begin{equation}\label{definition}
\text{metal:}\quad \frac{d\,R_{xx}}{d\, T}>0,\quad \text{insulator:}\quad \frac{d\,R_{xx}}{d\, T}<0\,,
\end{equation}
and inspect the temperature dependence of $R_{xx}$ and $\sigma_{xx}$.\,\footnote{A much strict definition of an insulator would be $\sigma(T=0)=0$. In the present paper we adopt a more realistic and phenomenological definition for an insulator in~\eqref{definition} typically used in the literature.} We now consider the high temperature limit where $T$ is the dominant scale in the problem. In this limit the temperature~\eqref{tem} at leading order is given by the simple expression
\begin{equation}\label{highTem}
T=\frac{3}{4\pi u_h}+\mathcal{O}(u_h^2)\,,
\end{equation}
and the corresponding diagonal resistivity and conductivity read
\begin{eqnarray}
R_{xx}=1-\frac{2\, u_h^2}{\alpha^2}(\rho^2-B^2-\frac{k}{2}\alpha^4)+\mathcal{O}(u_h^4)=1-\frac{9}{8\pi^2 \alpha^2}(\rho^2-B^2-\frac{k}{2}\alpha^4)T^{-2}+\mathcal{O}(T^{-4})\,,\label{highT}\\
\sigma_{xx}=1+\frac{2\, u_h^2}{\alpha^2}(\rho^2-B^2-\frac{k}{2}\alpha^4)+\mathcal{O}(u_h^4)=1+\frac{9}{8\pi^2 \alpha^2}(\rho^2-B^2-\frac{k}{2}\alpha^4)T^{-2}+\mathcal{O}(T^{-4})\,,
\end{eqnarray}
where we have used~\eqref{YVexpansion} as $u_h\rightarrow 0$ in the high temperature limit. It is then clear that there is a critical charge density $\rho_c=\sqrt{B^2+k \alpha^4/2}$. When $\rho<\rho_c$, $R_{xx}$ decreases monotonically with $T$ increased and displays insulating behavior, provided that the temperature is the largest scale in the problem. On the other hand, $R_{xx}$ increases with increasing $T$ when $\rho>\rho_c$, displaying metallic behavior. As a consequence, there is a metal-insulator transition by increasing the charge density at fixed magnetic field and disorder strength. On the other hand, the phase transition can also be triggered by increasing either the magnetic field $B$ or the disorder strength $\alpha$ at a fixed charge density. From~\eqref{highT} one can also understand why $Y'(X)<0$ ($k>0$) has a dramatic impact on the possibility to have a metal-insulator transition in the absence of magnetic field. 

After understanding the metal-insulator transition triggered by the charge density, magnetic field and disorder in our generic holographic setup, we would like to see the possible scaling behavior near the transition. In particular, we are interested in some robust features that could be potentially connected to experimental measurements. An interesting feature from~\eqref{highT} is that  the diagonal component scales with temperature at densities both below and above $\rho_c$. More precisely, $R_{xx}$ can be written in the form
\begin{eqnarray}\label{scaling}
R_{xx}\approx 1-\frac{9}{8\pi^2\alpha^2}(\rho^2-\rho_c^2)\frac{1}{T^2}=1\pm\frac{T_0^2(\rho)}{T^2}\,,
\end{eqnarray}
with $``+"$ for the insulating behavior and $``-"$ the metallic behavior. The scaling parameter $T_0$ is given by
\begin{equation}\label{scalingT}
T_0=\frac{3}{2\sqrt{2}\pi \alpha}|\rho^2-\rho_c^2|^{1/2}\,,
\end{equation}
and approaches zero at $\rho=\rho_c$. Therefore,  the $R_{xx}(T)$ curves for different $\rho$ can be made to overlap by the scaling parameter $T_0$ along the $T$ axis, which yields a collapse of the data onto two curves: an insulating branch for $\rho<\rho_c$ and a metallic branch for $\rho>\rho_c$.

Near the transition point, the dependence of the scaling parameter $T_0$ on the charge density is symmetric about $\rho_c$, and obeys a power law
\begin{equation}\label{scalingT0}
T_0=\frac{3\sqrt{\rho_c}}{2\pi \alpha}|\rho-\rho_c|^{1/2}=C\,|\rho-\rho_c|^{1/2}, \quad\quad C=\frac{3}{2\pi}\left(\frac{B^2}{\alpha^4}+\frac{k}{2}\right)^{1/4}\,,
\end{equation}
for both the insulating and metallic sides of transition. The power is exactly $1/2$ and is model independent, while the coefficient $C$ depends on the theory parameter $k$ and the state parameter $B^2/\alpha^4$. Interestingly, in the absence of magnetic field, $C$ is independent of the strength of disorder $\alpha$. Then, we have a quite robust scaling behavior with the power and the coefficient $C$ found to be essentially disorder independent.  It is also manifest that the metallic and insulating curves are mirror symmetry in the high temperature regime: $R_{xx}(\rho-\rho_c,T)=1/R_{xx}(\rho_c-\rho, T)$, suggesting that the mechanism responsible for electric transport in the insulating and metallic phases are related.

Actually, such kind of features have been reported in some two dimensional samples and materials~\cite{Kravchenko:1995,Kravchenko:1996,Popovic:1997,Coleridge:1997,Simmons:1998}. In particular, various experimental groups have demonstrated interesting scaling behaviors for resistivity near the transition point, which shows the collapse of data into two separated curves and displays remarkable mirror symmetry over a broad interval of temperatures. This observation has been interpreted as evidence that the transition
region is dominated by strong coupling effects characterizing the insulating phase~\cite{Dobrosavljevic:1997}.
Nevertheless, the dependence of $T_0$ near $\rho_c$ is a power law with  the power that is different from our holographic result $1/2$. A number of
experiments have yielded scaling exponents between 1.25 and 1.6~\cite{Kravchenko:1995,Kravchenko:1996,Popovic:1997,Coleridge:1997,Simmons:1998}. This might be due to the fact that our present holographic theory falls into a different universality class from those materials.
A natural extension is to consider a holographic setup that is asymptotically Lifshitz with a dynamical exponent $z$ which parametrizes the relative scaling of space and time. Then the power of the scaling behavior~\eqref{scalingT0} would be modified and could make the model compatible with experimental data. We will return to these points in our later discussion.

Another situation in experimental measurements is to fix the Landau filling factor $\nu=\rho/B$. From~\eqref{highT}, one observes that
\begin{equation}\label{landau}
R_{xx}=1-\frac{9 B^2}{8\pi^2 \alpha^2}\left(\nu^2-1-\frac{k}{2}\frac{\alpha^4}{B^2}\right)T^{-2}+\mathcal{O}(T^{-4})\,.
\end{equation}
Therefore, staying at the same filling factor $\nu>1$, one can observe a transition from the metallic state in the high $B$ case to the insulating state at lower $B$. This feature agrees qualitatively with the experimental observation in~\cite{Kravchenko:1995}.

So far we have discussed some generic features without being concerned with details of the holographic theory. We have a deeper understanding of the role of $Y'(X)<0$ in triggering a metal-insulator transition in the absence of magnetic field, and find a rigorous constraint on the theory parameter,~\eqref{constk}.  Some universal scaling behaviors near the phase transition are also examined. In particular, the transport is found to be scaled with a single parameter $T_0$ which approaches zero at the transition point.
In the following study we will consider two representative ``benchmark" models and will examine their temperature dependence in various cases. 

\subsection{Benchmark models}
We focus on the following two representative models:
\begin{align}
\label{emodel}\text{Exponential model}:\quad Y=e^{-\kappa\, X},\quad V(X)=\frac{1}{2m^2} X\,,\\
\label{lmodel}\text{Linear model:}\quad Y=1+\mathcal{K}\, X,\quad  V(X)=\frac{1}{2m^2} X\,.
\end{align}
Depending on the theory parameter $\kappa$ or $\mathcal{K}$, both models can describe either metallic or insulating phases.
The first model was studied in detail in~\cite{Baggioli:2016oqk}, where the electric transport properties were investigated and a disorder-driven metal-insulator transition was observed for $\kappa=0.5$. The linear model was introduced by the authors of~\cite{Gouteraux:2016wxj}, where the DC conductivity was discussed. It is able to have exactly vanishing conductivity at a finite value of $\mathcal{K}$, while in the first model the electric conductivity saturates to a small but finite value at large $\kappa$.

The constraints~\eqref{oldconstraint} imply that the theory parameter $\kappa$ of the Exponential model should not be negative, \emph{i.e.} $\kappa\geqslant 0$, and the requirement of no gradient instability imposes a further restriction on $\kappa$ ($\kappa\lesssim 0.5$ from Figure~6 of~\cite{Baggioli:2016oqk}). For the second model, $-1/6\leqslant\mathcal{K}\leqslant 0$ with the upper bound from~\eqref{oldconstraint} and the lower bound from  the zero density Schr\"{o}dinger potential analysis~\cite{Gouteraux:2016wxj}.
We should point out that for both models the allowed parameter space were examined in the absence of a magnetic field. Note that we have found a non-trivial constraint on $Y$ shown by~\eqref{constk}, from which one immediately obtains that $0\leqslant \kappa\leqslant1/6$ for the Exponential model~\eqref{emodel} and $-1/6\leqslant\mathcal{K}\leqslant 0$ for the Linear model~\eqref{lmodel}. Both are much more rigorous than the parameter space given in the original papers~\cite{Baggioli:2016oqk} and~\cite{Gouteraux:2016wxj}.\footnote{We point out that the authors of~\cite{Gouteraux:2016wxj} obtained a weaker constraint on $\mathcal{K}$, $-1/6\leqslant\mathcal{K}\leqslant1/6$. Notice that a positive $\mathcal{K}$ violates the requirement~\eqref{constraint}.} Interestingly, we will find that the new restriction~\eqref{constk} indeed imply a positive definite longitudinal conductivity for above two models. Nevertheless, whether there is a good insulating phase will depend on the non-linear details of the theory one is considering.

\subsubsection{The Exponential model}
For the Exponential model with $V$ and $Y$ given by~\eqref{emodel}, the background geometry reads
\begin{align}
&d s^{2}=\frac{1}{u^{2}}\left[-f(u) d t^{2}+\frac{1}{f(u)} d u^{2}+d x^{2}+d y^{2}\right]\,,\nonumber\\
&f(u)=\frac{ \sqrt{\pi} u^3}{4\alpha \sqrt{\kappa}}\left[B^{2} \left(\operatorname{erf}(\alpha \sqrt{\kappa} u)-\operatorname{erf}(\alpha \sqrt{\kappa} u_h)\right)+\rho^{2}(\operatorname{erfi}(\alpha \sqrt{\kappa} u)-\operatorname{erfi}(\alpha \sqrt{\kappa} u_h))\right]\,\nonumber,\\
&\qquad\qquad+\left(1-\frac{u^3}{u_h^3}\right)-\frac{1}{2}\alpha^2 u^3\left(\frac{1}{u}-\frac{1}{u_h}\right)\,,\\
&A_t(u)=\frac{\sqrt{\pi } \rho }{2 \alpha  \sqrt{\kappa }} \left(\text{erfi}\left(\alpha  \sqrt{\kappa } u_h\right)-\text{erfi}\left(\alpha  \sqrt{\kappa }
   u\right)\right)\,.\nonumber
\end{align}
We obtain the temperature
\begin{align}\label{etem}
T=\frac{3}{4 \pi  u_h}-\frac{\alpha^2 u_h}{8\pi}-\frac{B^2 u_h^3\, e^{-\kappa\alpha ^2 u_h^2}}{8 \pi }-\frac{\rho^2 u_h^3\, e^{\kappa\alpha^2 u_h^2}}{8\pi}\,,
\end{align} 
and the conductivities
\begin{align}
\label{esigxx}\sigma_{x x}=&\sigma_{y y}=\frac{\Omega  e^{-\kappa\alpha ^2 u_h^2}[\Omega+ e^{-\kappa\alpha ^2 u_h^2}(B^2 e^{-2\kappa\alpha ^2 u_h^2}+\rho^2)u_h^2]}{(\Omega+B^2 e^{-3\kappa\alpha ^2 u_h^2} u_h^2)^2+B^2\rho^2e^{-4\kappa\alpha ^2 u_h^2} u_h^4}\,,\\
\sigma_{x y}=&-\sigma_{y x}=\frac{B\rho e^{-3\kappa\alpha ^2 u_h^2} u_h^2[2\Omega+ e^{-\kappa\alpha ^2 u_h^2}(B^2 e^{-2\kappa\alpha ^2 u_h^2}+\rho^2)u_h^2]}{(\Omega+B^2e^{-3\kappa\alpha ^2 u_h^2} u_h^2)^2+B^2\rho^2e^{-4\kappa\alpha ^2 u_h^2} u_h^4}\,,\\
\Omega=&\frac{\alpha^2 e^{-\kappa\alpha ^2 u_h^2}}{2}[ e^{-\kappa\alpha ^2 u_h^2}-\kappa  u_h^4(B^2e^{-2\kappa\alpha ^2 u_h^2}-\rho^2)]\,.\nonumber
\end{align}
There are five parameters left in the DC transport: the temperature $T$, the charge density $\rho$, the magnetic field $B$, the disorder strength $\alpha$ and the parameter $\kappa$. 
We are interested in the behaviors of DC conductivities with respect to the disorder strength and magnetic field.

At first glance, the exponential form of coupling $Y=e^{-\kappa\, X}$ will not result in a negative conductivity, since $Y$ is positive definite for any choice of $X$. As one can see clearly, when $B=0$, the DC conductivity~\eqref{noBdc} for the Exponential model is indeed positive definite so long as $\kappa\geqslant 0$.
So, one expects that there would exist no negative longitudinal conductivity even in the presence of a magnetic field provided $\kappa$ is in the safe region given in~\cite{Baggioli:2016oqk}, \emph{i.e.} $0\leqslant\kappa\lesssim 0.5$. However, as we have argued in subsection~\ref{sub:const}, a negative longitudinal conductivity will appear when $\kappa>1/6$, and therefore $1/6<\kappa\lesssim0.5$ has to be excluded from the allowed parameter space. While the new constraint on $\kappa$ is obtained from the weak disorder analysis, to check it away from the weak disorder regime is necessary. 

\begin{figure}[H]
\centering
\includegraphics[width=0.48\linewidth]{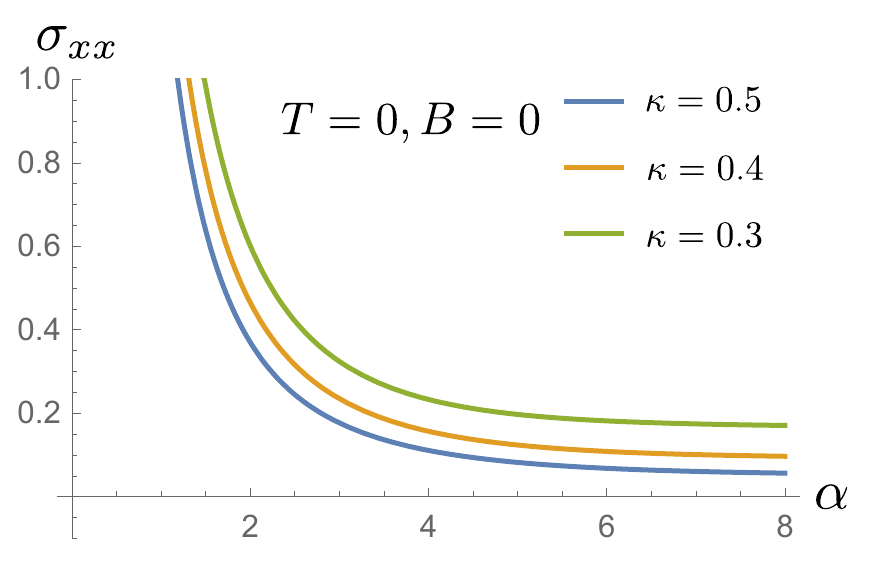}\quad
\includegraphics[width=0.48\linewidth]{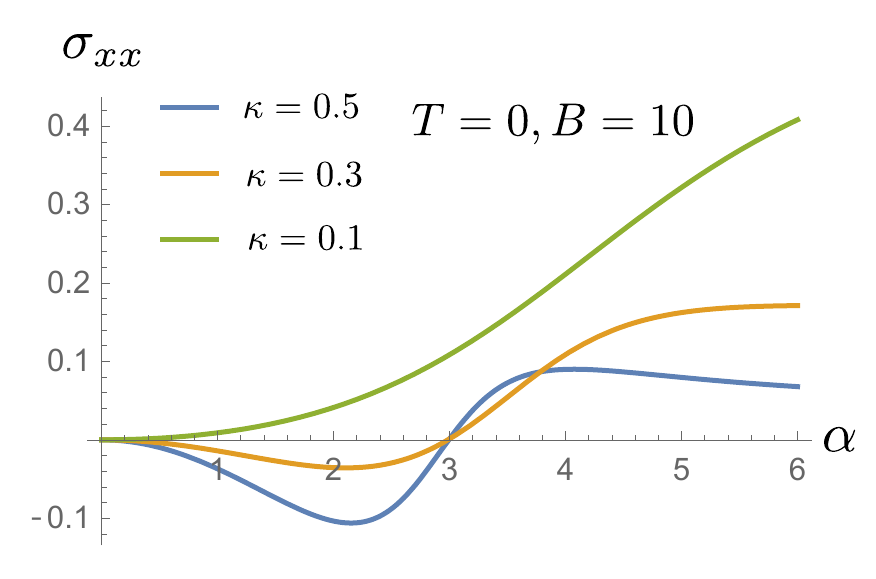}
\caption{\label{figevsalpha}Plots of the DC conductivity for the Exponential model without a magnetic field (left) and with a magnetic field (right). We fix $T=0$ and vary $\alpha$ for different choice of coupling constant $\kappa$. In the present of magnetic field, there is a region with negative $\sigma_{xx}$ developing when $\kappa$ is large but still lies in the healthy region proposed by~\cite{Baggioli:2016oqk}. We have worked in units with the charge density $\rho=1$.}
\end{figure}

\begin{figure}[H]\centering
\includegraphics[width=0.46\linewidth]{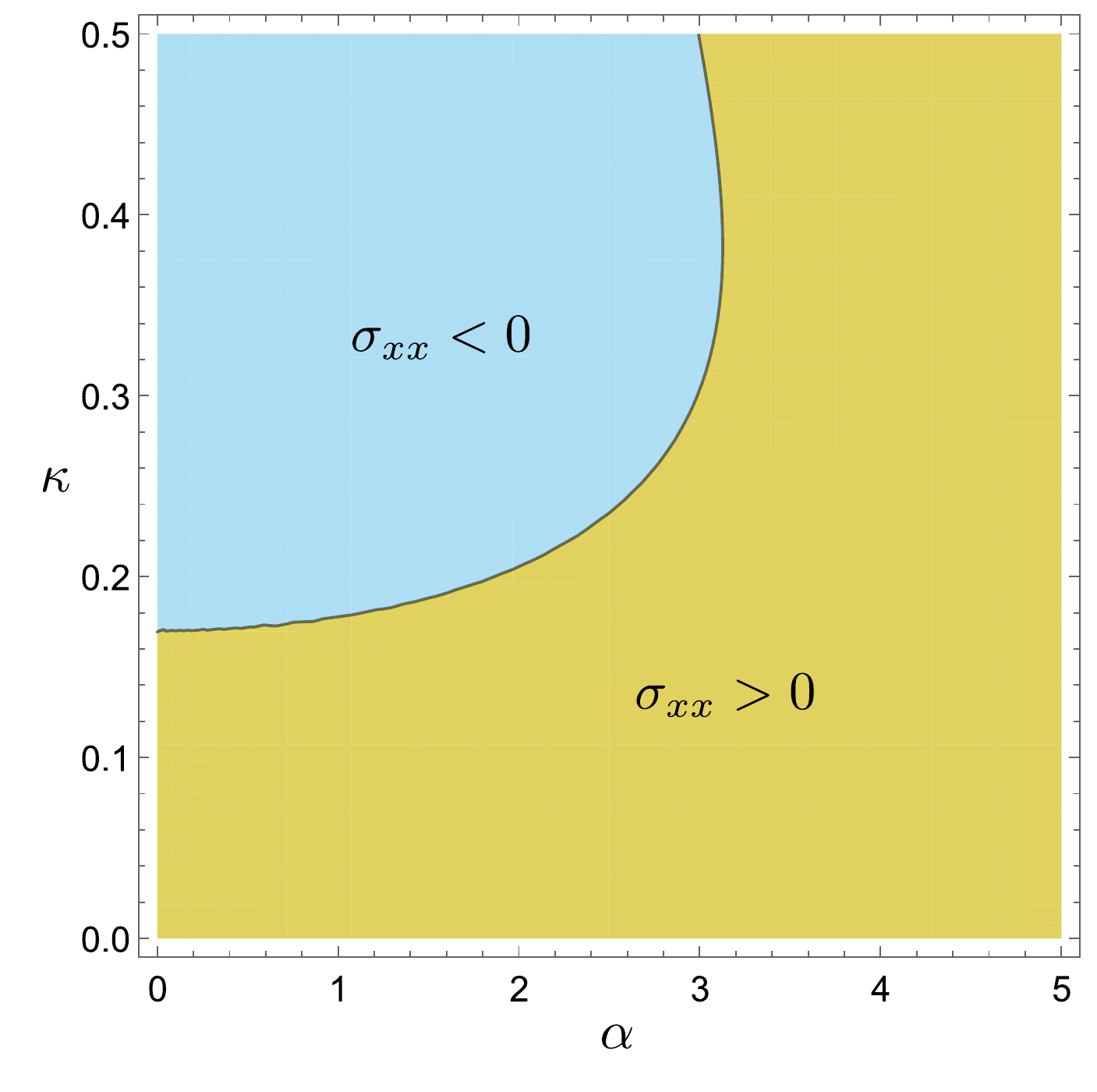}\quad\quad
\includegraphics[width=0.46\linewidth]{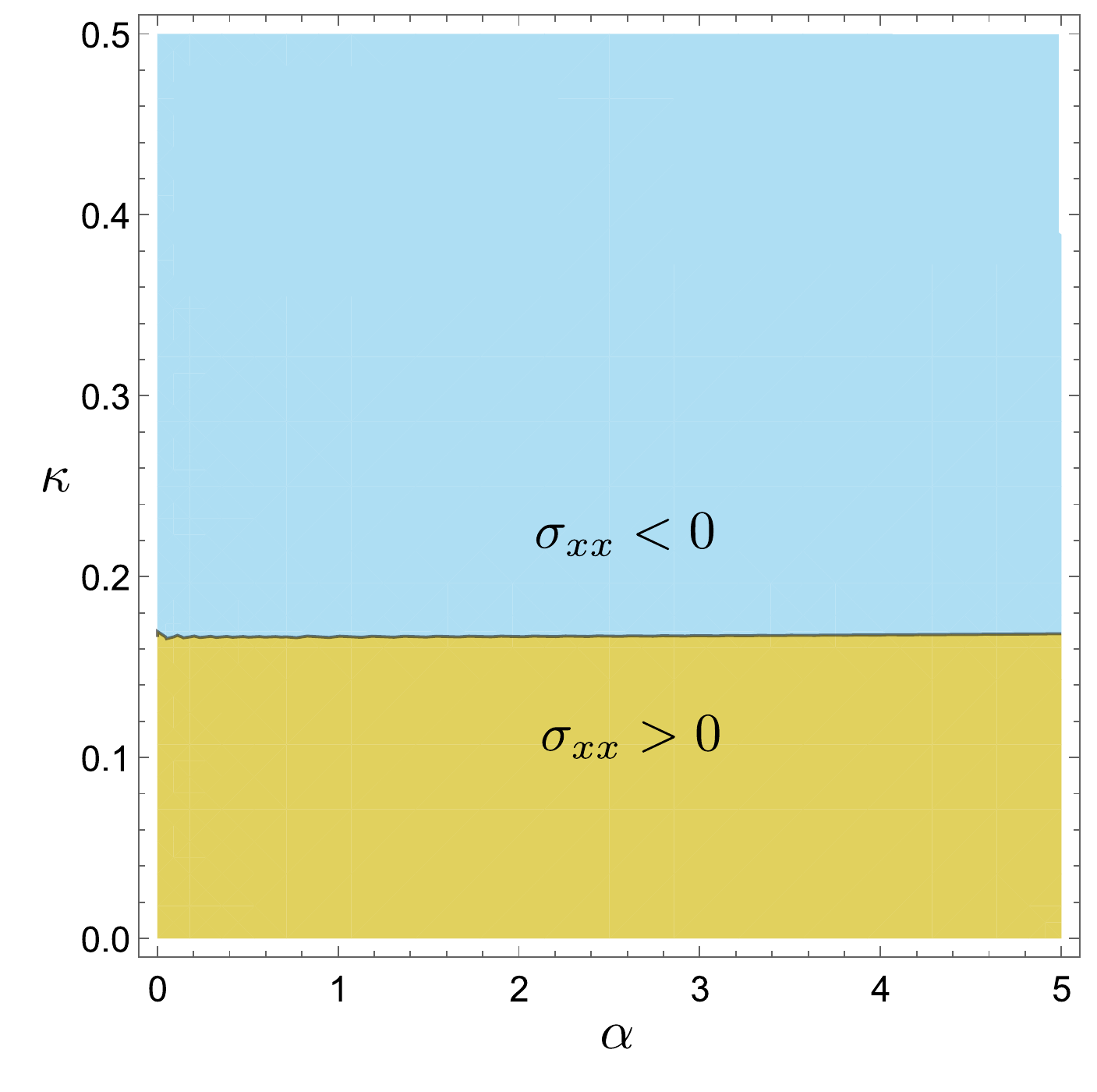}
\caption{\label{figevskappa} Distribution of the sign of $\sigma_{xx}$ as a function of the disorder strength $\alpha$ and the coupling constant $\kappa$ for the Exponential model. The left plot is for $(T=0, B=10)$ and the right one for $(T=0, B=1000)$. The blue regions denote negative conductivity and the yellow regions correspond to positive conductivity. To avoid a negative value of $\sigma_{xx}$, $\kappa$ should be smaller than $0.167$. We have worked in units with the charge density $\rho=1$.}
\end{figure}

We show $\sigma_{xx}$ as a function of the disorder strength $\alpha$ for different choices of $\kappa$ in ,\ref{figevsalpha}. As one can see, even within the safe region for $\kappa$ proposed in~\cite{Baggioli:2016oqk}, $\sigma_{xx}$ in a presence of magnetic field indeed becomes negative when $\kappa$ is large. In particular, the negative region first develops at small $\alpha$, which is consistent with our analysis from the weak disorder limit in subsection~\ref{sub:const}. Since the expression of $\sigma_{xx}$ in~\eqref{esigxx} is quite complicated, we are not able to fix the upper threshold value of $\kappa$ analytically. Nevertheless, as shown in ,\ref{figevskappa}, the upper limit of $\kappa$ read from our numerics is $\kappa\approx0.167$ which coincides quite well with the constraint~\eqref{constk} obtained from the weak disorder argument. Therefore, we obtain a new bound on $\kappa$ which satisfies all the consistency conditions and guarantees a non-negative longitudinal conductivity:
\begin{equation}\label{kappabd}
0\leqslant\kappa\leqslant1/6\,.
\end{equation}

\begin{figure}[H]\centering
\includegraphics[width=0.45\linewidth]{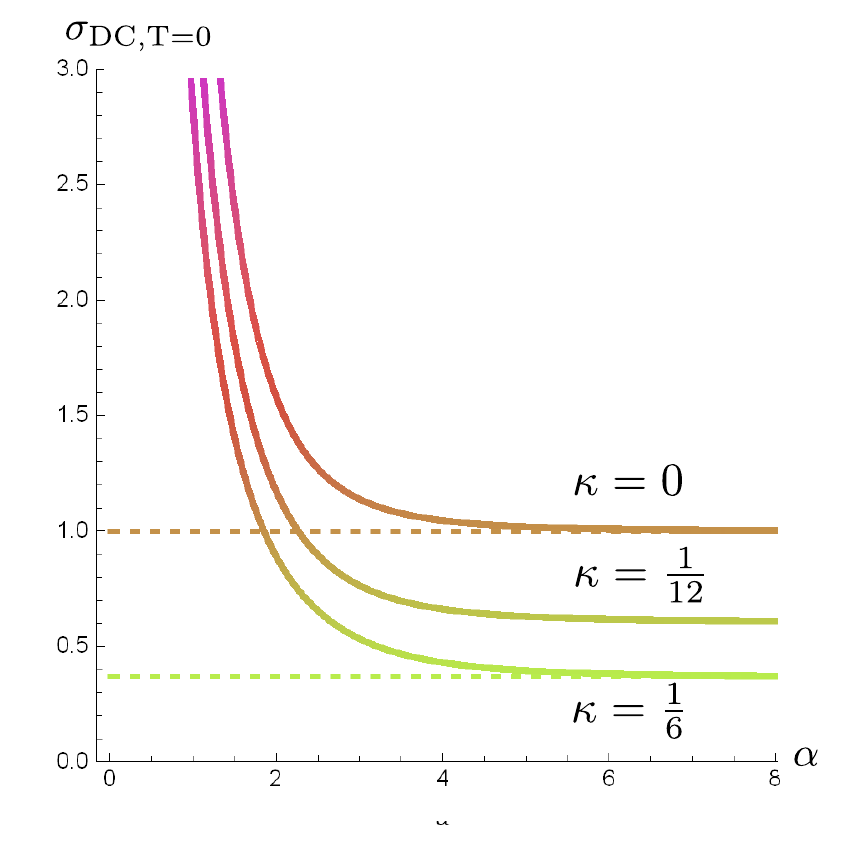}\quad\quad
\includegraphics[width=0.45\linewidth]{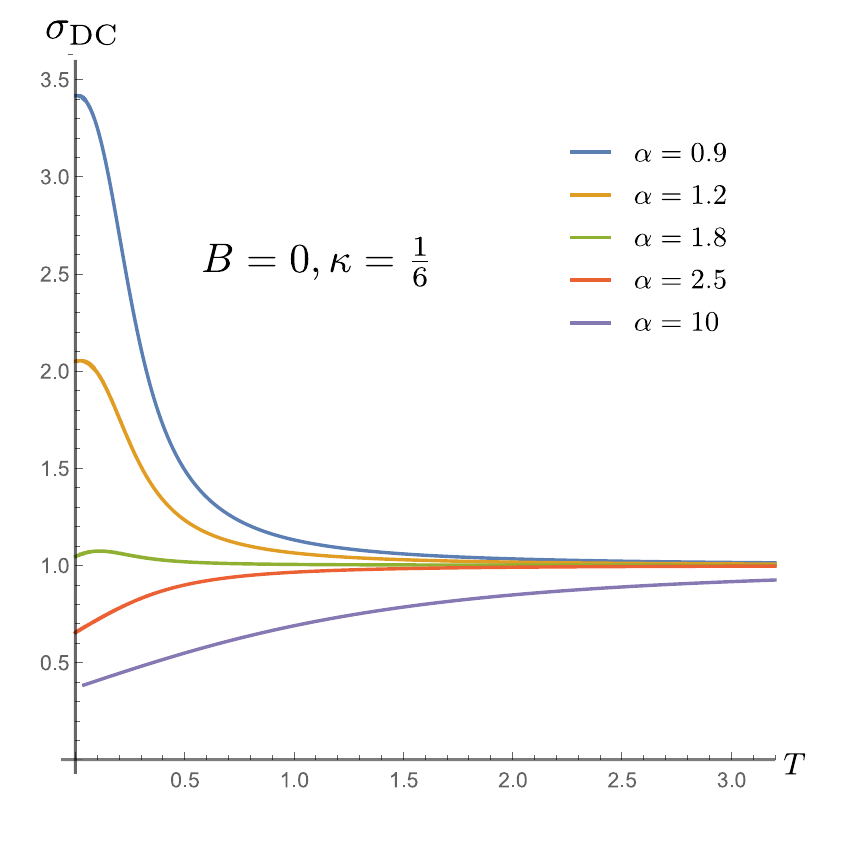}
\caption{Electric DC conductivity $\sigma_{DC}$ for the Exponential model in the absence of an applied magnetic field. Left: $\sigma_{DC}$ at zero temperature as a function of the disorder strength $\alpha$ for different theory parameter $\kappa$. Right: The temperature dependence of conductivity for $\kappa=1/6$ at different $\alpha$.  In zero magnetic field $\sigma_{DC}$ is bounded from below and its lowest value is about $0.368$.  We worked in units with the charge density $\rho=1$.}
\label{figeDC}
\end{figure}

The above analysis shows that the allowed range for $\kappa$ is significantly reduced compared to the safe region given in~\cite{Baggioli:2016oqk}. Therefore, it is necessary to reexamine the  metal-insulator transition in the absence of magnetic field for the new parameter range~\eqref{kappabd}. To see the nature of our holographic matter, we first check the DC conductivity at zero temperature following the discussion of~\cite{Baggioli:2016oqk}. 
While the insulator has $d\sigma_{DC}/dT>0$,\,\footnote{The insulating behavior has been defined as $d R_{xx}/dT<0$ in~\eqref{definition}. In the absence of magnetic field, one has $R_{xx}(B=0)=1/\sigma_{xx}(B=0)=1/\sigma_{DC}$. Therefore, the insulator can be equivalently defined as $d\sigma_{DC}/dT>0$ in the zero magnetic case.} we are able to distinguish a good insulator for which $\sigma_{DC}(T=0)\approx 0$ from a bad insulator where the electric conductivity saturates to a finite but relative large value, say $\sigma_{DC}(T=0)>0.1$ in the present study. In the left plot of Figure~\ref{figeDC}, we show $\sigma_{DC}$ at zero temperature as a function of the disorder strength $\alpha$ for different $\kappa$. The larger the value of $\kappa$ is, the lower the conductivity becomes. Note however that there is an upper bound on $\kappa$. As one can see, the lowest value of the DC conductivity is about $\sigma_{DC}\approx0.368$ ($\kappa=1/6, \alpha\rightarrow\infty$), so there is no good insulator at all. The temperature behavior of $\sigma_{DC}$ for different disorder strengths $\alpha$ is presented in the right plot of Figure~\ref{figeDC}. There exists a transition from a metallic phase to a bad insulator by increasing $\alpha$ in the system.

\begin{figure}[H]\centering
\includegraphics[width=0.46\linewidth]{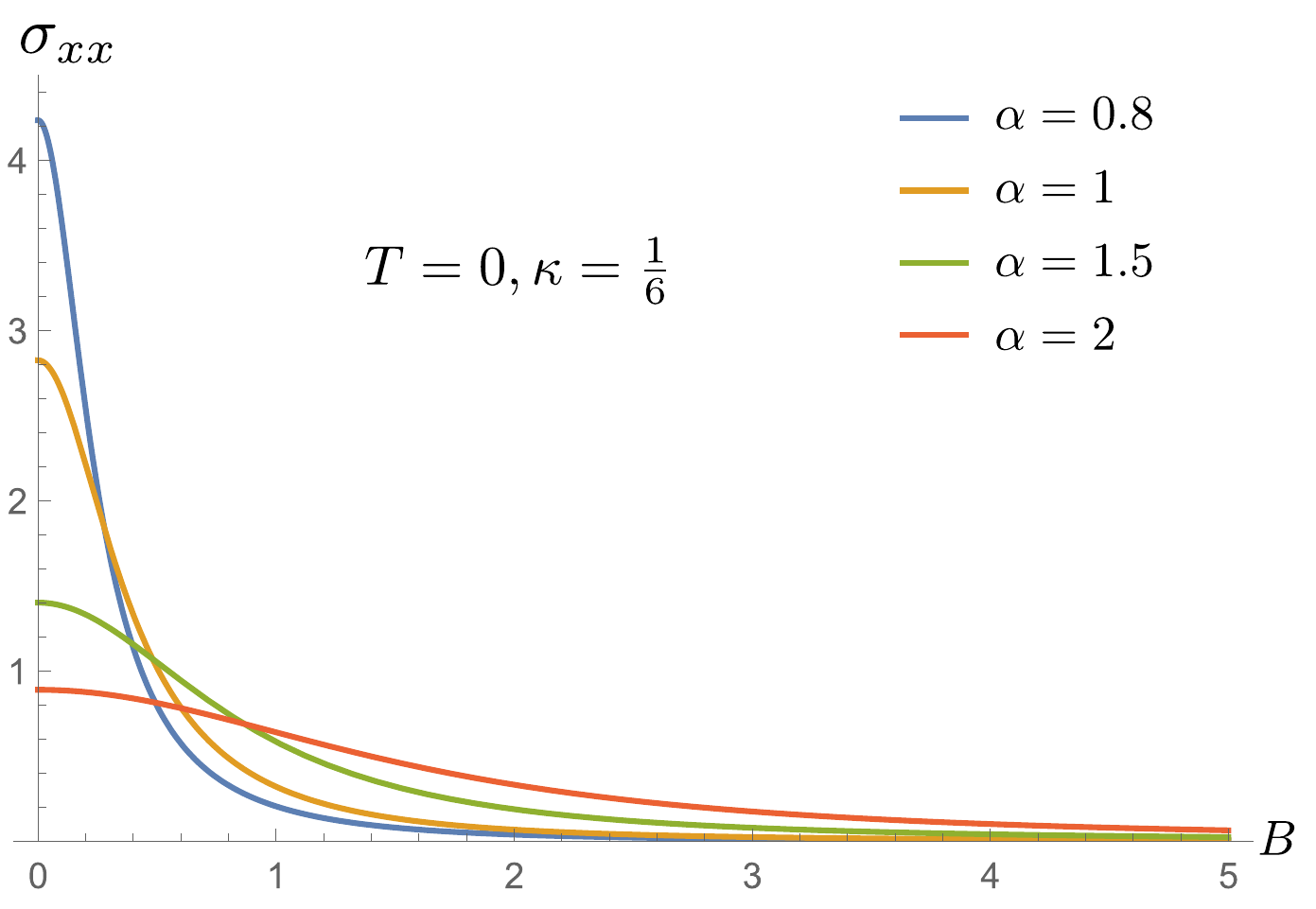}\quad
\includegraphics[width=0.45\linewidth]{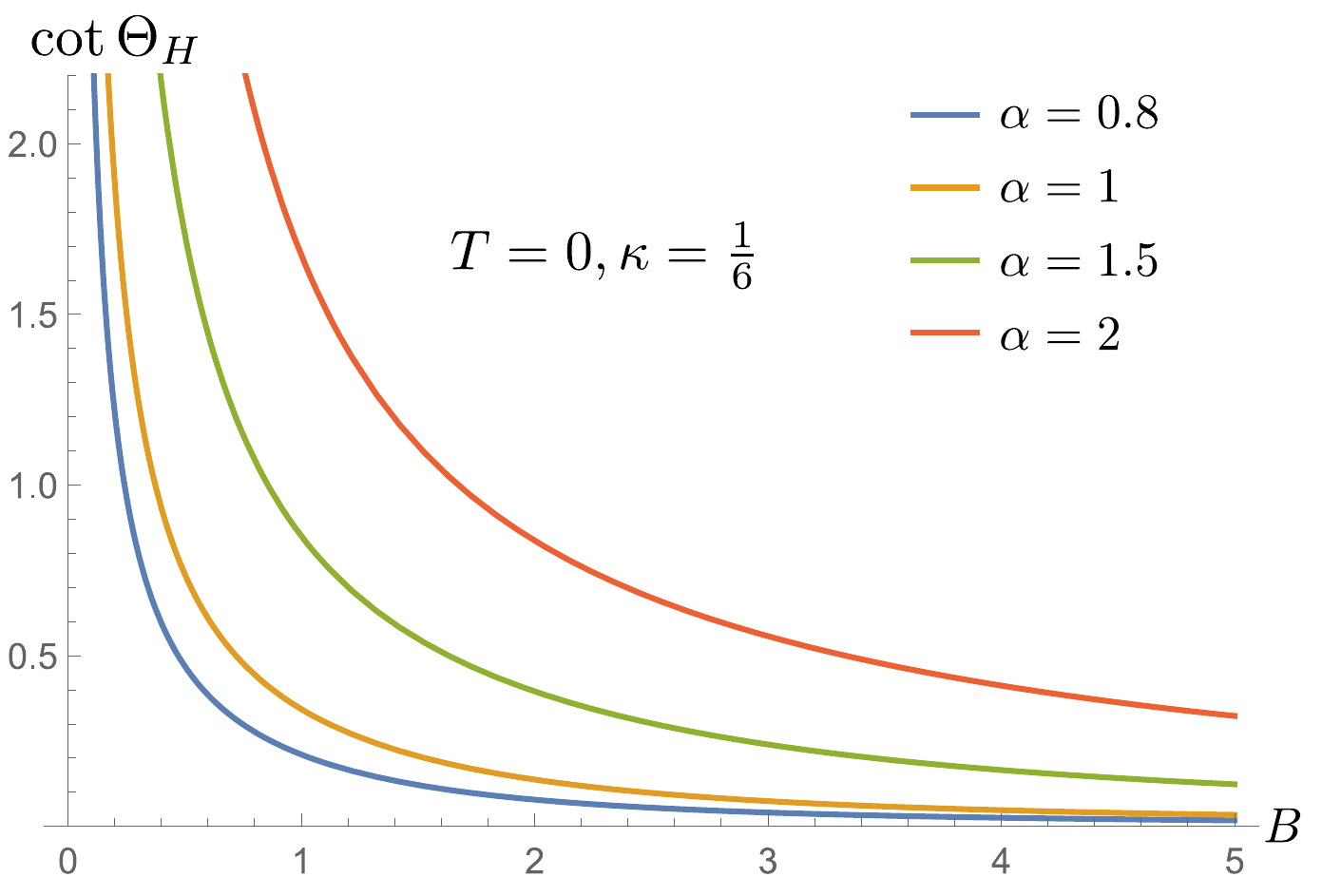}
\caption{\label{figevskappat} Magnetotransport for the Exponential model as a function of magnetic field.  Both the longitudinal conductivity $\sigma_{xx}$ (left) and the inverse Hall angle $\cot \Theta_H=\sigma_{xx}/\sigma_{xy}$ (right) decrease monotonically as $B$ is increased. We have fixed $T=0$ and $\rho=1$.}
\label{figevsB}
\end{figure}

The longitudinal conductivity $\sigma_{xx}$ with respect to the magnetic filed $B$ is presented in Figure~\ref{figevsB}. As one can see, for a fixed disorder strength $\alpha$ and temperature $T$, $\sigma_{xx}$ decreases as $B$ is increased. A heuristic understanding about this feature is as follows. Notice that we are considering a system with the density of charge carriers fixed. As $B$ is increased, more charge degrees of freedom are pushed to the orthogonal direction due to the Lorentz force. While $\sigma_{xx}$ keeps suppressed, one anticipates a more and more pronounced Hall conductivity $\sigma_{xy}$. In particular, as shown in~\eqref{clean}, in the clean limit without any disorder $\sigma_{xx}$ is vanishing and we are only left with a non-trivial Hall component. To see this feature clearly, we consider the inverse Hall angle $\cot \Theta_H=\sigma_{xx}/\sigma_{xy}$ which can be used to measure the relative magnitude of two conductivities.
As one can see from the right plot of Figure~\ref{figevsB}, the inverse Hall angle $\cot \Theta_H$ indeed decreases monotonically as $B$ is increased.
%

\subsubsection{The Linear model} 
\begin{figure}[H]\centering
\includegraphics[width=0.45\linewidth]{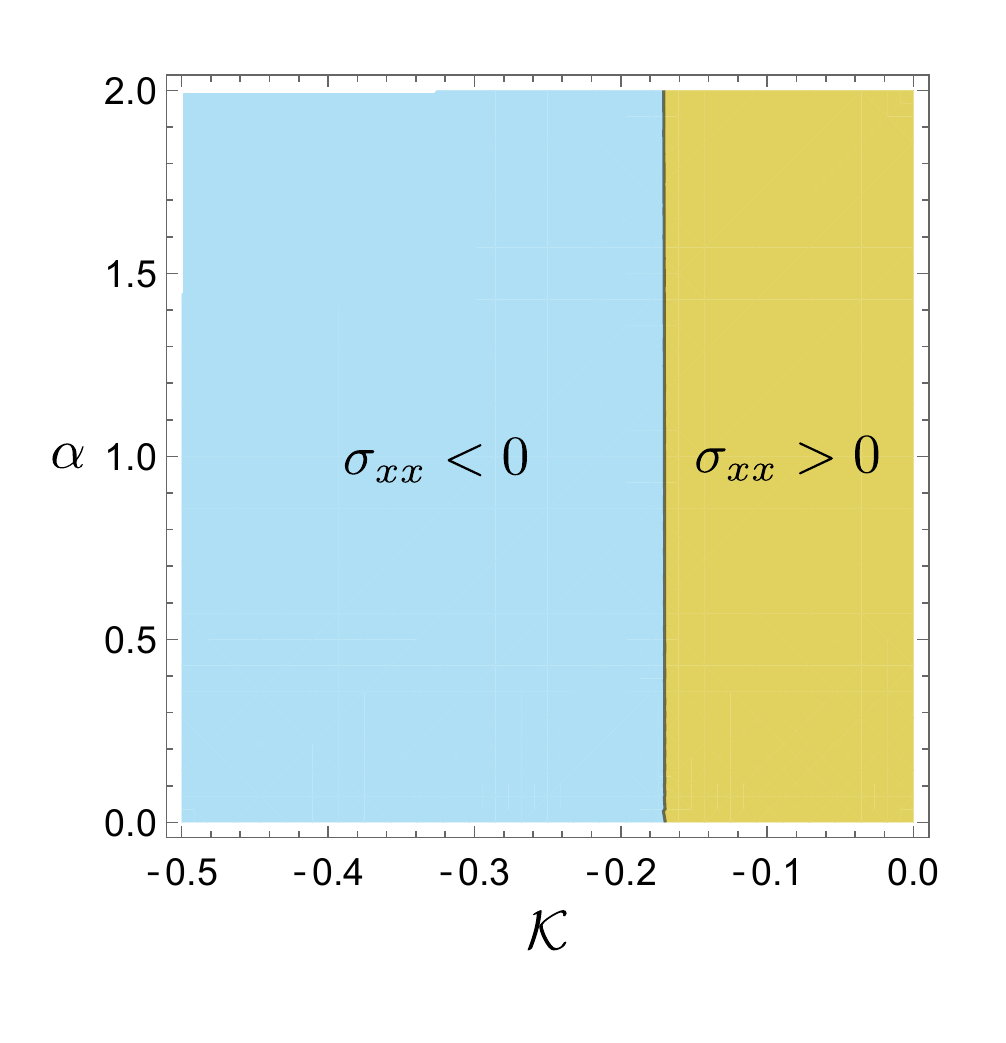}\quad\quad
\includegraphics[width=0.45\linewidth]{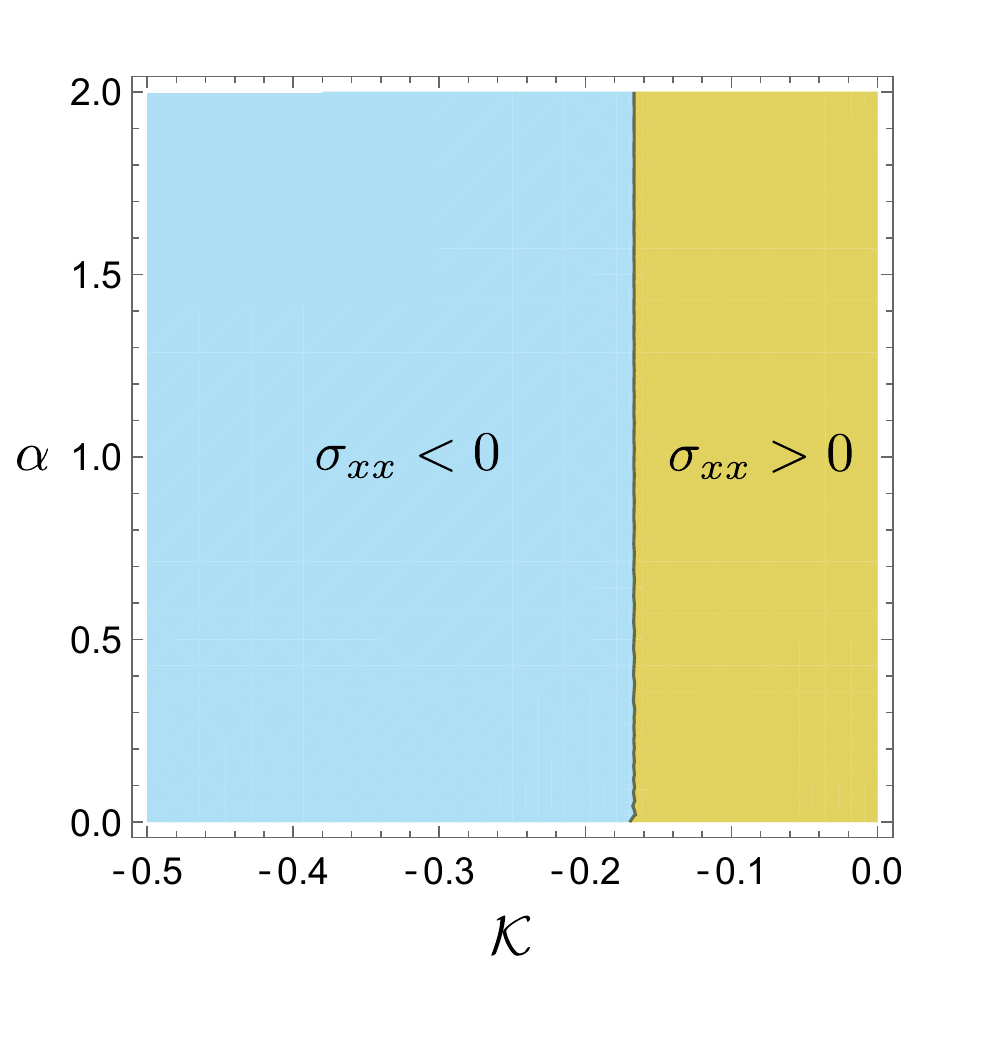}
\caption{\label{figlvsK} Distribution of the sign of $\sigma_{xx}$ as a function of the disorder strength $\alpha$ and the coupling constant $\mathcal{K}$ for the Linear model. The left plot is for $(T=0, B=10)$ and the right one for $(T=0, B=1000)$. $\sigma_{xx}$ becomes negative in blue regions. To avoid a negative longitudinal conductivity, the value of $\mathcal{K}$ should not be less than $-1/6$.}
\end{figure}

For the Linear model~\eqref{lmodel}, the background profiles for the blackening function and gauge potential are given by

\begin{align}
&d s^{2}=\frac{1}{u^{2}}\left[-f(u) d t^{2}+\frac{1}{f(u)} d u^{2}+d x^{2}+d y^{2}\right]\,,\notag\\
&f(u)=u^3\left[\frac{\rho^2(\tanh ^{-1}(\alpha  \sqrt{\mathcal{-K}} u)-\tanh ^{-1}(\alpha  \sqrt{\mathcal{-K}} u_h))}{2\alpha\sqrt{-\mathcal{K}}}+\frac{B^2}{6}(3(u-u_h)+\alpha^2\mathcal{K}(u^3-u_h^3))\right]\nonumber\\
&\qquad\qquad+\left(1-\frac{u^3}{u_h^3}\right)-\frac{1}{2}\alpha^2 u^3\left(\frac{1}{u}-\frac{1}{u_h}\right)\,,\\
&A_t(u)=\frac{\rho  \left(\tanh ^{-1}\left(\alpha  \sqrt{\mathcal{-K}} u_h\right)-\tanh ^{-1}\left(\alpha  \sqrt{\mathcal{-K}}
   u\right)\right)}{\alpha  \sqrt{\mathcal{-K}}}\notag\,.
\end{align}
The temperature reads
\begin{align}\label{ltem}
T=\frac{3}{4 \pi  u_h}-\frac{\alpha^2 u_h}{8\pi}-\frac{B^2 u_h^3 \left(1+\mathcal{K} \alpha ^2  u_h^2\right)}{8 \pi }-\frac{\rho ^2 u_h^3}{8 \pi  \left(1+\mathcal{K}  \alpha ^2 u_h^2\right)}\,,
\end{align} 
and the conductivities are given by
\begin{align}
\sigma_{x x}=&\sigma_{y y}=\frac{\Omega \left(1+\mathcal{K} \alpha ^2  u_h^2\right)[\Omega+ \left(1+\mathcal{K} \alpha ^2  u_h^2\right)( B^2 \left(1+ \mathcal{K} \alpha ^2  u_h^2\right)^2+\rho^2)u_h^2]}{(\Omega+B^2\left(1+\mathcal{K} \alpha ^2  u_h^2\right)^3 u_h^2)^2+B^2\rho^2\left(1+\mathcal{K} \alpha ^2  u_h^2\right)^4 u_h^4}\,,\\
\sigma_{x y}=&-\sigma_{y x}=\frac{B\rho \left(1+\mathcal{K} \alpha ^2  u_h^2\right)^3 u_h^2[2\Omega+ \left(1+\mathcal{K} \alpha ^2  u_h^2\right)(B^2 \left(1+\mathcal{K} \alpha ^2  u_h^2\right)^2+\rho^2)u_h^2]}{(\Omega+B^2\left(1+\mathcal{K} \alpha ^2  u_h^2\right)^3 u_h^2)^2+B^2\rho^2\left(1+\mathcal{K} \alpha ^2  u_h^2\right)^4 u_h^4}\,,\\
\Omega=&\frac{\alpha^2}{2}[\left(1+\mathcal{K} \alpha ^2  u_h^2\right)^2+\mathcal{K} u_h^4(B^2\left(1+\mathcal{K} \alpha ^2  u_h^2\right)^2-\rho^2)]\,.\nonumber
\end{align}
Based on our discussion at the beginning of this subsection, the allowed parameter space for $\mathcal{K}$ has been restricted to
\begin{align}\label{constK}
-1/6\leqslant\mathcal{K}\leqslant 0\,.
\end{align} 
As shown in Figure~\ref{figlvsK}, this parameter range for $\mathcal{K}$ does guarantee a positive definite longitudinal conductivity.
\begin{figure}[H]\centering
\includegraphics[width=0.41\linewidth]{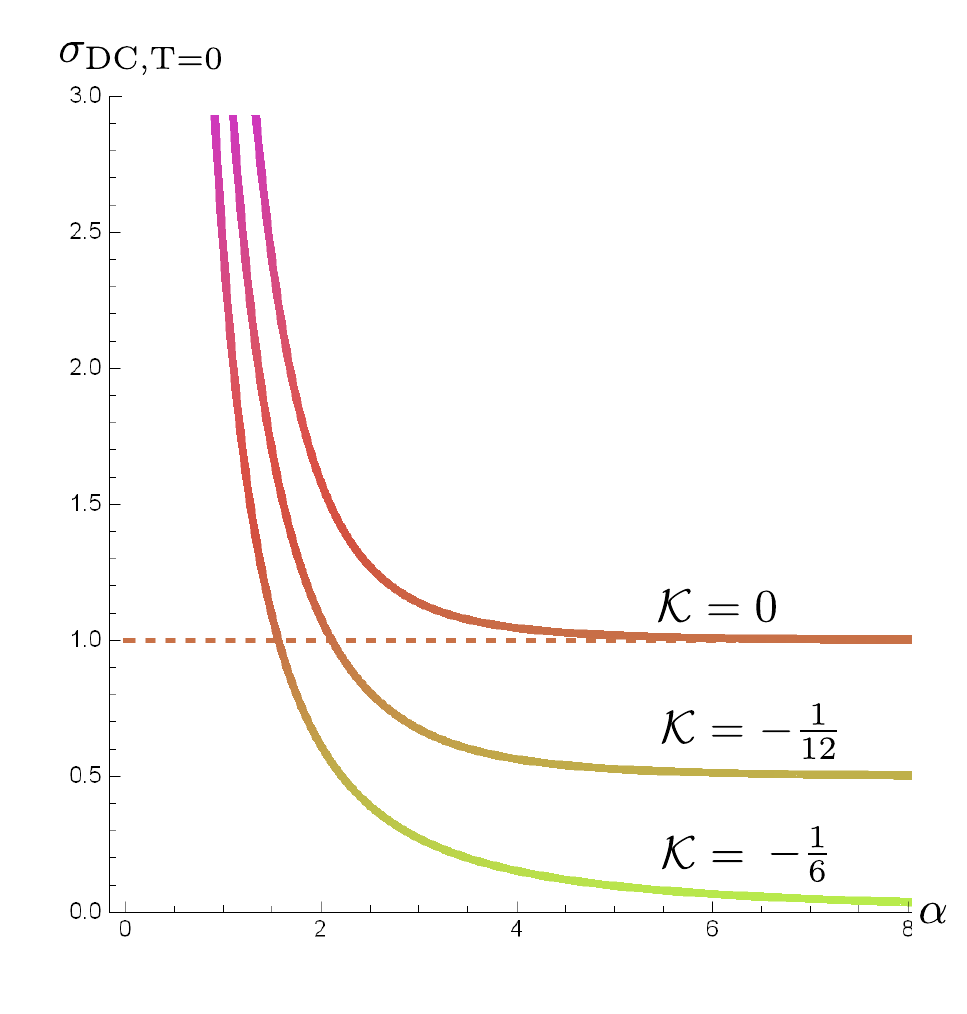}\quad\quad
\includegraphics[width=0.40\linewidth]{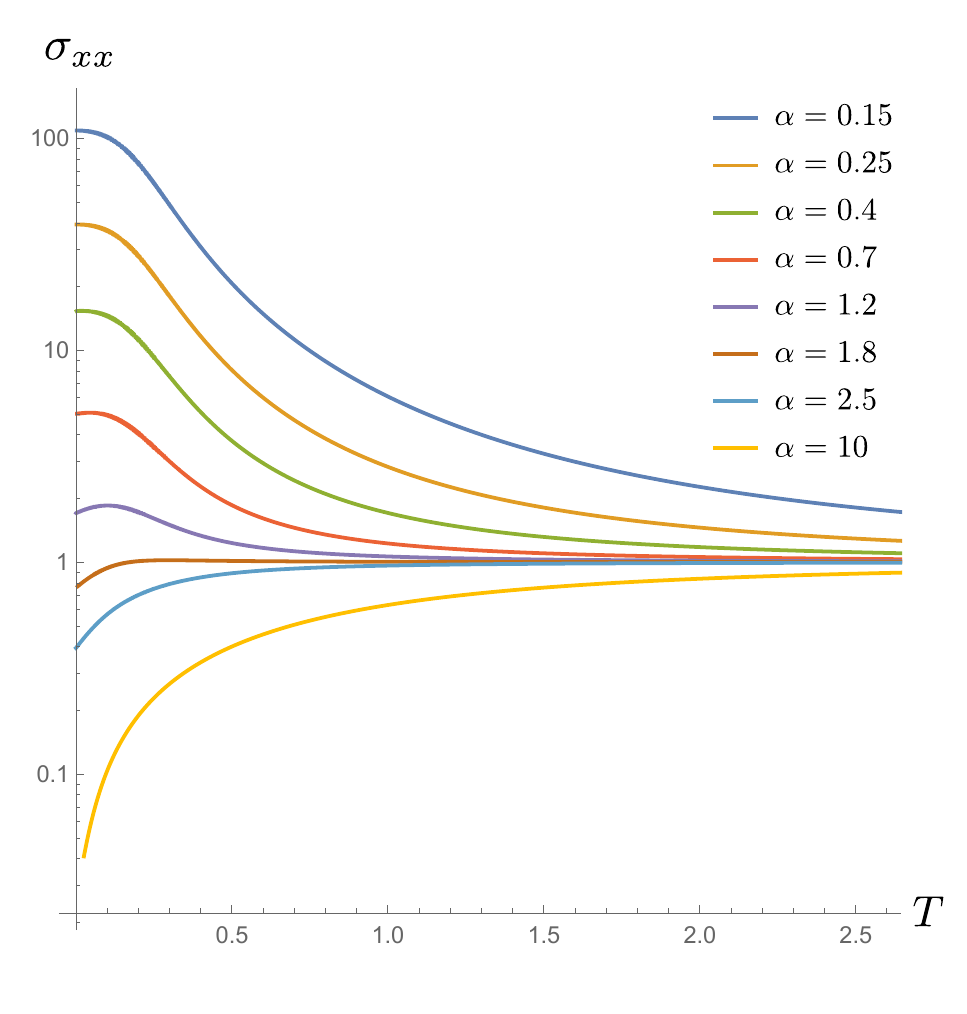}
\caption{Electric DC conductivity $\sigma_{DC}$ for the Linear model in zero magnetic field. Left: $\sigma_{DC}$ at zero temperature as a function of the disorder strength $\alpha$ for different $\mathcal{K}$. Right: The temperature dependence of conductivity at different $\alpha$. A metal-insulator transition driven by the disorder is manifest. We worked in the canonical ensemble with $\rho=1$.}
\label{figlDC}
\end{figure}

We present $\sigma_{DC}(T=0)$ as a function of $\alpha$ for different $\mathcal{K}$ in the left plot of Figure~\ref{figlDC}. In contrast to the Exponential model~\eqref{emodel}, the DC conductivity at zero temperature can get arbitrarily close to zero at large $\alpha$. In particular, $\sigma_{DC}(T=0)$ is exactly vanishing when $\mathcal{K}=-1/6$ in the infinity disorder strength limit $\alpha/\sqrt{\rho}\rightarrow \infty$. Away from this value, $\sigma_{DC}(T=0)$ is bounded from below by its asymptotic value at infinite disorder strength. As a consequence, we have a good insulator phase so long as $\mathcal{K}\rightarrow-1/6$. The temperature dependence of $\sigma_{DC}$ for different disorder strength $\alpha$ is shown in the right plot of Figure~\ref{figlDC}, from which there is  a clear metal-insulator transition driven by the disorder.

\subsection{Scaling for resistivity}
Our investigation, so far, suggests that there are holographic metal-insulator transitions that can be driven by the disorder, charge density as well as magnetic field.
As we have discussed in subsection~\ref{sub:scaling}, the resistivity $R_{xx}$ exhibits scaling behaviors near the transition point, showing the collapse of data into two separated curves in high temperature. In this part we will check the scaling behavior of resistivity for the Linear model~\eqref{lmodel} which allows a clear metal-insulator transition. We will show that the collapse of resistivity data into two separated curves holds over a broad interval of temperatures.

\begin{figure}[H]\centering
\includegraphics[width=0.46\linewidth]{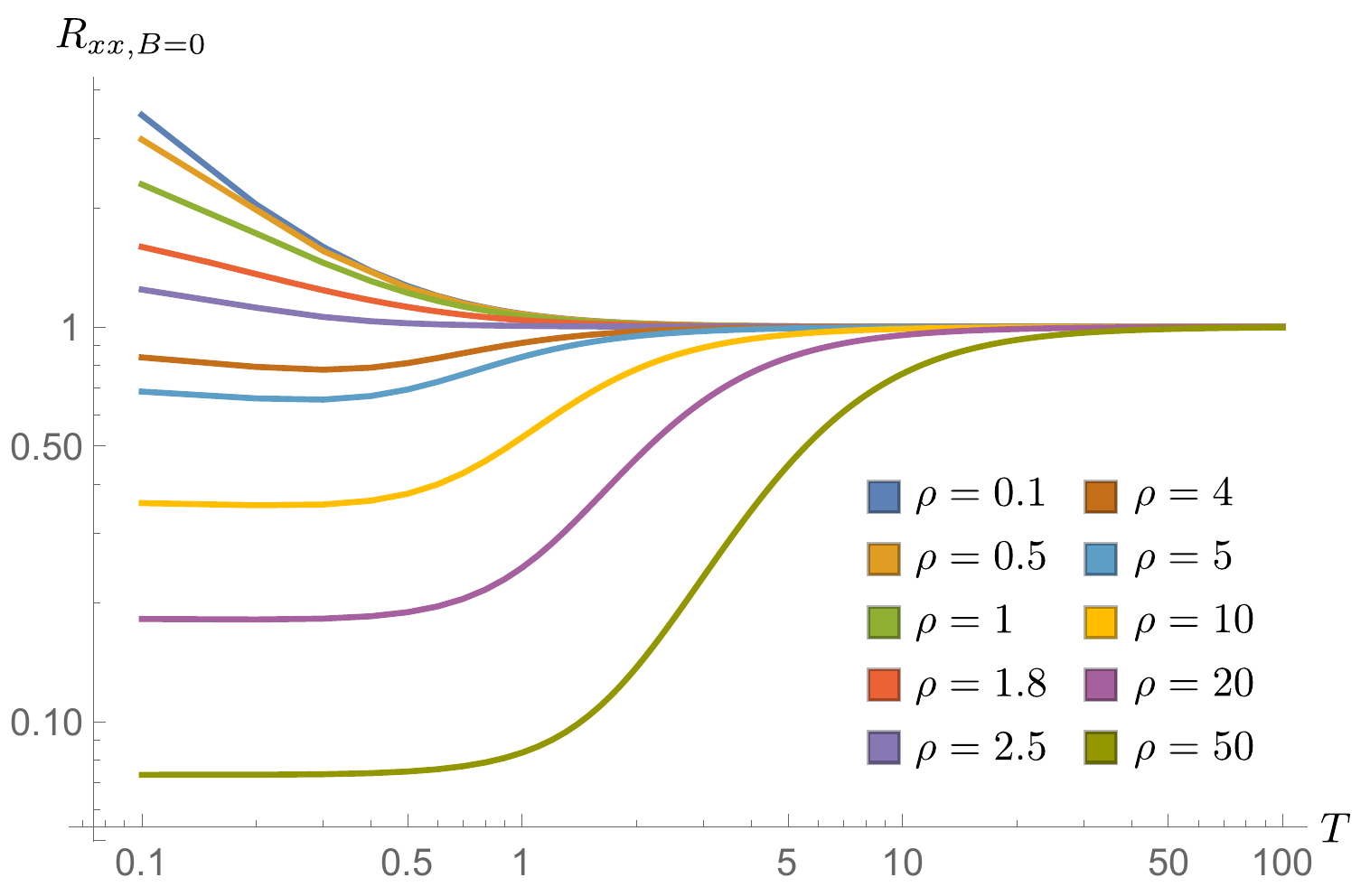}\quad\quad
\includegraphics[width=0.45\linewidth]{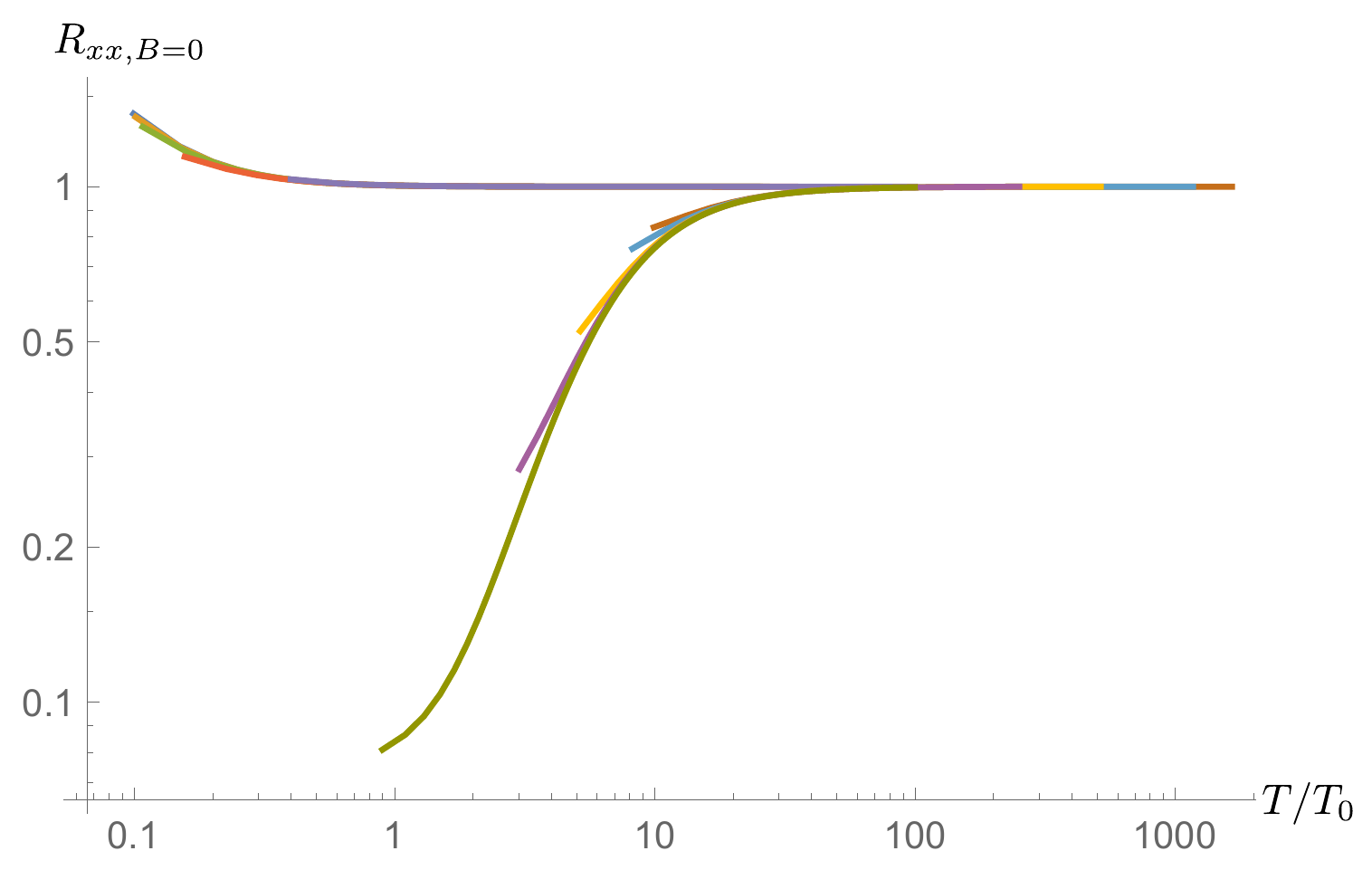}
\caption{Temperature dependence of the resistivity for different charge densities at zero magnetic field. Left: The metal-insulator transition by dialing the charge density $\rho$. Right: Scaling of resistivity with scaled temperature. $R_{xx}(T)$ dependence for different $\rho$ of the left plot can be made to overlap by scaling them along the $T$ axis with the scaling parameter $T_0$. Near the transition point $\rho_c\approx 2.6$, the scaling parameter $T_0\approx0.36 |\rho-\rho_c|^{1/2}$. We have fixed $\mathcal{K}=-1/6$ and $\alpha=3$.}
\label{fig:B0}
\end{figure}

Fixing the disorder strength $\alpha$ and dialing the charge density $\rho$, we present the resistivity curves $R_{xx}(T)$ for both metallic and insulating phases in Figure~\ref{fig:B0} at zero magnetic field. At low densities, the curves grow monotonically as the temperature decreases, exhibiting an insulating behavior. In contrast, at high densities, resistivity drops monotonically as the temperature decreases, characterizing a metallic behavior. However, for a small intermediate rang of charge density above some critical value $\rho_c$, the temperature behavior of $R_{xx}$ becomes slightly non-monotonic.  As shown in the right plot, the $R_{xx}(T)$ dependence for different $\rho$ can be made to overlap by scaling them along the $T$ axis in terms of the scaling parameter $T_0$. Near the phase transition, $T_0$ is found to be given by $T_0\sim|\rho-\rho_c|^{1/2}$, just as expected from~\eqref{scalingT0}.  Except for a small region of $\rho$ near $\rho_c$, the collapse of data into two separated curves over a broad interval of temperatures is manifest, in particular for the insulating side.

\begin{figure}[H]\centering
\includegraphics[width=0.46\linewidth]{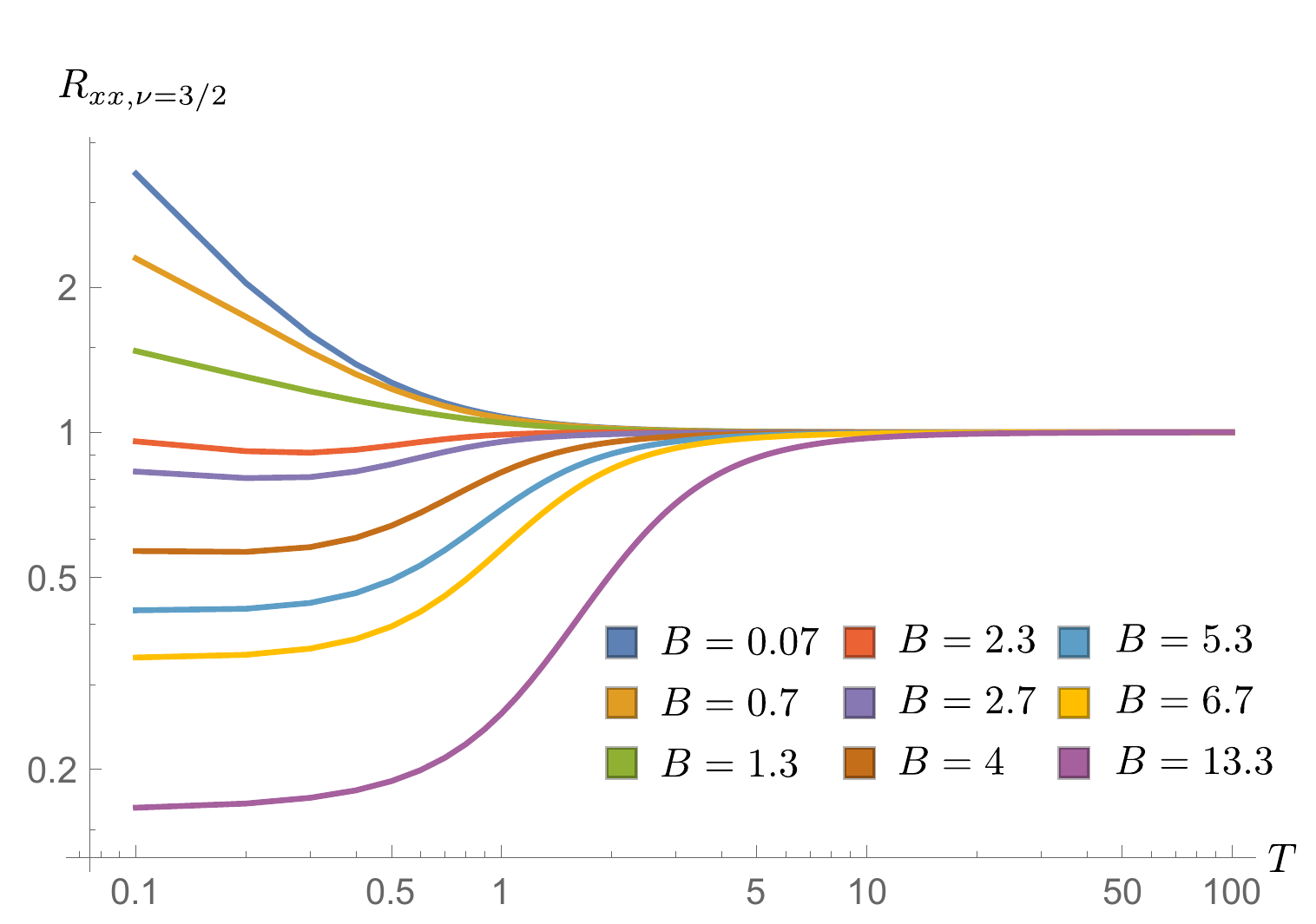}\quad\quad
\includegraphics[width=0.45\linewidth]{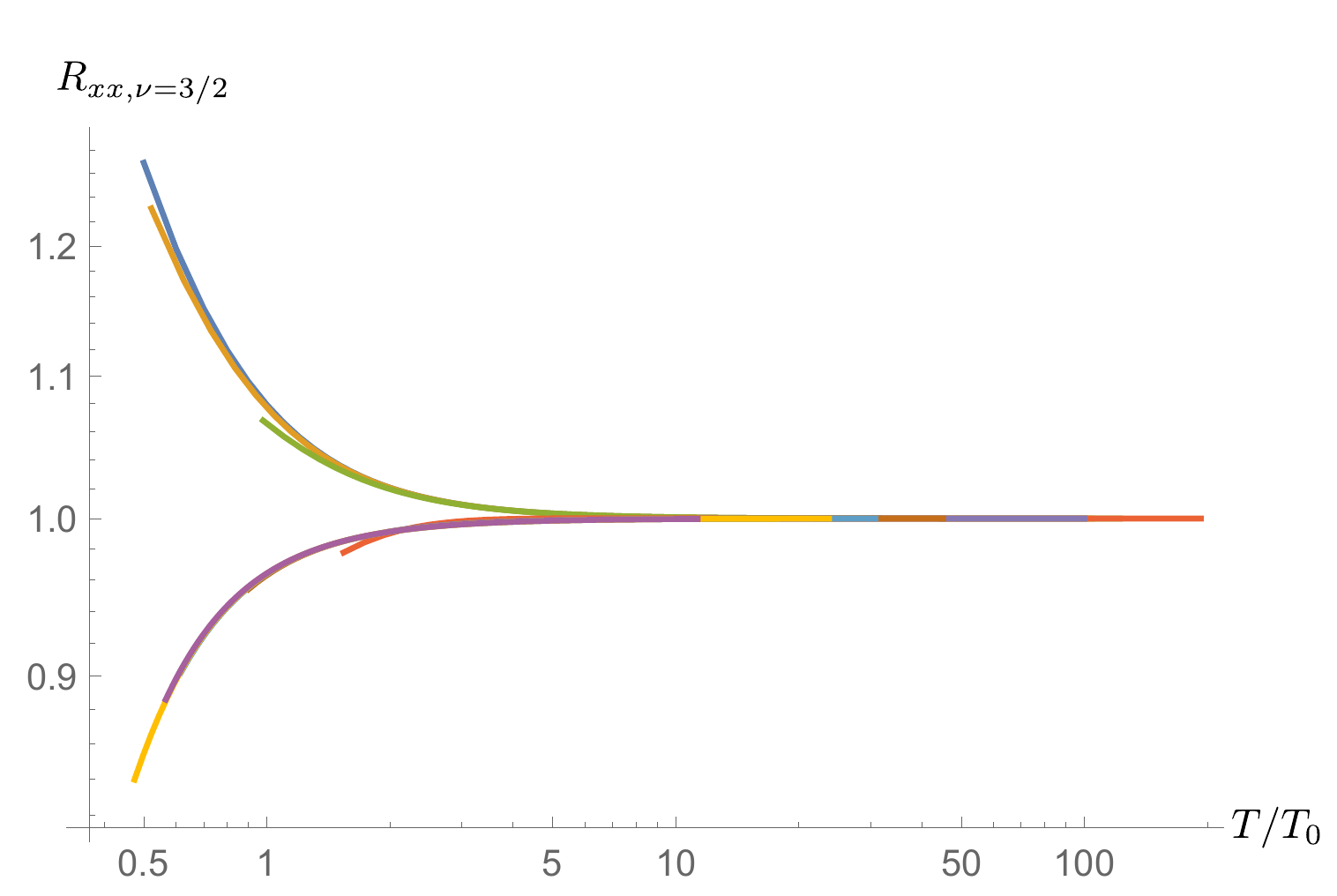}
\caption{Temperature dependence of the resistivity at Landau-Level filling factor $\nu=3/2$  for different charge densities. Left: The metal-insulator transition by dialing the magnetic field $B$. Right: Scaling of resistivity with scaled temperature $T/T_0$.  There exists scaling behavior for both metal and insulator in varying magnetic field, or equivalently charge density, while keeping $\nu=\rho/B$ fixed. We have fixed $\mathcal{K}=-1/6$ and $\alpha=3$.}
\label{fig:nu}
\end{figure}

In Figure~\ref{fig:nu} we show the temperature dependence of $R_{xx}$ in a magnetic field corresponding to a Landau-Level filling factor $\nu=3/2$. To produce these data of Figure~\ref{fig:nu}, we have varied both $\rho$ and $B$ such that $\nu$ remains constant. Similar to the case with $B=0$, we find the collapse of data into two separated curves in a perpendicular magnetic field with $\nu=3/2$, although $\rho_c$ is different. One expects to have similar feature when $\nu>1$, based on our discussion around~\eqref{landau}. This suggests that for our holographic quantum matter the metal-insulator transition at zero magnetic field and at fixed Landau-Level filling factor might be controlled by the same underline physics when viewed in terms of resistivity. 

\begin{figure}[H]\centering
\includegraphics[width=0.45\linewidth]{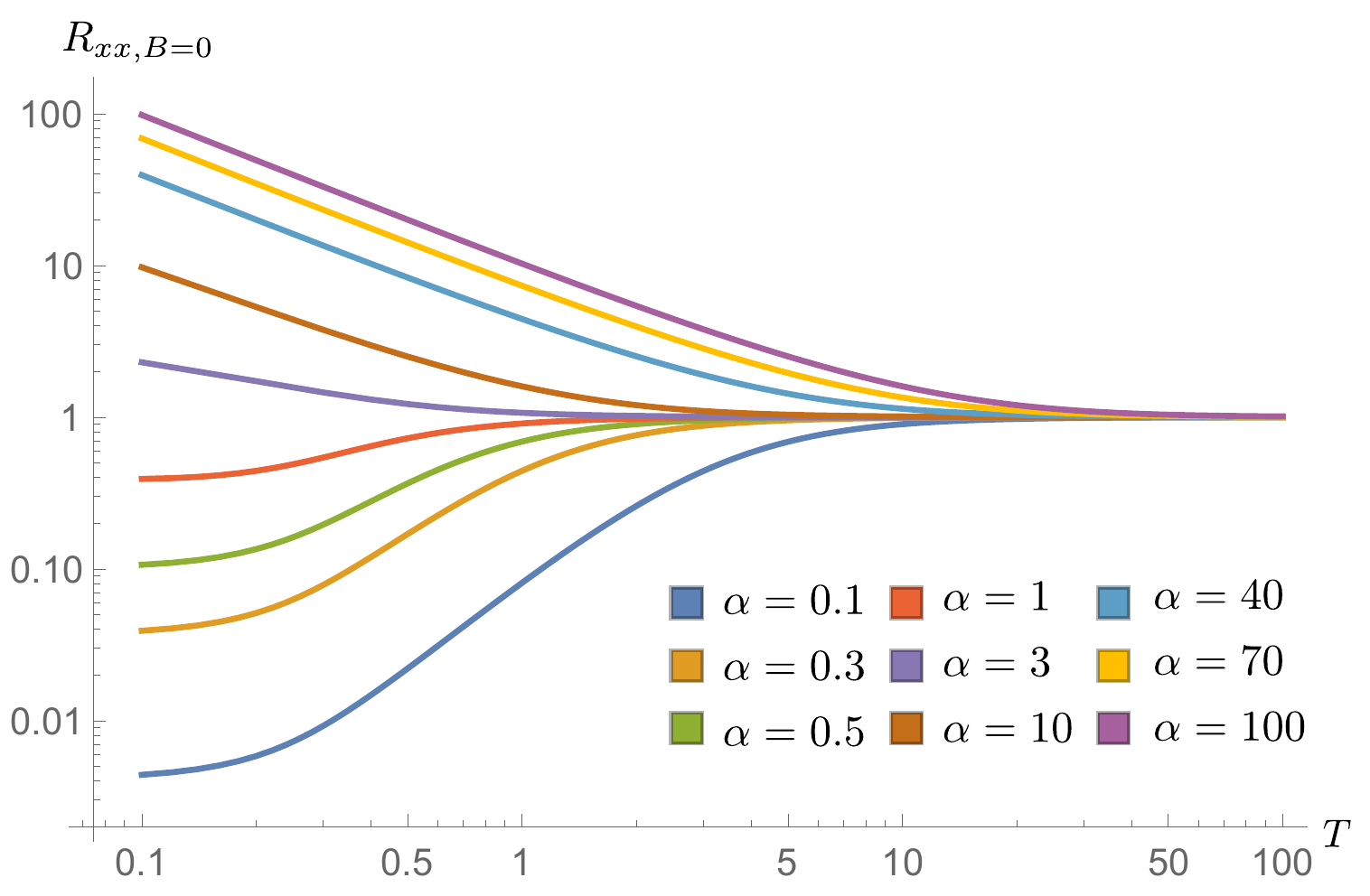}\quad\quad
\includegraphics[width=0.45\linewidth]{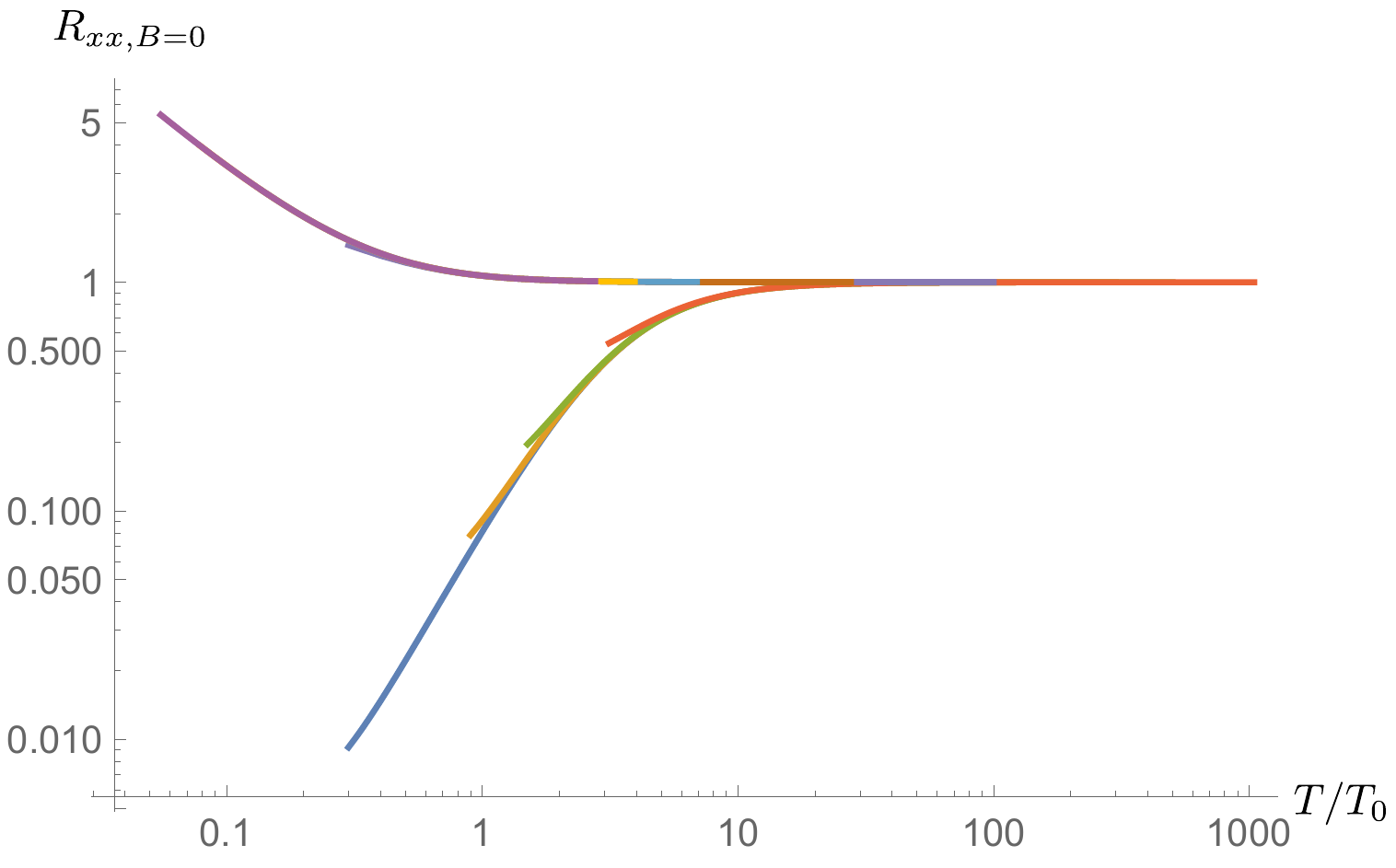}
\caption{Temperature dependence of the resistivity versus disorder strength at zero magnetic field. Left: The metal-insulator transition driven by the disorder strength $\alpha$, which corresponds to the right plot of Figure~\ref{figlDC} with $R_{xx}(B=0)=1/\sigma_{DC}$. Right: Scaling of resistivity with scaled temperature $T/T_0$. The collapse of data into two separated curves both in the metallic and insulating sides is manifest. We have fixed $\mathcal{K}=-1/6$ and $\rho=1$.}
\label{fig:alpha}
\end{figure}

For completeness, we check the case with respect to the disorder strength $\alpha$. The temperature dependence of $R_{xx}(B=0)$ is presented in Figure~\ref{fig:alpha}. A strong metallic temperature dependence of the resistivity is observed at disorder strength above some critical value, while insulating behavior is seen at disorder strength below the critical value. At the critical disorder strength, there appears to be a transition from a metallic-like phase to a strongly localized one. It is clear from the right plot that the $R_{xx}(T)$ dependence for different $\alpha$ can be made to overlap by scaling them along the $T$ axis in terms of the scaling parameter $T_0$. 

We close this section by remarking that the scaling behavior of resistivity in our holographic setup breaks down at sufficiently low temperatures. Indeed, the higher the temperature is, the better the collapse of resistivity data into two separated curves.\,\footnote{The low temperature behavior of $R_{xx}$ depends on the details of coupling functions $Y$ and $X$. In the present study we focus on the Linear model as a benchmark example. It is possible to have a better scaling behavior at low temperatures by considering  a different choice of coupling functions.}  In this sense, the scaling behavior of resistivity data is due to the UV $AdS_4$ fix point rather than the IR fix point. The observation of scaling behavior of resistivity in two-dimensional metal-insulator transition was argued to be followed directly from the quantum critical point associated with such a transition~\cite{Dobrosavljevic:1997}. Nevertheless, it seems no quantum critical point in our holographic quantum matter.


\section{AC Conductivity and Phase Diagram}\label{Sec:optical}
We have shown that the positive definitiveness of the longitudinal DC conductivity imposes further constraint on the theory parameters. In particular, for the Exponential model~\eqref{emodel}, the allowed range for the coupling constant $\kappa$ is reduced significantly. As a consequence, the DC conductivity in the absence of magnetic field is bounded from below, and thus one can not get a good insulating phase for which $\sigma_{DC}(T=0)$ is very small or eventually zero. In contrast, for the Linear model~\eqref{lmodel}, within the allowed parameter range for the coupling $\mathcal{K}$~\eqref{constK}, it can give rise to a manifest disorder-like phenomenology that includes a very clear disorder-driven metal-insulator transition. To gain better physical intuition about the transition, we now consider the optical conductivity, from which one could see how the spectral weight transfers as the disorder effect becomes more and more important. Then we construct the phase diagrams for the two models in the temperature-disorder plane, and study the behaviors of specific heat and charge susceptibility across different phases.

\subsection{Optical conductivity}
The behavior of the optical conductivity in presence of a magnetic field is complicated, which obscures the understanding of the spectral weight transfer due to the disorder.\footnote{In the presence of a magnetic field there exists a peak at a finite value of frequency due to the cyclotron resonance~\cite{Hartnoll:2007ip,Kim:2015wba}. It will be interesting to investigate the interplay between the disorder and cyclotron resonance, and we leave this issue for future research.  The AC conductivity in the presence of lattice symmetry breaking has been widely investigated in the literature, see \emph{e.g.} ~\cite{Donos:2019tmo,Baggioli:2020edn,Baggioli:2016oqk,Donos:2012js,Andrade:2017cnc,Donos:2013eha} with homogeneous lattices and~\cite{Rangamani:2015hka,Ling:2014saa,Horowitz:2012ky,Ling:2013nxa,Jokela:2017ltu} with inhomogeneous lattices.} So we focus on the electric response by turning off the magnetic field. Following the standard procedure in holography, we turn on small perturbations around a background:
\begin{equation}
\begin{split}
\delta A_x&= a_x(u,\omega)\, e^{-i \omega t},\qquad\;\;\, \delta \phi_x= \chi_x(u,\omega)\, e^{-i \omega t}\,,\\
 \delta g_{tx}&= h_{tx}(u,\omega)\, e^{-i \omega t},\qquad \delta g_{u x}= h_{ux}(u,\omega)\, e^{-i \omega t},
\end{split}
\end{equation}
where we have restricted ourselves to the homogeneous perturbations by fixing the momentum $k=0$. To fix the redundancy due to the diffeomorphism  invariance, one can either use the gauge fixing or work in the gauge invariant variables. In our case, we will adopt the radial gauge by setting $h_{ux}=0$.
In order to compute the retarded Green's function $G^R_{J^xJ^x}(\omega)$ for the U(1) current $J^x$, we impose the ingoing boundary condition near the horizon, and turn off the source of $h_{tx}$ and $\chi_x$ in the UV where $u\rightarrow0$. Then, from the UV expansion for $a_x$
\begin{equation}
a_x(u,\omega)=a_x^{(0)}(\omega)+a_x^{(1)}(\omega) u+\mathcal{O}(u^2),
\end{equation}
we obtain
\begin{equation}
\sigma_{xx}(\omega)\equiv\frac{1}{i\,\omega}G^R_{J^xJ^x}(\omega)=\frac{1}{i\,\omega}\frac{a_x^{(1)}(\omega)}{a_x^{(0)}(\omega)}\,,
\end{equation}
as a function of frequency $\omega$.

\begin{figure}[H]\centering
\includegraphics[width=0.43\linewidth]{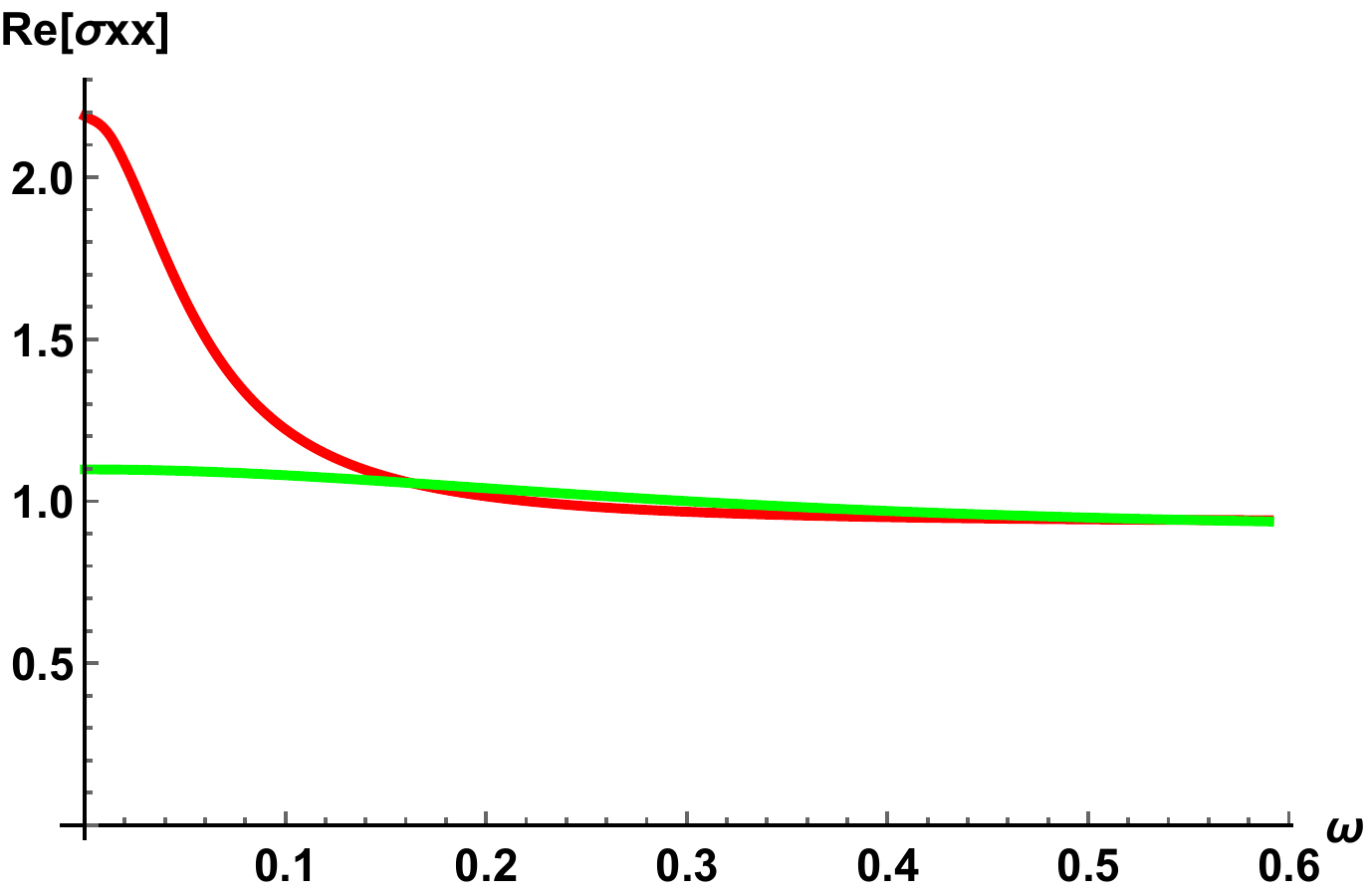}\quad\quad
\includegraphics[width=0.43\linewidth]{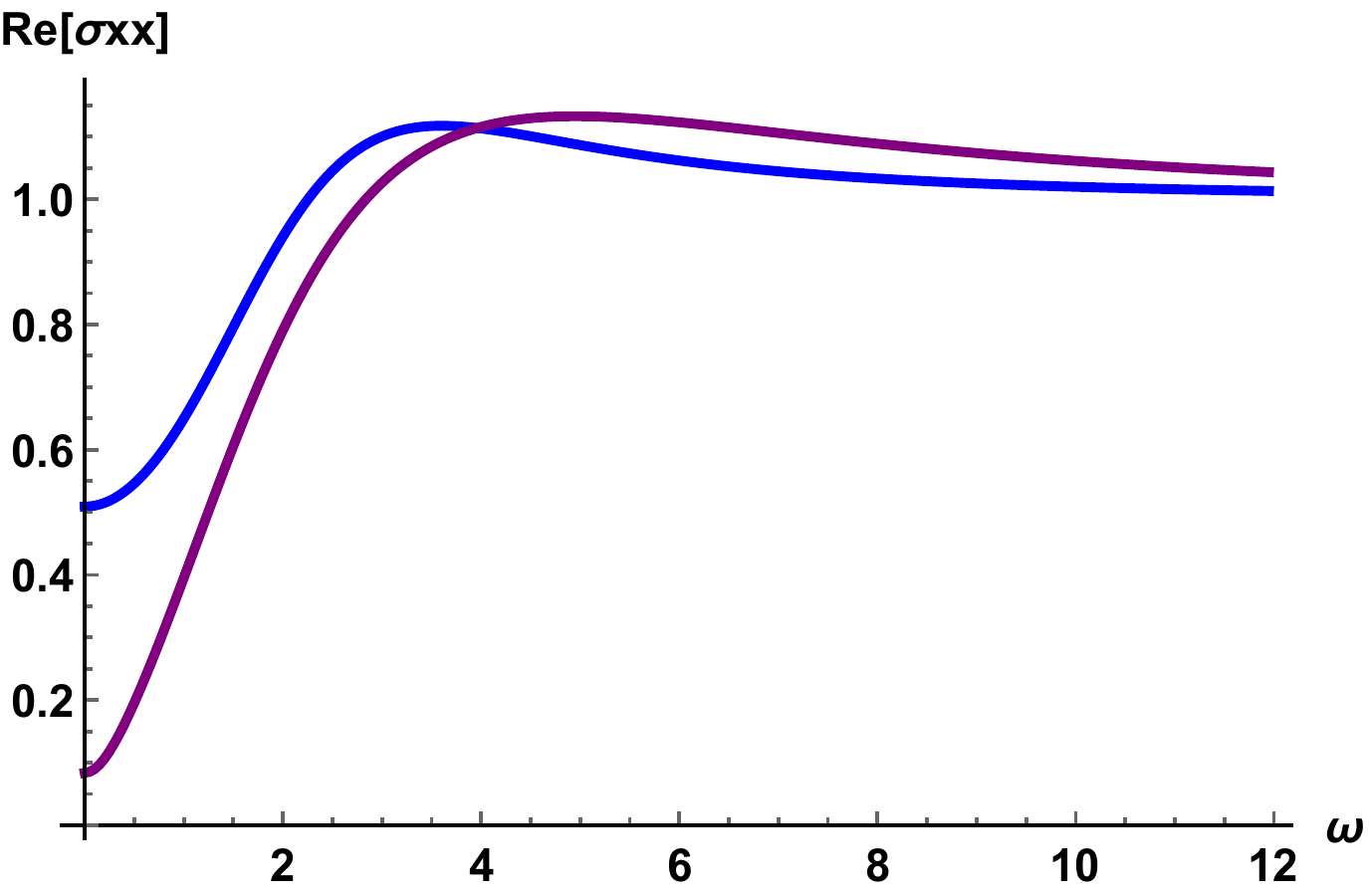}
\includegraphics[width=0.43\linewidth]{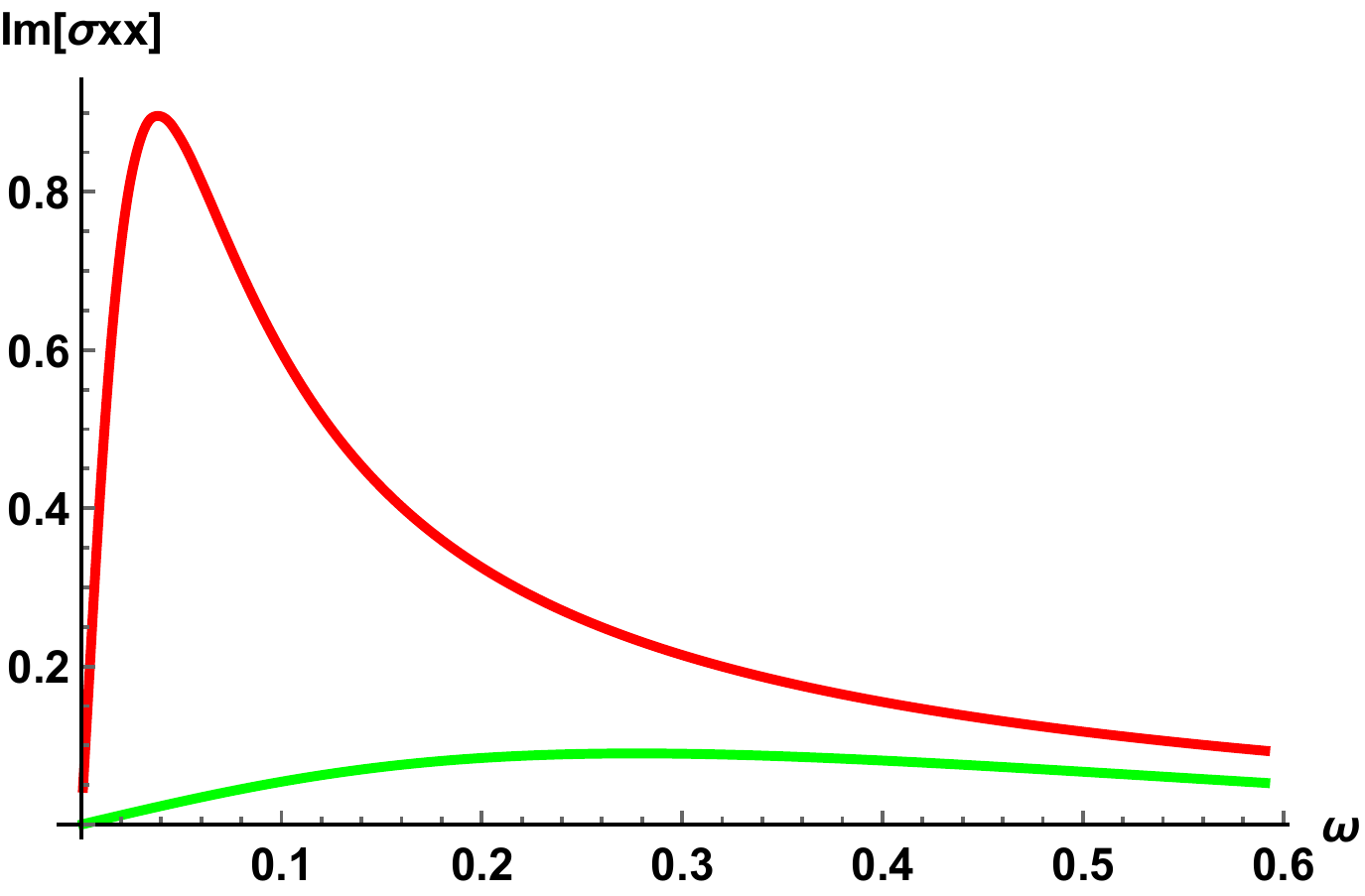}\quad\quad
\includegraphics[width=0.43\linewidth]{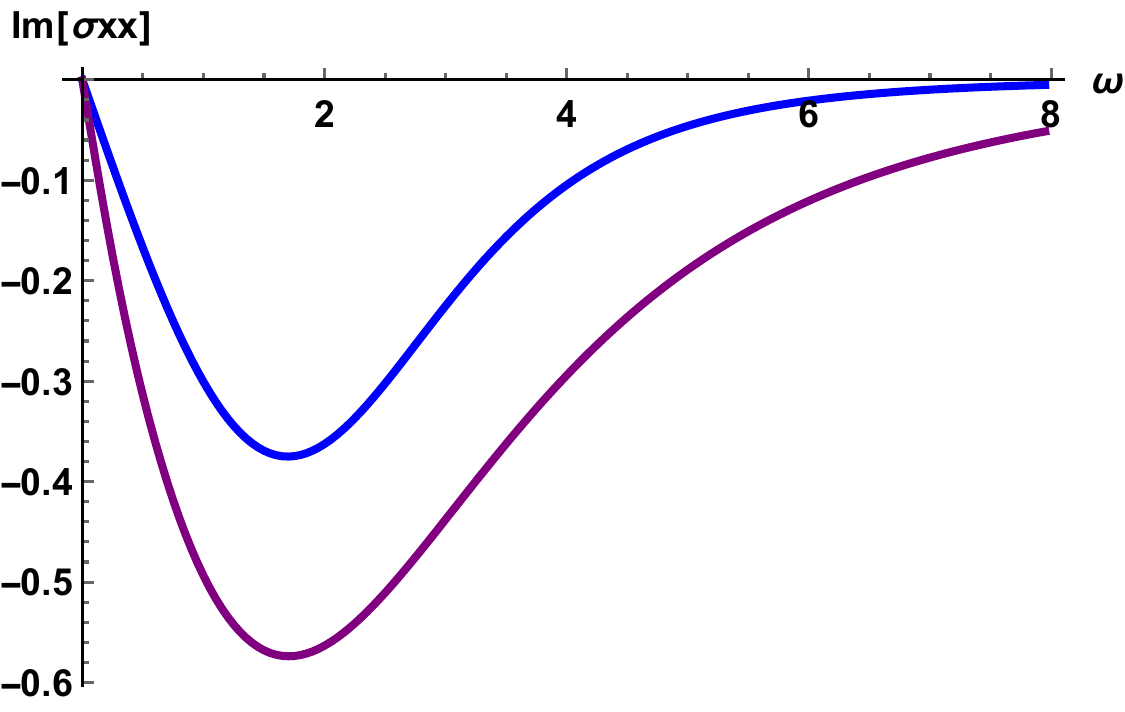}
\caption{Representative examples of the AC electric conductivity $\sigma_{xx}$ for the Linear model~\eqref{lmodel} with unitary charge density $\rho=1$. There are four phases: (a) good metal (red) with $(\alpha=0.6, T=0.5)$, (b) incoherent metal (green) with $(\alpha=1.5, T=0.5)$, (c) bad insulator (blue) with $(\alpha=4.5, T=0.3)$ and (d) good insulator (purple) with $(\alpha=7.8, T=0.05)$. As the strength of disorder effect becomes strong, the Drude peak of the good metal becomes broader and disappears, and there is a transition to incoherent metal (green). For sufficiently strong disorder effect, a growing peak in the mid-infrared develops (blue and purple), and the system becomes an insulator.
We choose $\mathcal{K}=-1/6$ which is the threshold value of the parameter range of $\mathcal{K}$.}
\label{figlAC}
\end{figure}

We now present our numerical results. Some representative examples of the AC electric conductivity are shown in Figure~\ref{figlAC} for the Linear model~\eqref{lmodel}. When the disorder strength is weak, one has a coherent metallic phase where there is a sharp Drude peak at small $\omega$. As the strength of disorder effect increases, the width of the peak in real part increases, while the maximum value of the peak decreases. There will be no more Drude peak and one arrives at an incoherent metallic phase where there is no clear and dominant localized long lived excitation, see green curves  of Figure~\ref{figlAC} . When the disorder is strong enough, a ``particle-like" peak at low $\omega$ is replaced by a ``vortex-like" dip, and the spectral weight transfers to the mid-infrared, resulting in an insulating behavior.  The stronger the disorder effect, the more spectral weight transfer to the mid-infrared. As a consequence, the DC conductivity keeps decreasing, and then triggers the transition from bad insulator to good insulator. Therefore, there is a clear disorder-driven transition from a coherent metal with a sharp Drude peak to a good insulator with a tiny or vanishing DC conductivity at zero temperature. 

For the Exponential model~\eqref{emodel}, within the allowed range of $\kappa$~\eqref{kappabd}, we can obtain similar optical conductivity features presented by Figure~\ref{figlAC}, except the one corresponding to the good insulator. The amount of spectral weight transformation to higher $\omega$ is limited by the bound~\eqref{kappabd} on $\kappa$. Within this simple model, the spectral weight transfer efficiency is not sufficiently compared to the Linear model, resulting a value of DC conductivity that is too large to be identified as good insulator.

\subsection{Phase diagram, specific heat and charge susceptibility}
\begin{figure}[H]\centering
\includegraphics[width=0.40\linewidth]{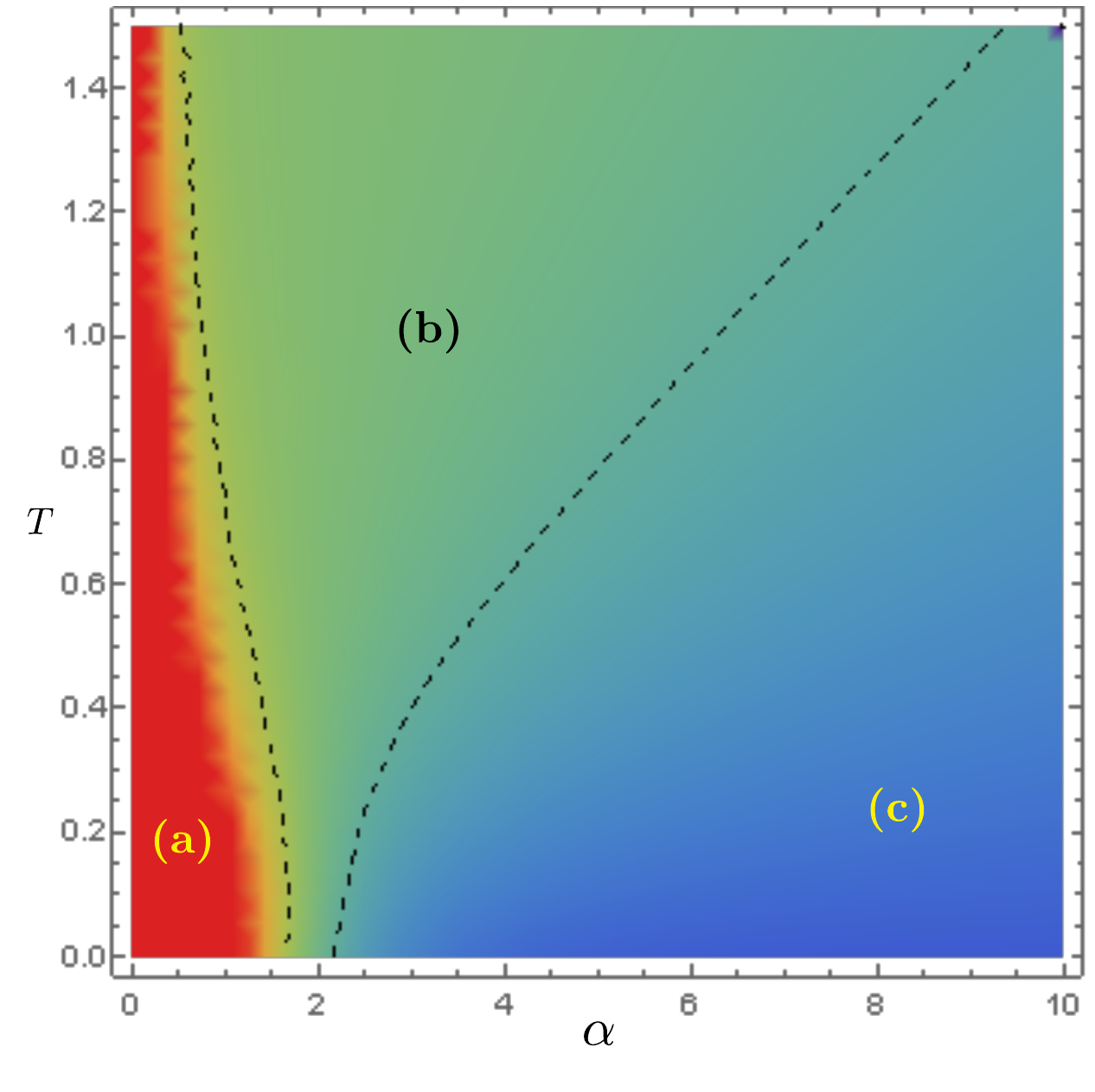}\ \ \ \ \ 
\includegraphics[width=0.05\linewidth]{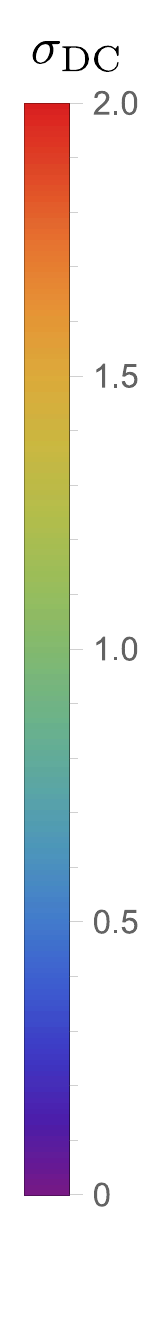}
\includegraphics[width=0.40\linewidth]{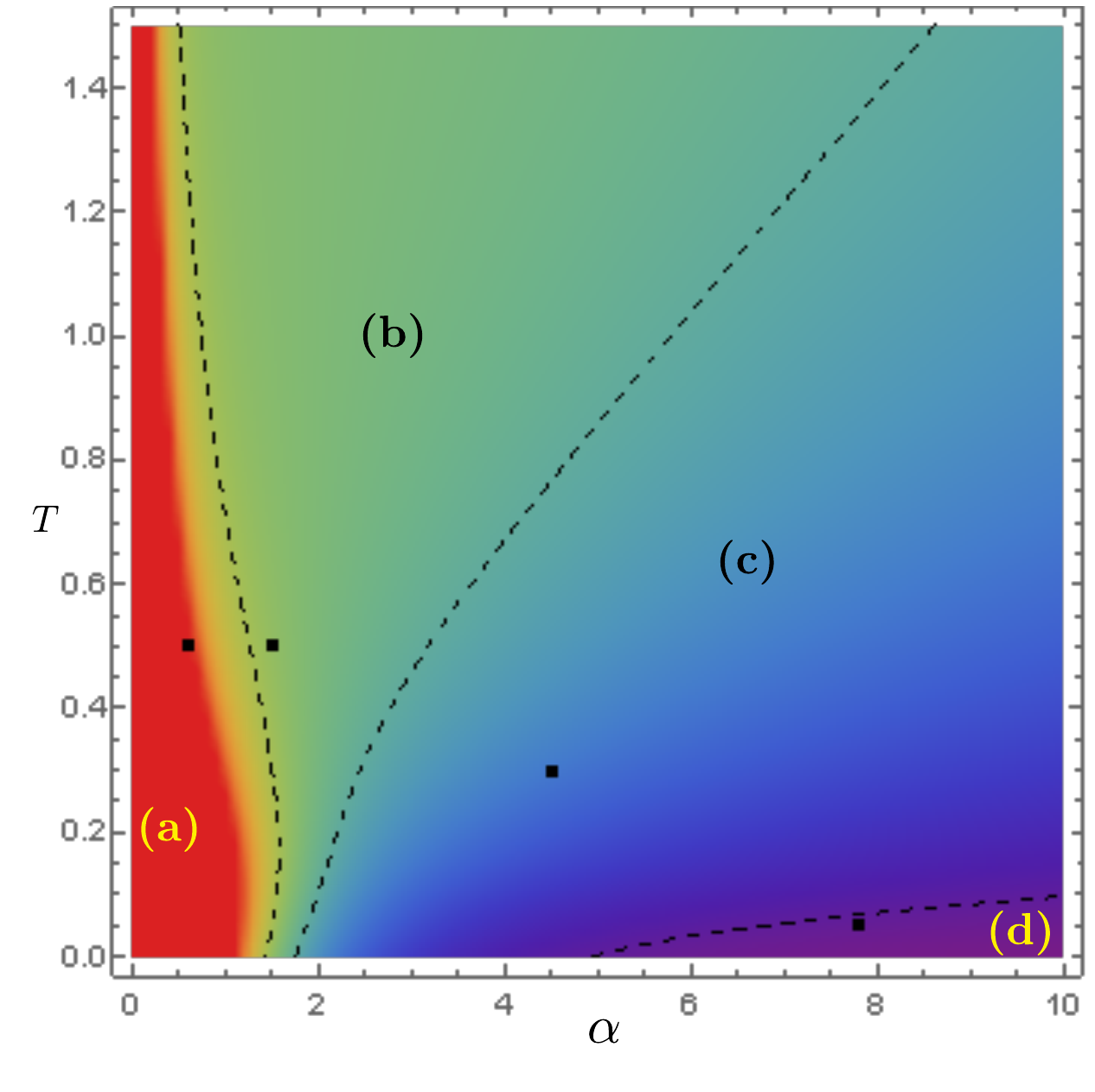}
\caption{Phases diagrams for the Exponential model with $\kappa=1/6$ (left) and the Linear model with $\mathcal{K}=-1/6$ (right), in the absence of an applied magnetic field. Four regions are denoted by (a) good metal, (b) incoherent metal, (c) bad insulator and (d) good insulator, respectively. Following~\cite{Baggioli:2016oqk}, the phase boundaries are denoted by the dashed lines corresponding to $\sigma_{DC}=0.1, 0.8, 1.2$. There is no region (d) for good insulator in the left plot. Each dark point in the right plot corresponds to the optical conductivity of Figure~\ref{figlAC}. We have worked in the canonical ensemble with the charge density $\rho=1$.}
\label{figphase}
\end{figure}

The phase diagrams for the two models in the temperature-disorder plane are constructed in Figure~\ref{figphase}. For the Linear model~\eqref{lmodel}, the phase diagram incorporates all four phases of matter for large disorder effect, i.e. $\mathcal{K}\rightarrow-1/6$. Both the quantum phase transition and the finite temperature crossover are manifest in the right plot of Figure~\ref{figphase}. In contrast, for the other model presented by the left plot of Figure~\ref{figphase}, there is no good insulator phase in the allowed range of $\kappa$~\eqref{kappabd}. For each phase of matter a representative example of optical conductivity has been shown in Figure~\ref{figlAC}.

Although there are as many as four different phases in the temperature-disorder phase diagram of Figure~\ref{figphase}, there is no genuine thermodynamic phase transition in the present background geometries. All phases share the same symmetries of the underlying theory, and thus beyond a simple Ginzburg-Landau description.
The free energy $\mathcal{F}$ can be computed by adding standard holographic counterterms and is behaved continuously and smoothly during the metal-insulator transition. This feature can be understood from the behavior of the thermal entropy. For the background geometry~\eqref{solution}, the entropy density is obtained from the area of the black hole $s=2\pi/{u_h^2}$. It is easy to check that the entropy density $s=-\partial \mathcal{F}/\partial T$ behaves smoothly in the temperature-disorder plane, thus there is no thermodynamic phase transition in Landau's theory. Nevertheless, it would be helpful to see if there is any other local quantity that is able to distinguish different phases of Figure~\ref{figphase}. 

An interesting probe to the nature of the phase of matter is the specific heat  which characterizes the ability of the material to regulate the temperature within the materials. A good insulator typically has a higher specific heat because it takes time to absorb more heat before it actually heats up  to transfer the heat. In contrast, a good conductor has a lower specific heat, requiring very little heat energy to heat the materials. This means that the heat will be conducted rapidly, and so it will have a high conductivity due to its low specific heat. The specific heat $c_V$ at constant density is determined by $c_V=T(\partial s/\partial T)_\rho$ with $s$ the entropy density. 

\begin{figure}[H]\centering
\includegraphics[width=0.42\linewidth]{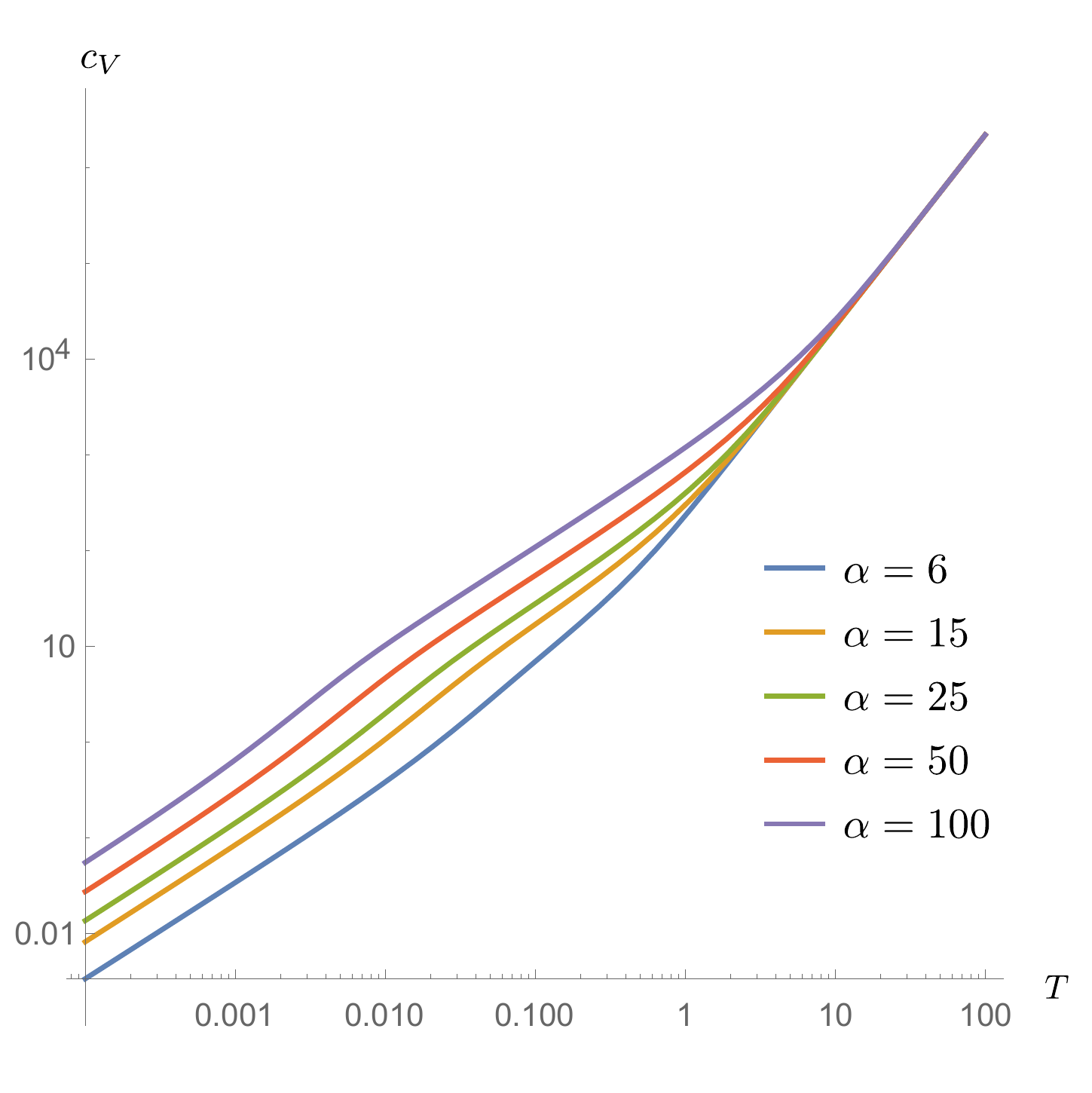}\quad
\includegraphics[width=0.41\linewidth]{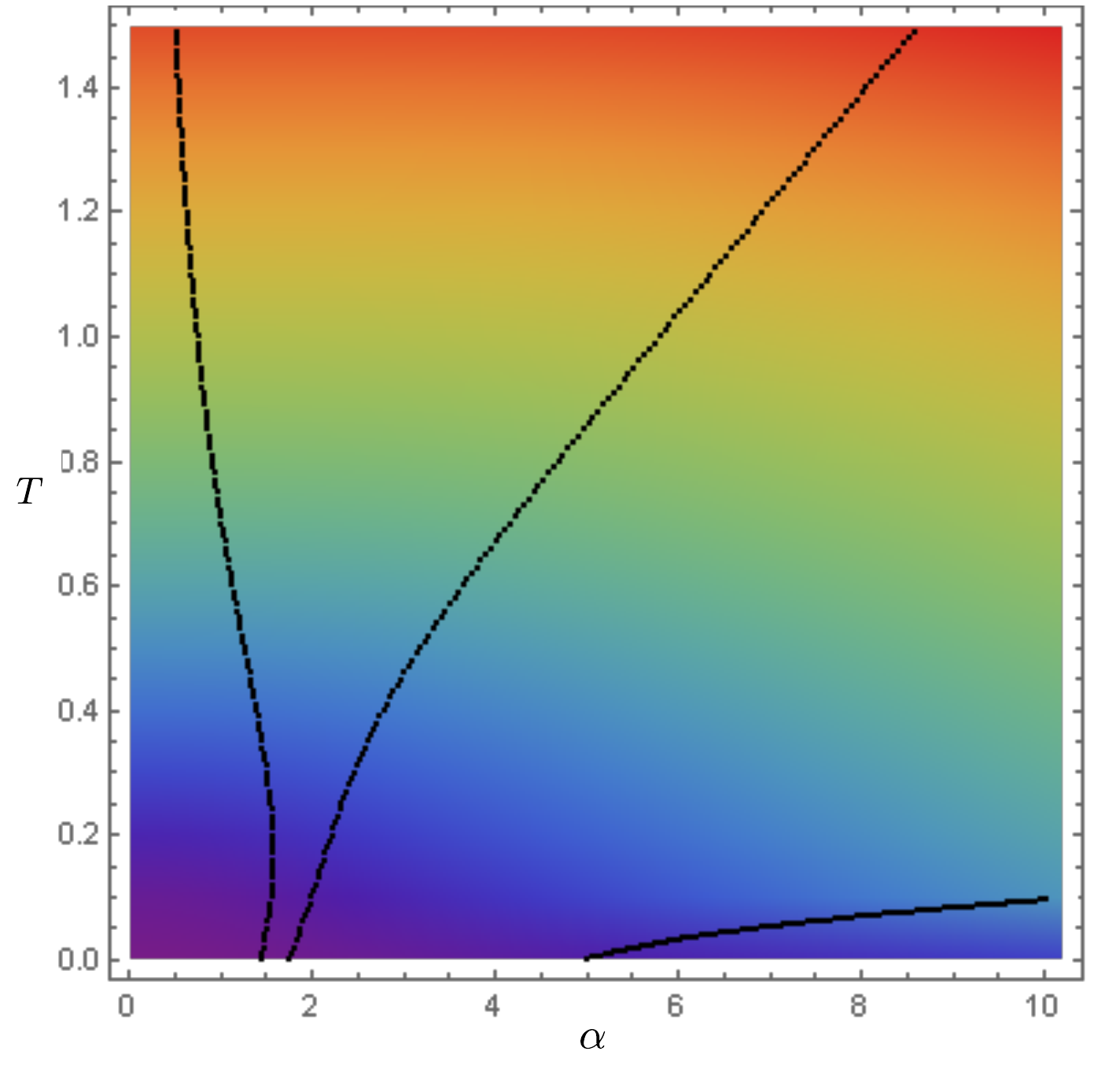}
\includegraphics[width=0.063\linewidth]{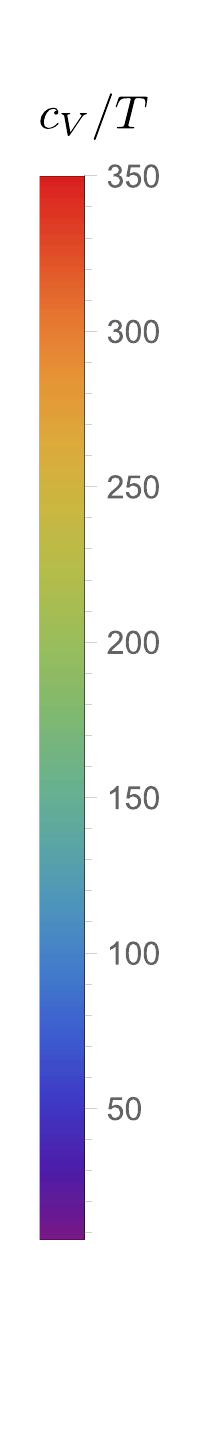}
\caption{Specific heat $c_V$ for the Linear model~\eqref{lmodel} at zero magnetic field. The behavior of $c_V$ as a function of $T$ is present in the left plot, and the density plot of $c_V/T$ in the temperature-disorder plane is shown in the right plot. Black dotted lines correspond to $\sigma_{DC}=0.1, 0.8, 1.2$ of Figure~\ref{figphase}, \emph{ i.e.} the phase boundaries for various phases.
We have fixed $\mathcal{K}=-1/6$ and $\rho=1$. }
\label{figcv}
\end{figure}

It is obvious that $c_V\sim T^2$ at high temperature, since $T\sim1/u_h$ in the high temperature limit~\eqref{highTem}. So we focus on the low temperature behavior of $c_V$. For specific, we consider the Linear model~\eqref{lmodel} which have all four phases we are interested in. Then, using~\eqref{ltem} and setting $B=0$, we find the entropy density as a function of temperature
\begin{equation}\label{entropyT}
s(T)=s_0+\frac{4\sqrt{2 \pi s_0}(s_0+2\pi\alpha^2\mathcal{K})}{\sqrt{24+\pi\alpha^4(1-6\mathcal{K})}}\,T +\mathcal{O}(T^2)\,,
\end{equation}
where $s_0=\pi(\sqrt{24+\alpha^4(1+6\mathcal{K})^2}+\alpha^2(1-6\mathcal{K}))/6$ is the entropy density at zero temperature and the charge density has been fixed to unit, $\rho=1$. Note that we always have a non-vanishing zero temperature entropy $s_0$ as $-1/6\leqslant\mathcal{K}\leqslant0$.
Then the specific heat at low temperature is given by
\begin{equation}
c_V=\frac{4\sqrt{2 \pi s_0}(s_0+2\pi\alpha^2\mathcal{K})}{\sqrt{24+\pi\alpha^4(1-6\mathcal{K})}}\,T+\mathcal{O}(T^2)\,.
\end{equation}
One immediately observes that the low-temperature specific heat scales as $T$ no matter the systems is in a metallic or insulating phase.
We point out that the linear-$T$  specific heat is reminiscent of a gas of fermions where a shell of thickness $T$ of occupied states above the Fermi surface contributing an energy $T$ each.\,\footnote{As a contrast, for a gas of free bosons in two spatial dimensions, the low temperature specific heat scales as $T^2$ (a sphere of volume $T^2$ of occupied states in momentum space, each with energy $T$).} The behavior of $c_V$ is present in Figure~\ref{figcv}. The temperature scaling behaviors of $c_V$ at low temperature ($\sim T$) and high temperature ($\sim T^2$) are confirmed by the left plot. One also finds that $c_V$ increases as $\alpha$ is increased, which means the insulating phase has a higher specific heat as we anticipate.
Compared with the phase diagram in the right plot of Figure~\ref{figphase}, the density plot of $c_V$ divided by $T$ in the temperature-disorder plane (right plot of Figure~\ref{figcv}) exhibits a significantly different behavior. Therefore, the specific heat is not a good probe to the metal-insulator transition.

\begin{figure}[H]\centering
\includegraphics[width=0.5\linewidth]{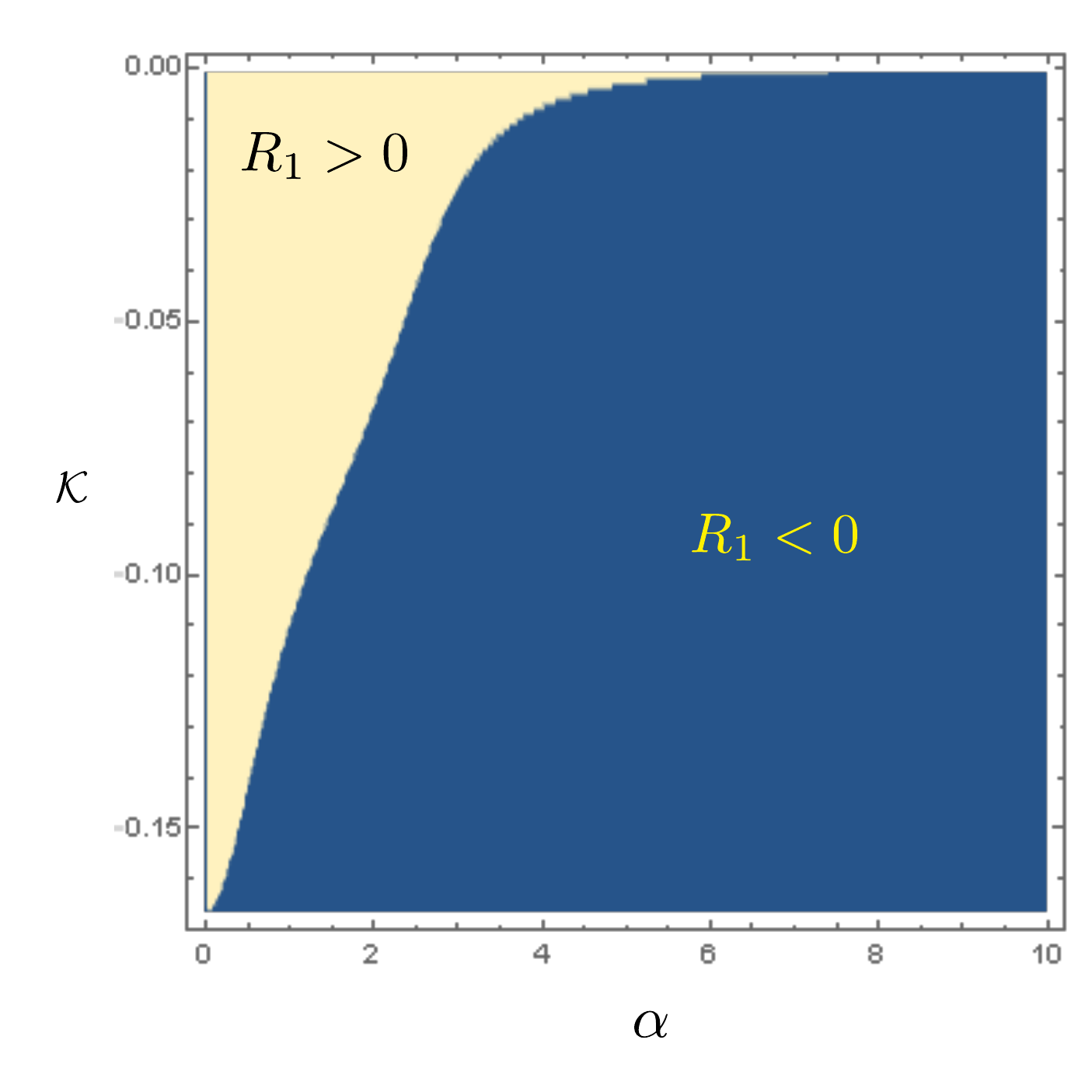}
\caption{Distribution of the sign of $R_1$ in~\eqref{RxxT} as a function of the disorder strength $\alpha$ and the coupling constant $\mathcal{K}$ for the Linear model~\eqref{lmodel}. We have set $B=0$ and worked in units with $\rho=1$.}
\label{fig:R1}
\end{figure}

 Another interesting feature is the linear of the entropy in~\eqref{entropyT} as well as the linear specific heat of Figure~\ref{figcv}, which clearly follows from the $AdS_2$ geometry near the extremal horizon. It was suggested by the authors of~\cite{Davison:2013txa} that a linear $T$ resistivity of cuprates’ strange metal is connected to a linear entropy density. We now check the temperature dependence of the resistivity at the same temperature regime. We find that at low temperatures the resistivity behaves as
\begin{align}\label{RxxT}
R_{xx}=R_0+R_1\,T+\mathcal{O}(T^2)\,,
\end{align}
where $R_0$ is a residual resistivity at zero temperature and $R_1$ is a constant that depends on $\alpha$ and $\mathcal{K}$. Therefore, there could be a linear $T$ behavior of $R_{xx}$ but with a residual resistivity $R_0$ at sufficiently low temperatures. Moreover, its coefficient $R_1$ can change sign, as seen from Figure~\ref{fig:R1}. Therefore, it suggests that the linear temperature dependence of the entropy does not guarantee a strange metal behavior of the cuprates. This issue is easy to understand by noting that the arguments of~\cite{Davison:2013txa} are not applicable in the present model. The inclusion of axions leads to a temperature dependent shear viscosity to entropy ratio~\cite{Hartnoll:2016tri}, thus $R_{xx}$ does not scale like the entropy density.

Since the specific heat can be in principle independent of the charge degrees of freedom, one might be not surprised that $c_V$ is not a good probe to the metal-insulator transition. Another observable that is closely related to charge carriers is the static charge susceptibility $\chi=\left(\partial \rho/\partial \mu\right)_{T}$ which measures the equilibrium response of the charge density to a change in the chemical potential $\mu$.  In the holographic setup, $\mu$ can be read off from the UV data via $\mu=A_t(u=0)$. For the Linear model~\eqref{lmodel}, we obtain
\begin{align}
\chi=\frac{\rho}{\mu}\left(1-\frac{32 \rho ^2 \sinh ^5(\xi ) \cosh (\xi )}{\xi \left[\rho ^2 \cosh (6 \xi )+\cosh (2 \xi ) \left(8 \alpha ^4 \mathcal{K} (6 \mathcal{K}-1)-9 \rho
   ^2\right)+8 \left(\alpha ^4 \mathcal{K} (6 \mathcal{K}+1)+\rho ^2\right)\right]}\right)^{-1} \,,
\end{align}
where $\xi=\alpha  \sqrt{-\mathcal{K}} \mu /\rho$ and we have turned off the magnetic field $B$. When $\alpha\rightarrow0$, one recovers the charge susceptibility for the Reissner-Nordstr\"{o}m black hole:
\begin{align}
\chi_{RN}=\frac{3 \rho  \left(\mu ^4+2 \rho ^2\right)}{\mu  \left(\mu ^4+6 \rho ^2\right)}\,,
\end{align}
with $\mu=\rho\, u_h$ in the present coordinate system~\eqref{solution}.

\begin{figure}[H]\centering
\includegraphics[width=0.42\linewidth]{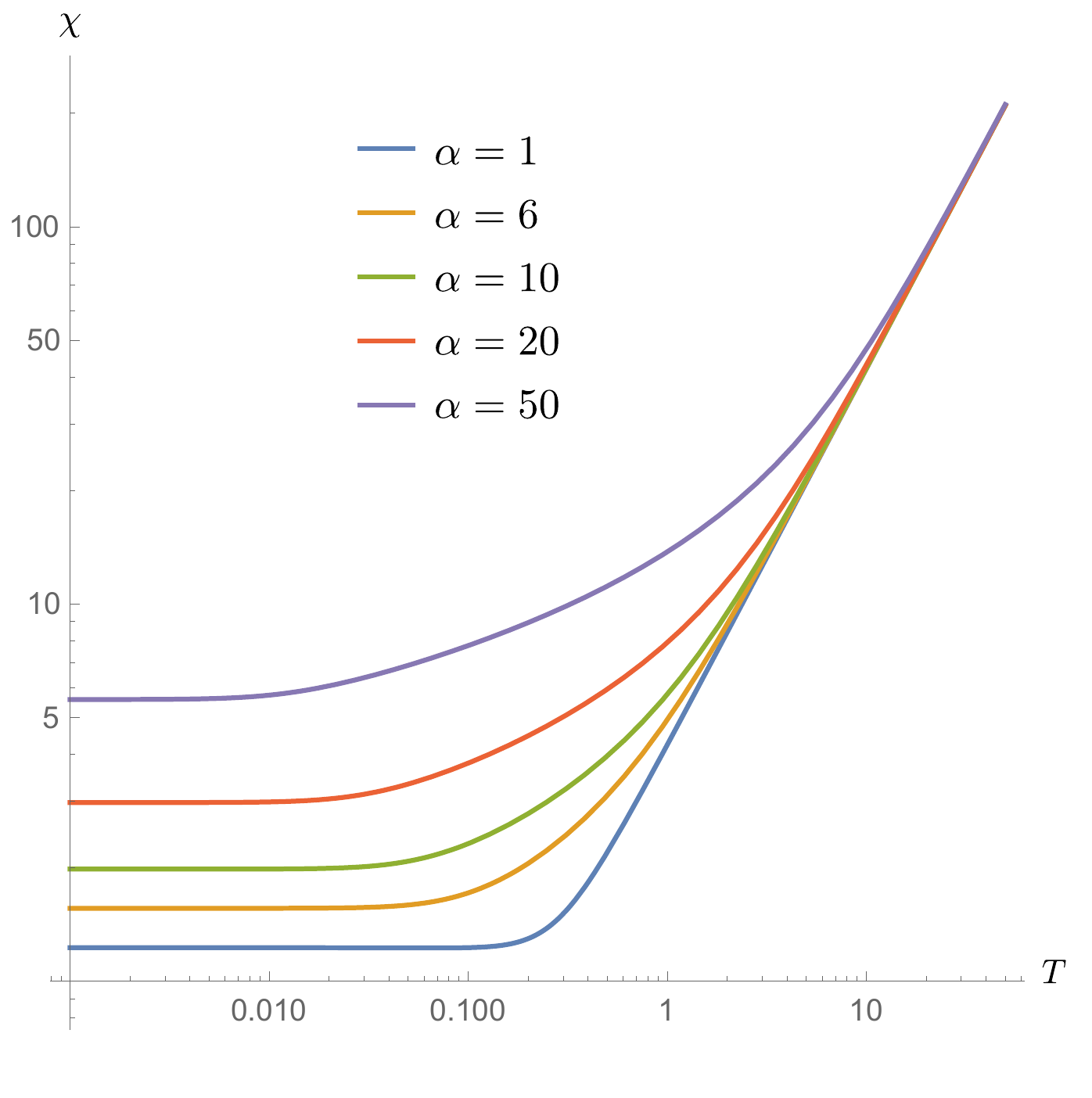}\quad
\includegraphics[width=0.41\linewidth]{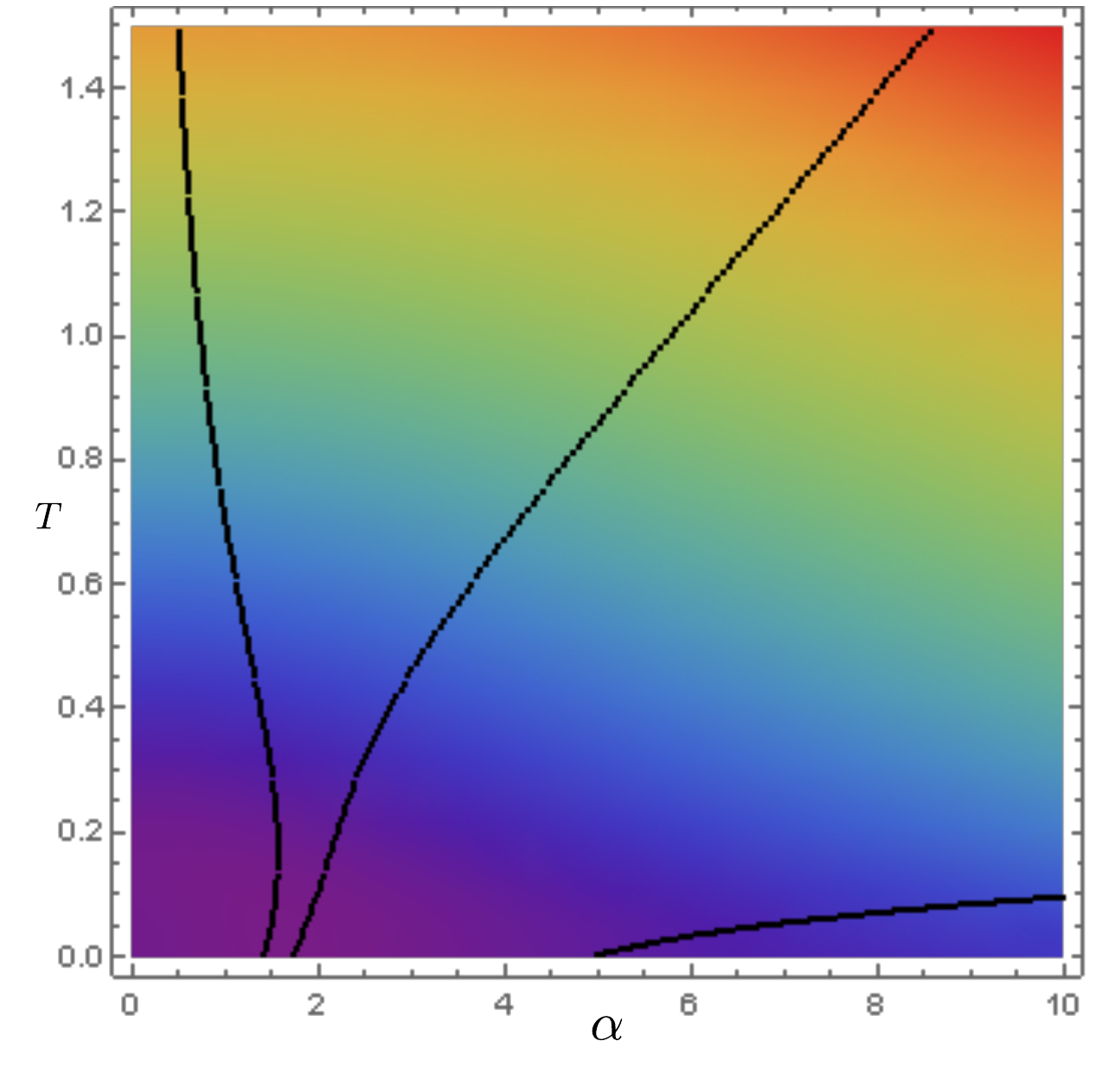}
\includegraphics[width=0.044\linewidth]{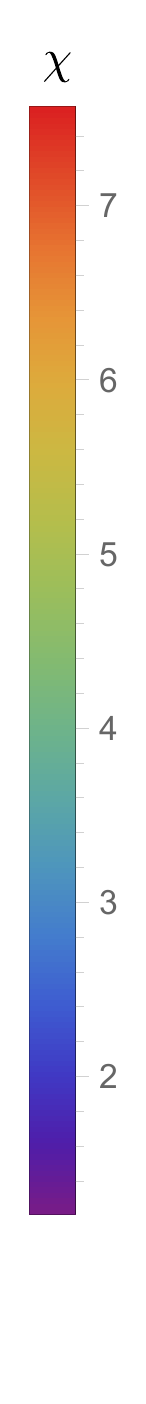}
\caption{ Static charge susceptibility $\chi$ for the Linear model~\eqref{lmodel} at zero magnetic field. The behavior of $\chi$ as a function of $T$ is present in the left plot, and the density plot of $\chi$ in the temperature-disorder plane is shown in the right plot. Black dotted lines denote the phase boundaries of metallic and insulating phases of Figure~\ref{figphase}. We have fixed $\mathcal{K}=-1/6$ and $\rho=1$. }
\label{figchi}
\end{figure}

The behavior of charge susceptibility for different temperature $T$ and disorder strength $\alpha$ is shown in Figure~\ref{figchi}. It is clear from the left plot that $\chi$ has a $T$-linear scaling at high temperatures and goes to a constant at low temperatures. One also finds that $\chi$ increases as the disorder strength $\alpha$ is increased, suggesting the insulating phase has a larger charge susceptibility. The density plot of $\chi$ in the temperature-disorder plane (right plot of Figure~\ref{figchi}) exhibits a significantly different behavior from the phase diagram of Figure~\ref{figphase}. So $\chi$ is not a good probe to characterize the metal-insulator transition. Compared with the density plots of $c_V/T$ and $\chi$, it is noteworthy that both share very similar structure in the temperature-disorder plane. In particular, $c_V/T$ and $\chi$ have the same temperature scaling at high and low temperatures. This suggests that in our present holographic quantum matter, the process of heat transfer is dominated by the charge degrees of freedom.

\section{Complexity}\label{Sec:Complexity}
In the framework of effective holographic theories for condensed matter we have studied a minimal example of a metal-insulator transition that can be driven by disorder, charge density as well as magnetic field.
A natural question is if there is some other probe to characterize different phases or phase transitions.
As we have just shown, the local quantities, such as free energy and specific heat, are not able to distinguish such phase transition. 
There are increasing evidences that the non-local observables from quantum information play a key role in strongly coupled quantum systems. One wonders if the complexity, as a non-local observable from quantum computation, could be a probe to the quantum phase transition as well as the finite temperature crossover in the present holographic theory.~\footnote{It has been shown that complexity can detect quantum phase transitions and shows signatures of revivals in a topological system~\cite{Ali:2018aon}. The behavior of complexity near a holographic phase transition has been studied \emph{e.g.} in~\cite{Momeni:2016ekm,Yang:2019gce,Ghodrati:2018hss,Guo:2019vni}.}

In a discrete quantum circuit system, the complexity measures how difficult it is to obtain a particular target state from a certain reference state. This concept has recently been generalized to continuous systems, such as complexity geometry~\cite{Brown:2016wib,Brown:2017jil,Nielsen:2006,Yang:2018tpo}, path-integral optimization~\cite{Caputa:2017urj,Bhattacharyya:2018wym}  and Fubini-study metric~\cite{Chapman:2017rqy}. There are two widely studied proposals to compute the complexity in holography: one is known as complexity-volume (CV) duality~\cite{Stanford:2014jda} and the other as complexity-action (CA) duality~\cite{Brown:2015bva,Brown:2015lvg}. In this section, we analyze the behavior of complexity during the disorder-driven metal-insulator transition. We focus on the CV conjecture and leave the CA conjecture and field theoretic methods for the future. Our main subject is the complexity of formation.

The complexity of formation is defined by the complexity of a thermal state from a vacuum state. In the holographic setup, it corresponds to the maximal volume of the codimension-one surface connecting the codimension-two time slices (denoted by $t_L$ and $t_R$) at two AdS boundaries:

\begin{equation}
\delta C=C_{TFD}-C_{vac}=\mathop{Max}\limits_{\partial_\Sigma=t_L\cup t_R}\left[\frac{\delta V(\Sigma)}{G_N L}\right]\,.
\end{equation}
Here $\delta V$ is the difference of the volume  between the AdS black hole and AdS vacuum, and $\delta C$ roughly measures the difficulty to build thermal field double (TFD) state from the vacuum state. 

For the background geometry~\eqref{solution}, the black brane is Reissner-Nordstr\"{o}m like with $u_{h}$ the outer horizon. The maximal volume is given by the $t_L=t_R$ slice in bulk, \emph{i.e.} the co-dimensional one surface connecting the two boundaries through the outer bifurcation horizon in the Penrose diagram. 
So the volume integral for the black brane simplifies to 
\begin{equation}
V_{B}=2\Omega \int_{0}^{u_{h}} \frac{du}{u^{3} \sqrt{f(u)}}\,,
\end{equation}
with $\Omega$ the area of the spatial geometry when both $u$ and $t$ are fixed. Note that there is a UV divergence associated with the asymptotic boundary $u\rightarrow 0$. This UV divergence is removed by considering the contribution from the volume of the AdS vacuum:
\begin{equation}
V_{0}=2\Omega \int_{0}^{\infty} \frac{du}{u^{3}}=2\Omega\left(\int_{0}^{u_{h}} \frac{du}{u^{3}}+\frac{1}{2 u_{h}^{2}}\right)\,.
\end{equation}
Then we obtain the difference of the volume
\begin{equation}\label{deltaV}
\begin{split}
\delta V=V_B-V_{0}&=2\Omega\left[\int_{0}^{u_h}\frac{1}{u^3}\left(\frac{1}{\sqrt{f(u)}}-1\right)du-\frac{1}{2 u_h^2}\right]\,,\\
&=\frac{2\Omega}{u_{h}^{2}}\left[ \int_{0}^{1} \frac{d\tilde{u}}{\tilde{u}^{3}}\left(\frac{1}{\sqrt{\tilde{f}(\tilde{u})}}-1\right)-\frac{1}{2}\right]\,,
\end{split}
\end{equation}
where we have introduced the dimensionless variable $\tilde{u}=u/u_{h}$ and $\tilde{f}(\tilde{u})=f(u)$ that is convenient for numerical analysis.

\begin{figure}[H]\centering
\includegraphics[width=0.48\linewidth]{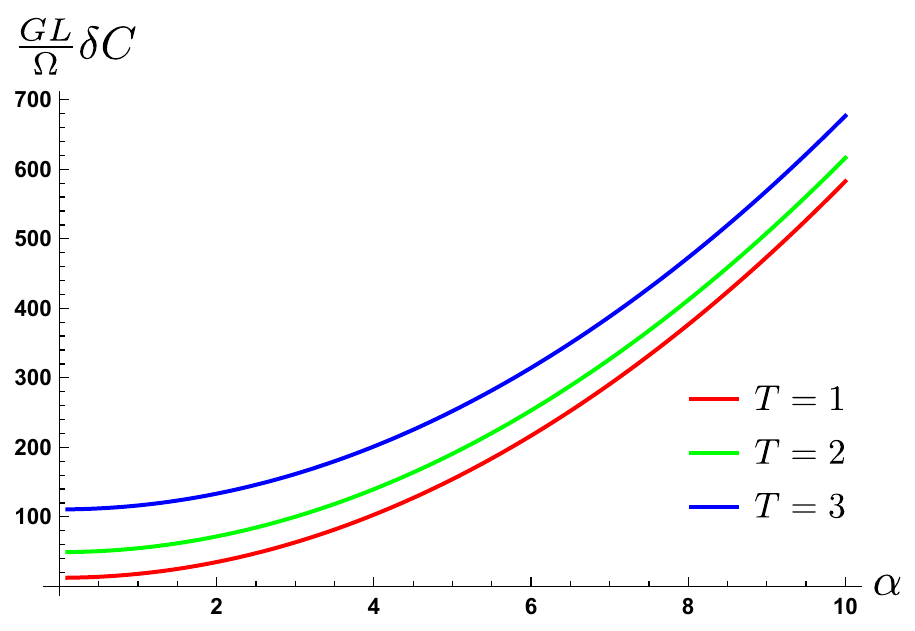}\quad
\includegraphics[width=0.48\linewidth]{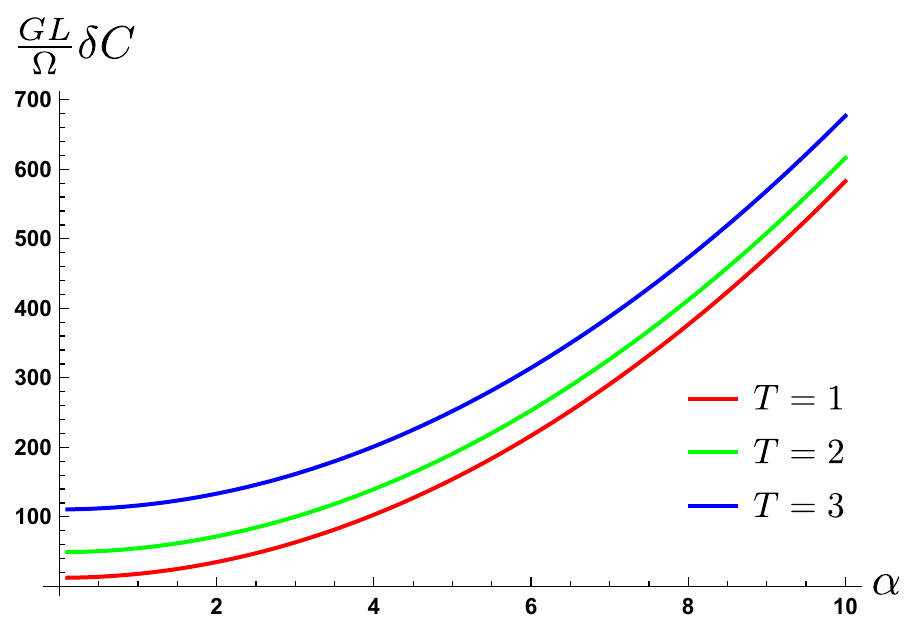}
\caption{Complexity of formation under the variation of $\alpha$ for the Exponential model with $\kappa=1/6$ (left) and the Linear model with $\mathcal{K}=-1/6$ (right). Different colors correspond to different temperatures. For other parameters we choose $\rho=1$ and $B=0$.}
\label{fig:deltaC}
\end{figure}

To study the behavior of complexity in different phases, we fix the temperature and vary the disorder strength $\alpha$. The complexity behavior is shown in Figure~\ref{fig:deltaC}. Both the Exponential model and the Linear model share very similar behavior. We find that the complexity of formation increases smoothly as we increase $\alpha$. So under the phase transition from a metallic state to an insulating state, the difficulty to build a thermal state from vacuum state
becomes larger. A heuristic picture is as follows. One anticipates that the wave-function of the charge carriers becomes localized towards the insulating phase. Therefore, it is harder for a ``gate" to couple different degrees of freedom and we need more gates to reach the desired state.  To have a deeper understanding of this behavior, one needs some microscopic mechanism of the metal-insulator transition and concrete quantum mechanical theory of complexity. 

\begin{figure}[H]\centering
\includegraphics[width=0.46\linewidth]{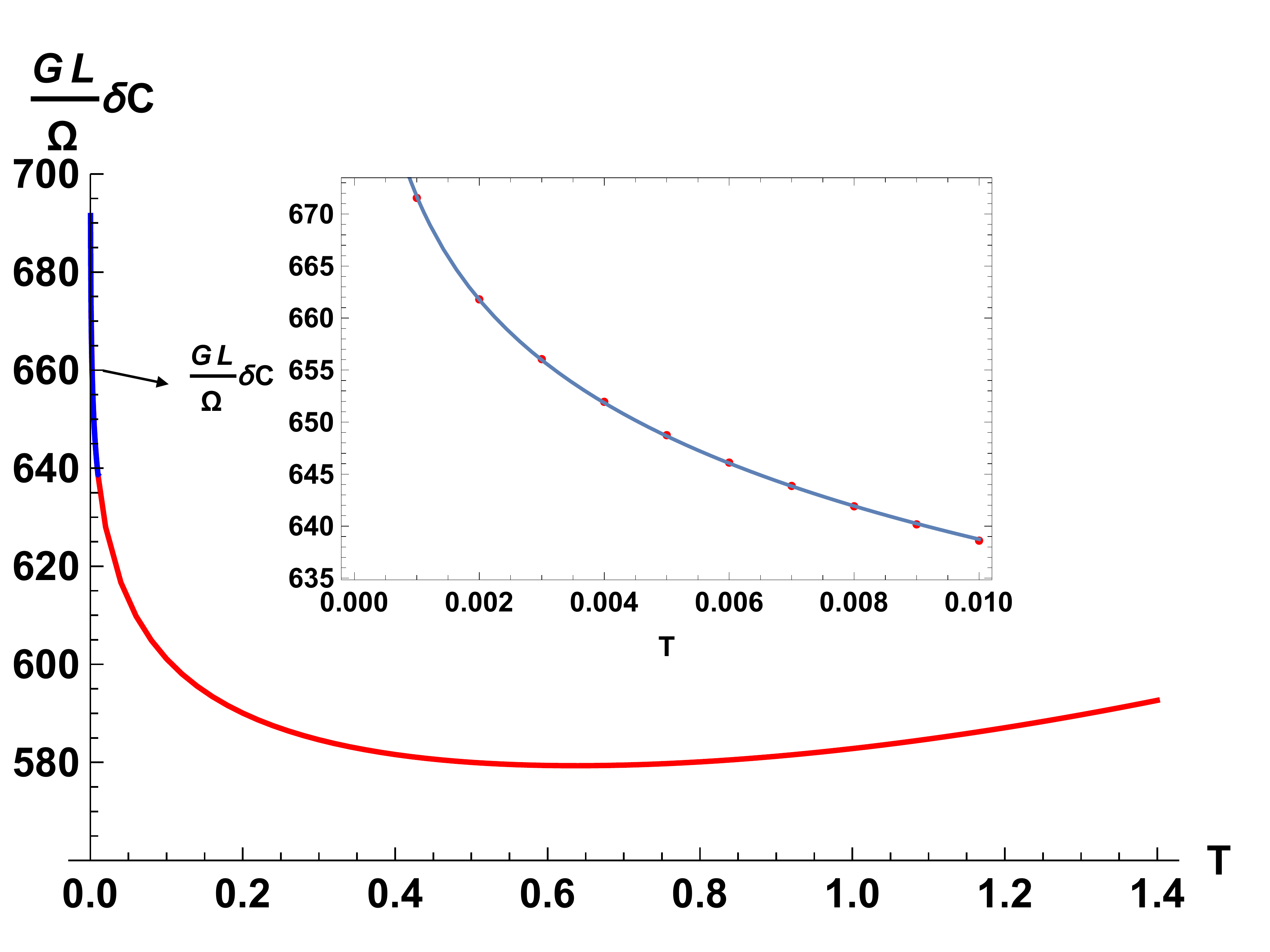}\quad\quad
\includegraphics[width=0.46\linewidth]{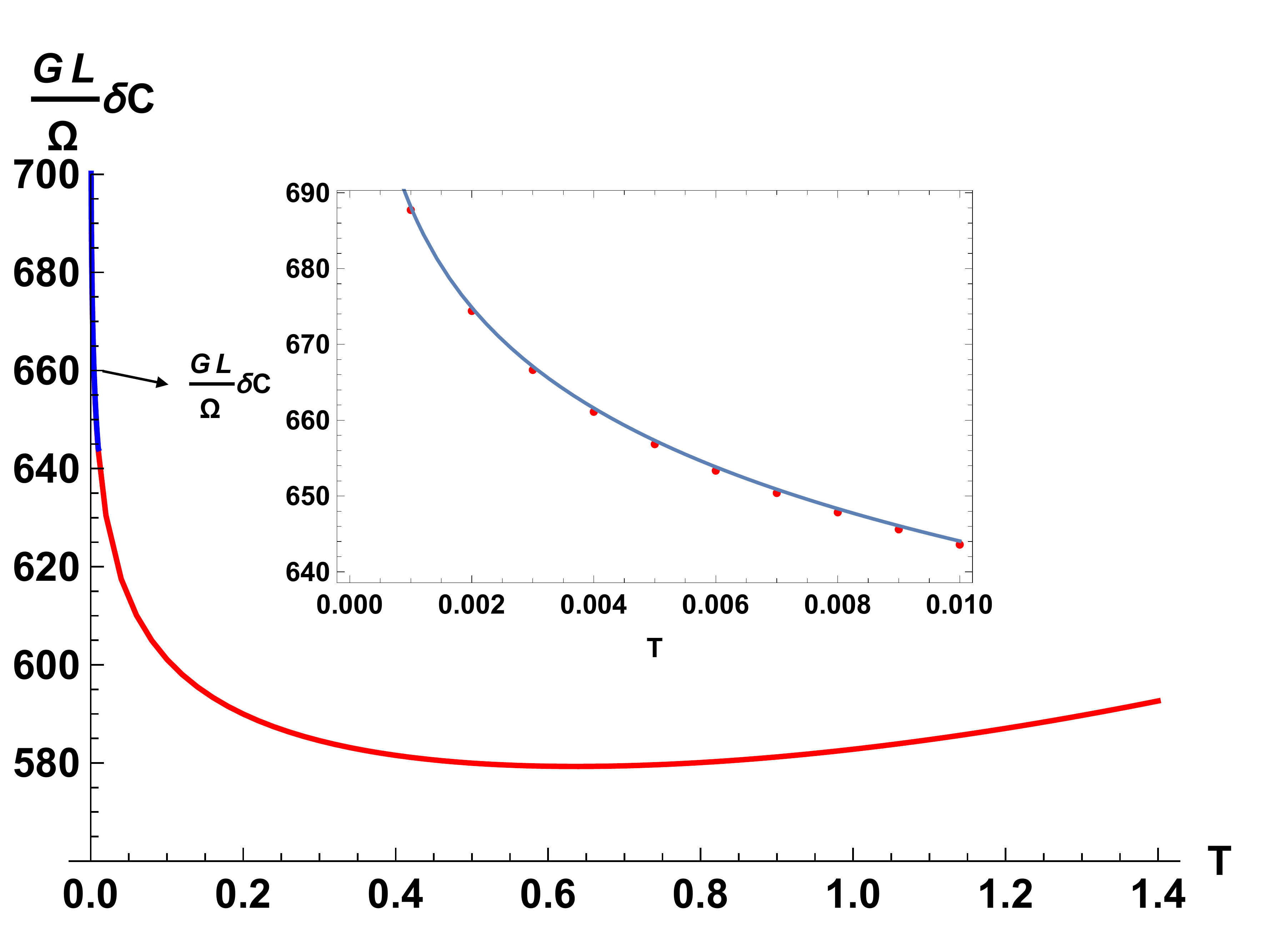}
\caption{Complexity of formation as a function of temperature for the Linear model with $\mathcal{K}=-1/6$. Left:  The charged case with $\rho=1$. Right: The neutral case with $\rho=0$. Inset: $\delta C$ diverges as $\ln(1/T)$ at low temperatures. For the charged case $\frac{G L}{\Omega}\delta C\approx 572.8-14.3\ln(T)$ and for the neutral case $\frac{G L}{\Omega}\delta C\approx 555.3-19.2\ln(T)$. For other parameters we choose $\alpha=10$ and $B=0$.}
\label{fig:3rdlaw}
\end{figure}

A more complicated situation arises at sufficiently low temperatures. Instead of a cancellation of the UV divergence, there is also a new IR divergence due to the infinitely long throat of the extremal geometry. To be more specific, we denote the horizon for the extremal background as $u_0$.  For the extremal case $f'(u_0)=0$ and near the extremal horizon one has
\begin{equation}\label{expandT0}
f(u)=\frac{1}{2}f''(u_0)(u-u_0)^2+\mathcal{O}((u-u_0)^3)\,.
\end{equation}
Therefore the near horizon geometry takes the form of $AdS_{2}\times R^{2}$.
Then we obtain from~\eqref{deltaV} that 
\begin{equation}
\delta V(T=0)=-\frac{2\sqrt{2}\Omega}{u_0^3\sqrt{f''(u_0)}}\ln(u_0-u)+\dots\,,
\end{equation}
near the extremal $AdS_2$ region in the far IR. So there is a logarithmic IR divergence for the $T=0$ case. While for finite $T$ case, the volume near the horizon region is finite. This means that the corresponding ``extremal" states at zero temperature are infinitely complex compared to the finite temperature states, and therefore no physical physical process is able to produce the extremal states in a finite amount of time. This feature is known as ``Third Law of Complexity" that was proposed in~\cite{Carmi:2017jqz}. The authors of~\cite{Carmi:2017jqz} considered a particular charged black hole geometry. It can be easily seen that this behavior also happens for the neutral case as long as the disorder is included, for which the near horizon expansion~\eqref{expandT0} still holds. The temperature dependence of $\delta C$ as a function of temperature for both charged (left panel) and neutral (right panel) cases is presented in Figure~\ref{fig:3rdlaw}. Form our numerics, we find that the complexity of formation diverges as $\ln(1/T)$ as the temperature goes to zero. So our study gives a further test for the complexity third law in a neutral background. 

Since $\delta C$ diverges logarithmically at zero temperature limit, it is not able to characterize the metal-insulator quantum phase transition at zero temperature. Note that we have defined the complexity of formation from the AdS vacuum state. It is possible and interesting to define a reference state other than the AdS vacuum to make the corresponding complexity of formation to be free from the IR divergence. In order for the new complexity to be able to characterize the metal-insulator transition, it should be at least sensitive to the disorder strength $\alpha$.  We leave the study of this possibility for the future.

\begin{figure}[H]\centering
\includegraphics[width=0.42\linewidth]{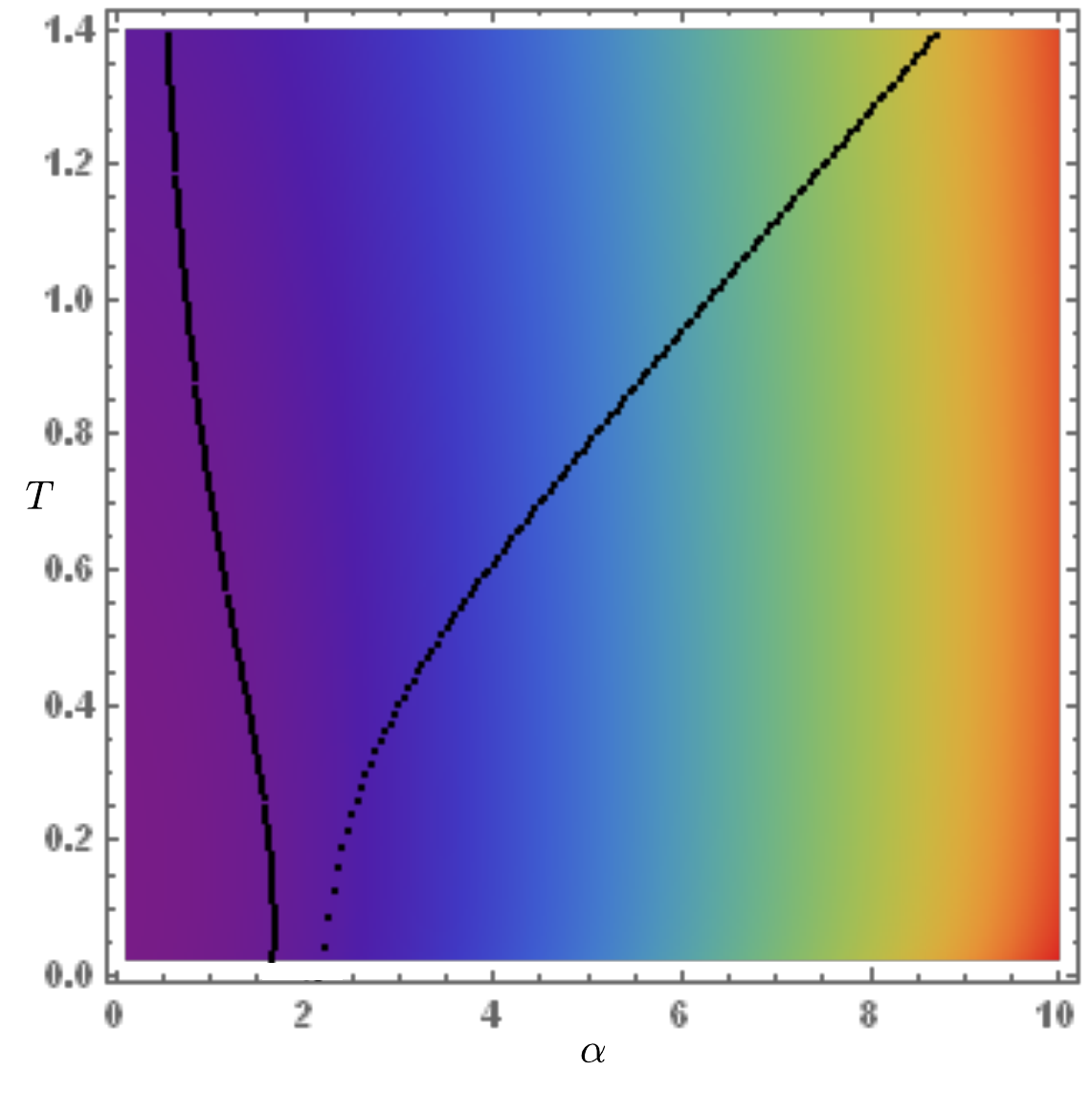}\quad
\includegraphics[width=0.055\linewidth]{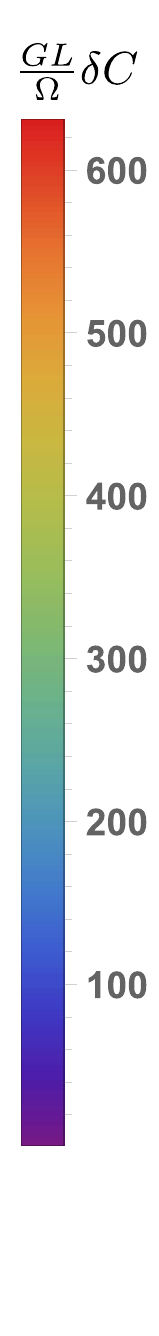}
\includegraphics[width=0.42\linewidth]{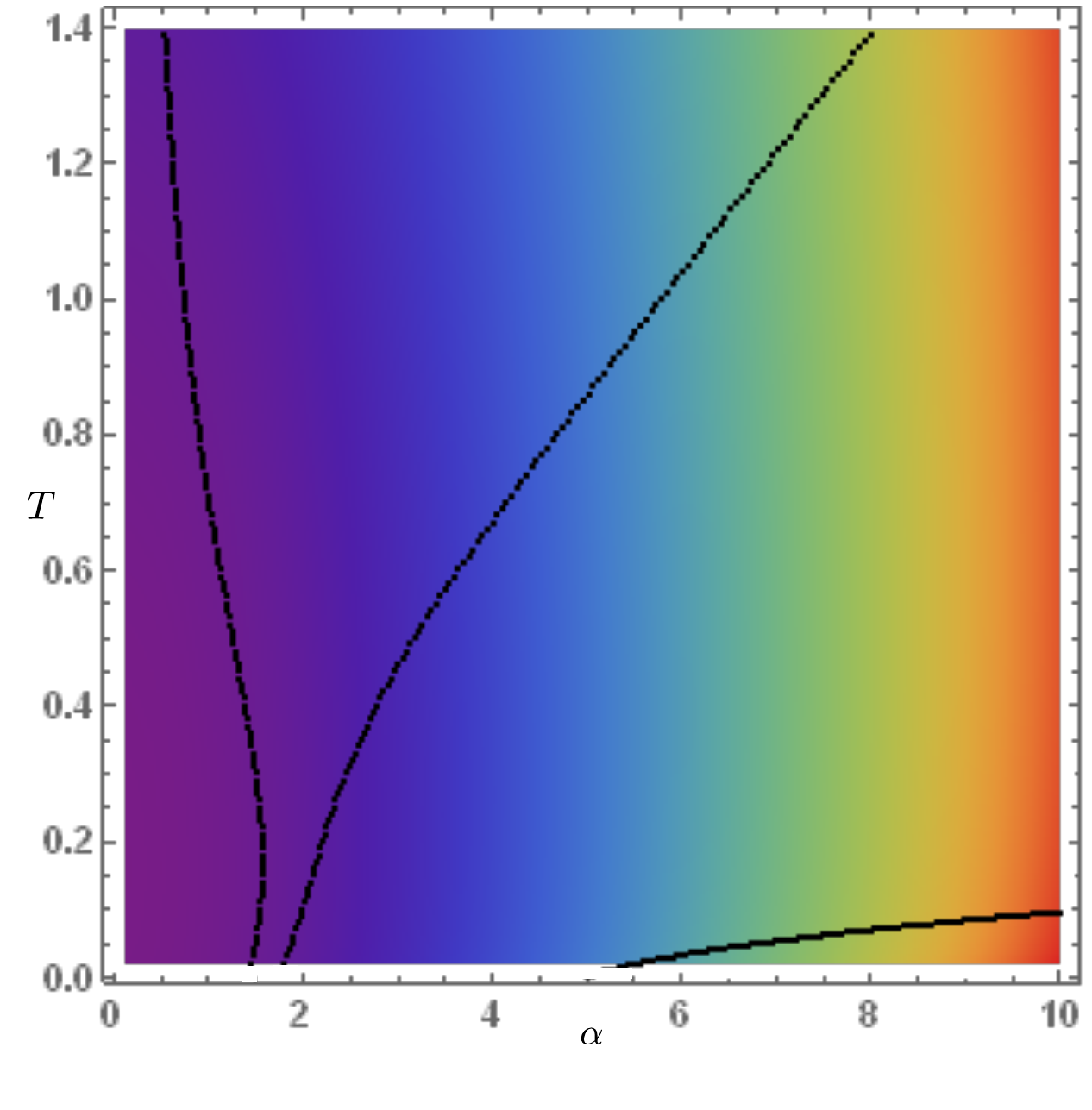}
\caption{Complexity of formation under the variation of $\alpha$ and $T$ for the Exponential model with $\kappa=1/6$ (left) and the Linear model with $\mathcal{K}=-1/6$ (right). The phase boundaries are denoted by the dashed lines corresponding to $\sigma_{DC}=0.1, 0.8, 1.2$ of Figure~\ref{figphase}. We consider the parameter space with $T/\sqrt{\rho}>0.02$ to avoid the divergence of $\delta C$ at low temperatures. For other parameters we choose $\rho=1$ and $B=0$.}
\label{fig:Cdensity}
\end{figure}

Next, it is also interesting to investigate the complexity behavior across the phase space at finite temperature. We show the complexity of formation with respect to temperature $T$ and disorder strength $\alpha$ in Figure~\ref{fig:Cdensity}. To avoid the divergence of $\delta C$ at zero temperature, we consider the parameter space with $T/\sqrt{\rho}>0.02$. One can see that $\delta C$ changes smoothly across different phase boundaries and is more sensitive to the disorder strength than the temperature. However, its patten is quite different from the behavior of conductivity in Figure~\ref{figphase}, and therefore not a good probe to the metal-insulator phase transition.

From above analysis we find that complexity of formation changes continuously across phase transitions, and there is no clear characteristic behavior of complexity in different phases. Although the complexity has different values in different phases, it seems not to be a good probe to the metal-insulator phase transition. In particular, $\delta C$ diverges in low temperature limit, and thus can not be a probe for ground state physics. Whether there is a refined version of complexity that is a good probe to the phase transition is an interesting open problem. Furthermore, it will be also interesting to examine the complexity growth rate during the phase transition and to check if the Lloyd’s bound is violated or not. We shall leave them as a future work.

\section{Conclusion and Discussion}\label{Sec:Conclusion}

We have investigated magnetotransport in a minimal holographic setup of a two dimensional metal-insulator transition and uncovered some interesting features, shedding light on this interesting transition and the physical mechanism that drives it.

Thanks to the homogeneous and isotropic of the geometry, we can solve the background equations of motion exactly and obtain the generic DC conductivities by means of black hole horizon data only, see~\eqref{sigxx} and~\eqref{sigxy}. Without referring to the details of coupling functions $Y(X)$ and $V(X)$, we are able to obtain some non-trivial constraint on the theory parameter by considering the weak disorder limit $\alpha\rightarrow 0$. More precisely, to avoid a negative diagonal conductivity $\sigma_{xx}$, the parameter $k$ in the weak-disorder expansion of $Y(X)$~\eqref{YVexpansion} must fall within the range~\eqref{constk}. For the Exponential model~\eqref{emodel}, the value of $\kappa$ was chosen to be $\kappa=0.5$ in~\cite{Baggioli:2016oqk}, corresponding to $k=0.5$, which is obviously outside the allowed parameter range. Indeed, as we showed explicitly in ,\ref{figevsalpha}, there is an unphysical negative region for $\sigma_{xx}$ in presence of a background magnetic field. We reexamined this model with $\kappa$ in the new parameter range~\eqref{constk} and found that there is no good insulator phase at all. In this sense, the Exponential model is not at good example for describing holographic metal-insulator transitions. Another model with a linear coupling~\eqref{lmodel} still works well in the presence of a magnetic field. As the strength of disorder is increased, the presence of a metal-insulator transition is manifest, see  Figure~\ref{figlDC}. From the optical conductivity in Figure~\ref{figlAC}, one finds a clear disorder-driven transition from a coherent metal with a sharp Drude peak to a good insulator with a tiny or vanishing DC conductivity at zero temperature. We have constructed the temperature-disorder phase diagram in Figure~\ref{figphase}. While in the present study we have focused on the transport, there may be stronger constraints which come not only from transport coefficients, but from the full two point functions and in particular the positivity of the spectral density. A complete analysis on the spectral densities is necessary to argue that these models are solid, which beyond the scope of this work.~\footnote{We thank Elias Kiritsis for raising this point.}

We have showed explicitly why $Y'(X)<0$ plays the key role in the metal-insulator transition by considering the high temperature limit. Scaling of an appropriate physical observable is one of the hallmarks of a phase transition. In the present work we uncovered the scaling behavior for the resistivity $R_{xx}$ near the phase transition, see Figures~\ref{fig:B0},~\ref{fig:nu} and~\ref{fig:alpha} driven by the charge density, magnetic field and disorder strength, respectively.  $R_{xx}(T)$ curves for different values of charge density $\rho$ can be made to overlap by the scaling parameter $T_0$ along the $T$ axis, which yields a collapse of the data onto two curves: an insulating branch for $\rho<\rho_c$, and a metallic branch for $\rho>\rho_c$. The parameter $T_0$ approaches zero at the critical charge density $\rho_c$, and increases as a power law $T_0\sim|\rho-\rho_c|^{1/2}$ both in metallic $(\rho>\rho_c)$ and insulating $(\rho<\rho_c)$ regions. We also found that the metallic and insulating curves are mirror symmetry in the high temperature regime: $R_{xx}(\rho-\rho_c,T)=1/R_{xx}(\rho_c-\rho, T)$. Our observations means that the mechanism responsible for the temperature dependence of conductivity on both insulating and metallic sides of the transition would be the same. We have also found similar scaling behavior for $R_{xx}$ in a magnetic field corresponding to a Landau-Level filling factor $\nu=3/2$. It suggests that in our holographic matter the metal-insulator transition at zero magnetic field and at fixed Landau-Level filling factor might be controlled by the same physical mechanism, or it would originate with some fundamental feature that is common to both.

In condensed matter physics, the observed scaling and mirror symmetry have been considered as a consequences of a simple analysis assuming that a $T=0$ quantum critical point describes the metal-insulator transition~\cite{Dobrosavljevic:1997}. While it seems no quantum critical point in our present holographic setup, it is helpful to compare our result to the scaling theory of localization in~\cite{Dobrosavljevic:1997}.
In particular, within quantum critical region the power-law exponent is $T_0\sim|\rho-\rho_c|^{z\nu_1}$ with $z$ the dynamical critical exponent and $\nu_1$ the correlation length exponent~\cite{Dobrosavljevic:1997}. A number of experiments have yielded scaling exponents that are different from our holographic model, suggesting that they should be in different universality classes. The scaling exponent $1/2$ of $T_0$ from our holographic setup is partially due to that the bulk geometry is asymptotically AdS.
A straightforward way to obtain a different scaling exponent in holography is to work with geometries that violate hyperscaling--describing an anomalous scaling of the free energy parametrized by $\theta$--and/or exhibit non-relativistic Lifshitz scaling with the dynamical critical exponent $z$. One could also consider different types of nonlinear electrodynamics which take into account nonlinear interactions between the charged degrees of freedom, in particular, the Dirac-Born-Infeld action.

There are four different phases in the temperature-disorder phase diagram as shown by Figure~\ref{figphase}: good metal, incoherent metal, bad insulator and good insulator. However, there is no genuine thermodynamic phase transition and  all phases share the same symmetries of the underlying theory, thus beyond the Landau classification. We have tried to see if there is any quantity that is able to distinguish these different phases. For local observable, we considered the specific heat $c_V$ and static charge susceptibility $\chi$. We found that $c_V/T$ and $\chi$ share very similar pattern in the temperature-disorder plane and change smoothly across different phases (see Figures~\ref{figcv} and~\ref{figchi}), but they exhibited a significantly different behavior from the conductivity in Figure~\ref{figphase}. For non-local candidate, we examined the behavior of complexity using the CV conjecture.  As shown in Figure~\ref{fig:Cdensity}, the complexity of formation also changes smoothly across different phase boundaries with its patten quite different from the conductivity in Figure~\ref{figphase}. Therefore it is not a good probe to the metal-insulator transition. It is still an open question to find a good probe to the metal-insulator transitions.\,\footnote{Instead of the complexity, another important non-local observable is the entanglement entropy, for which its gravity dual is known as Ryu-Takayanagi formula~\cite{Ryu:2006bv}. It has been shown to be a good probe to characterize the properties of phase transitions within the holographic scenario, see \emph{e.g.}~\cite{Albash:2012pd,Cai:2012sk,Cai:2012nm,Ling:2015dma,Kuang:2014kha,Bai:2014tla,Baggioli:2020cld}.} We also showed that there is a logarithmic IR divergence for the complexity of formation at low temperatures (see Figure~\ref{fig:3rdlaw}). This kind of divergence also happens for the neutral case once the disorder is considered. Our study provided a further test for the complexity third law~\cite{Carmi:2017jqz} in a neutral background. While the complexity third law has been confirmed from the holographic viewpoint, demonstrating it from the dual field theory side (such as tensor network) is also a very interesting direction.

Our present study focused on a dual system in two spatial dimensions, it is interesting to consider the case in three dimensions and to see if there is a similar behavior near the metal-insulator transition. Furthermore, insight from holography has also been given into various bounds and possible universality, it would be helpful to test if the present model against the aforementioned conjectured bounds~\cite{Figueroa:2020tya,Baggioli:2020ljz}. In our present study the disorder was introduced through a ``mean field" approach, where translational symmetry is broken, but the spacetime geometry is homogeneous. This is the simplest way to incorporate momentum relaxation, while there are some features concretely differ from the inhomogeneous setups (\emph{e.g.} commensurability~\cite{Andrade:2015iyf}).
It would be interesting to extend our studies to more complicated holographic systems which break translations without retaining the  homogeneity of the background such as~\cite{Cremonini:2019fzz,Cremonini:2016rbd,Cai:2017qdz,Andrade:2017ghg}. So far we limited ourselves to the electric conductivity, it is also worth studying the thermal response and the mechanical response.  We leave the study of all those issues and phenomenological consequences for the future.

\section*{Acknowledgements}
We would like to thank Matteo Baggioli, Kotetes Panagiotis, Sen Zhou, Sera Cremonini, Elias Kiritsis, Wei-Jia Li and Zhuo-Yu Xian for helpful conversations. The work was supported in part by the National Natural Science Foundation of China Grants No.11947302 and No.11991052.

\appendix

\section{DC Conductivity}\label{Sec:App}
In order to study the transport properties of the dual system we follow the method developed in~\cite{Donos:2014uba,Blake:2014yla}. The bulk perturbations take the form
\begin{equation}
\begin{split}
&\delta A_{x}=-E_{x} t+a_{x}(u),\quad \delta A_{y}=-E_{y} t+a_{y}(u), \\
&\delta g_{t x}=h_{t x}(u),\qquad \qquad\, \delta g_{t y}=h_{t y}(u)\,,\\
&\delta g_{xu}=h_{xu}(u),\qquad\quad\;\;\,  \delta g_{yu}=h_{yu}(u)\,,\\
&\delta \phi_{x}=\chi_{x}(u),\qquad\qquad\;\;\;  \delta \phi_{y}=\chi_{y}(u)\,,
\end{split}
\end{equation}
with $E_x (E_y)$ the electric field along the $x(y)$ direction. The linearized Maxwell equations~\eqref{eommax} can be shown to imply that 
\begin{equation}
\partial_u J^x= \partial_u J^y=0\,,
\end{equation}
with
\begin{equation}\label{current}
\begin{split}
J^x & =   \sqrt{-g}  Y[X] F^{u x}=[a_x'(u)+u^2 B\, h_{yu}(u)]f(u)Y(\alpha^2 u^2)-u^2\rho\, h_{tx}(u)\,, \\
J^y & =   \sqrt{-g}  Y[X] F^{u y}=[a_y'(u)-u^2 B\, h_{xu}(u)]f(u)Y(\alpha^2 u^2)-u^2\rho\, h_{ty}(u) \,.
\end{split}
\end{equation}
The asymptotic behaviors near the AdS boundary, $u=0$, show that $J^x$ and $J^y$ are nothing more than the electric currents in the dual field theory.
Since they are conserved along the radial direction, they can be calculated anywhere in the bulk. The strategy is to evaluate them at the horizon $u=u_h$, where the constraint of regularity imposes a non-trivial relation between the currents and the electric fields. In particular, the requirement for the fields to be regular at the horizon implies the following relation:
\begin{equation}\label{irconstraint}
\begin{split}
a_x'(u)=\frac{E_x}{f(u)},\quad\quad h_{tx}(u)=f(u) h_{xu}(u)\,,\\
a_y'(u)=\frac{E_y}{f(u)},\quad\quad h_{ty}(u)=f(u) h_{yu}(u)\,,
\end{split}
\end{equation}
with $\chi_{x}(u)$ and $\chi_{y}(u)$ finite at the horizon. 

Plugging~\eqref{irconstraint} into the $xu$ and $yu$ components of the linearized Einstein's equations~\eqref{eomgravity}, one finds that $h_{tx}(u_h)$ and $h_{ty}(u_h)$ are fixed by $E_x$ and $E_y$. Finally, we are able to compute $J^x$ and $J^y$ from~\eqref{current} using the horizon data and obtains
\begin{align}
\left(\begin{array}{l}{J^{x}} \\ {J^{y}}\end{array}\right)=\left(\begin{array}{ll}{\sigma_{x x}} & {\sigma_{x y}} \\ {\sigma_{y x}} & {\sigma_{y y}}\end{array}\right)\left(\begin{array}{l}{E_{x}} \\ {E_{y}}\end{array}\right)\,,
\end{align} 
where the longitudinal and Hall conductivities are given by
\begin{align}
\label{appsigxx}\sigma_{x x}=&\sigma_{y y}=\frac{\Omega(u_h) Y(\alpha^2 u_h^2)[\Omega(u_h)+ Y(\alpha^2 u_h^2)(B^2 Y(\alpha^2 u_h^2)^2+\rho^2)u_h^2]}{[\Omega(u_h)+B^2Y(\alpha^2 u_h^2)^3 u_h^2]^2+B^2\rho^2Y(\alpha^2 u_h^2)^4 u_h^4}\,,\\
\label{appsigxy}\sigma_{x y}=&-\sigma_{y x}=\frac{B\rho Y(\alpha^2 u_h^2)^3 u_h^2[2\Omega(u_h)+ Y(\alpha^2 u_h^2)(B^2 Y(\alpha^2 u_h^2)^2+\rho^2)u_h^2]}{[\Omega(u_h)+B^2Y(\alpha^2 u_h^2)^3 u_h^2]^2+B^2\rho^2Y(\alpha^2 u_h^2)^4 u_h^4}\,,
\end{align}
with $\Omega(u_h)=\alpha^2[m^2V'(\alpha^2 u_h^2) Y(\alpha^2 u_h^2)^2+\frac{u_h^4}{2}(B^2Y(\alpha^2 u_h^2)^2-\rho^2)Y'(\alpha^2 u_h^2)]$. The resistivity matrix $R$ is obtained by inverting the conductivity matrix $\sigma$:
\begin{equation}
R_{x x}=R_{y y}=-\frac{\sigma_{xx}}{\sigma_{xx}^2+\sigma_{yy}^2},\quad R_{x y}=-R_{y x}=-\frac{\sigma_{xy}}{\sigma_{xx}^2+\sigma_{yy}^2}\,.
\end{equation}
%



\end{document}